\documentclass[11pt]{article}
\pdfoutput=1

\usepackage[utf8]{inputenc}
\usepackage{multirow}
\usepackage{amsmath, amsfonts, amssymb}
\usepackage{comment}
\usepackage{graphicx}
\usepackage{psfrag}
\usepackage{amsthm}
\usepackage[usenames,dvipsnames,svgnames,table]{xcolor}
\usepackage{enumerate}
\usepackage{arydshln}
\usepackage{soul}
\usepackage{slashed}
\usepackage{subfig}
\usepackage{mathrsfs}
\usepackage{a4wide}
\usepackage{tikz}
\usepackage{tikz-cd}
\usetikzlibrary{shapes.geometric}
\usepackage{tcolorbox}

\usepackage{color}
\definecolor{dark-gray}{gray}{0.20}
\definecolor{gray}{gray}{0.30}
\definecolor{light-gray}{gray}{0.80}
\definecolor{dark-red}{rgb}{0.7,0,0}
\definecolor{dark-green}{rgb}{0.1,0.4,0}
\definecolor{dark-blue}{rgb}{0.3,0.3,0.7}
\definecolor{light-blue}{rgb}{0.8,0.8,1}
\definecolor{swamp}{RGB}{240, 199, 197}
\definecolor{landscape}{RGB}{180, 250, 199}
\definecolor{undecided}{RGB}{252, 252, 197}

\usepackage{pifont}

\usepackage{setspace}

\newcommand{\beq}{\begin{equation}}  \newcommand{\eeq}{\end{equation}}
\newcommand{\bal}{\begin{aligned}}   \newcommand{\eal}{\end{aligned}}
\newcommand{\be}{\begin{equation}}
\newcommand{\ee}{\end{equation}}

\def\be{\begin{equation}}
\def\ee{\end{equation}}
\def\bea{\begin{eqnarray}}
\def\eea{\end{eqnarray}}

\def\simleq{\; \raise0.3ex\hbox{$<$\kern-0.75em
      \raise-1.1ex\hbox{$\sim$}}\; }
   \def\simgeq{\; \raise0.3ex\hbox{$>$\kern-0.75em
      \raise-1.1ex\hbox{$\sim$}}\; }

\numberwithin{equation}{section}

\usepackage{jheppub}
\usepackage{hyperref}
\usepackage{cleveref}

\hypersetup{
	colorlinks=true,
	linkcolor=dark-blue,
	citecolor=dark-red,
	urlcolor=dark-green,
	linktoc=page
}

\theoremstyle{remark}

\crefname{appendix}{Appendix}{Appendices}

\title{\centering Tensionless String Limits in 4d Conformal Manifolds}

\author{Jos\'e Calder\'on-Infante$^1$ and}
\author{Irene Valenzuela$^{1,2}$}
\affiliation{$^{1}$CERN, Theoretical Physics Department, 1211 Meyrin, Switzerland}
\affiliation{$^{2}$Instituto de F\'isica Te\'orica UAM-CSIC and Departamento de F\'isica Te\'orica, Universidad Aut\'onoma de Madrid, Cantoblanco, 28049 Madrid, Spain}

\abstract{ 
Drawing on insights from the Swampland program, we initiate a classification of infinite distance limits in the conformal manifolds of 4d SCFTs. Each limit is characterized by a Hagedorn-like behavior of the large $N$ density of states, which we argue holographically correspond to different tensionless string limits. We focus on 4d large $N$ SCFTs with simple gauge groups, which exhibit an overall free limit at infinite distance within the conformal manifold. In this class of theories, only three types of weak-coupling limits arise. They are distinguished by the exponential rate $\alpha$ of the anomalous dimension of the higher-spin tower, which we find to be determined by the ratio of the central charges $a/c$. We compute the large $N$ partition function at the free point for all these SCFTs, and derive a universal expression for the Hagedorn temperature as a function of $\alpha$ (or, equivalently, of $a/c$), regardless of the gauge group or matter content. This Hagedorn-like density of states suggests that these weak-coupling limits correspond holographically to the tensionless limits of three different strings: the critical Type IIB string and two non-critical strings that arise exclusively in non-Einstein gravitational theories. Our findings are consistent with the Emergent String Conjecture when applied to theories with Einstein gravity at low energies. We also use our results to present a new argument for the absence of scale separation in the holographic AdS bulk dual of these 4d SCFTs. This argument is based on the existence of a bona fide 't Hooft limit, or equivalently, on satisfying the sharpened lower bound for the Distance Conjecture. 
}

\begin{document}
\hypersetup{pageanchor=false}
\makeatletter
\let\old@fpheader\@fpheader
\preprint{CERN-TH-2024-165 \\ \vspace*{-0.8cm} \hfill  IFT-UAM/CSIC-24-141}

\makeatother

\maketitle

\newcommand{\remove}[1]{\textcolor{red}{\sout{#1}}}
\newpage
\section{Introduction}

The pursuit of defining the space of consistent quantum theories lies at the heart of both the conformal bootstrap and Swampland programs. The conformal bootstrap seeks to map out the space of possible conformal field theories (CFTs) through consistency conditions such as symmetry, causality, and unitarity. On the other hand, the Swampland program aims to determine the criteria under which effective field theories (EFTs) can be consistently coupled to quantum gravity. These two approaches converge in the study of quantum gravity in anti-de Sitter (AdS) background, where the AdS/CFT correspondence allows the exploration of these theories either from the bulk perspective, aligned with the Swampland framework, or from the boundary perspective, using conformal bootstrap techniques.

\medskip

One particularly fruitful intersection of these programs arises when considering infinite distance limits in moduli spaces, which, within the context of AdS/CFT, correspond to infinite distance limits in the conformal manifold. The Swampland program provides powerful insights into the behavior of theories at these limits through the Distance Conjecture \cite{Ooguri:2006in}. Conversely, conformal bootstrap techniques offer rigorous tools for proving such conjectures within the framework of CFTs. In this work, we will see how the intuition gained from the Swampland program can in fact lead to novel discoveries about CFTs, particularly in the context of infinite distance limits in conformal manifolds.

From the Swampland perspective, the Distance Conjecture posits that an infinite tower of states becomes exponentially light as one approaches an infinite field distance limit. This statement has been translated into CFT language in \cite{Baume:2020dqd,Perlmutter:2020buo}, leading to the formulation of the CFT Distance Conjecture \cite{Perlmutter:2020buo}, which further refines the Swampland Distance Conjecture by specifying the nature of the tower of states. For CFTs in more than two spacetime dimensions, it predicts the existence of a tower of higher-spin (HS) operators whose conformal dimensions approach the unitarity bound exponentially fast with the Zamolodchikov distance, as the latter becomes parametrically large. This is confirmed by all known 4d CFTs, since infinite distance points in the conformal manifold correspond to free theories, for which a higher-spin symmetry emerges. Remarkably, part of this CFT Distance Conjecture, namely that theories with HS symmetry must lie at infinite distance in the conformal manifold, has been recently proven in \cite{Baume:2023msm} using conformal field theory techniques. A similar phenomenon occurs in two-dimensional CFTs, albeit with low-spin operators \cite{Kontsevich:2000yf,Roggenkamp:2003qp,Acharya:2006zw,soibelman}, and a recent work provides also a proof for the analogous part of the conjecture, namely that a point in which the scalar gap vanishes must be at infinite distance in the conformal manifold \cite{Ooguri:2024ofs}.

\medskip

This work explores the fruitful interplay between Swampland ideas and CFT techniques to learn new aspects about these infinite distance limits. We take the first steps towards a classification of infinite distance limits within 4d conformal manifolds, by focusing on gauge theories with simple gauge group and admitting a large $N$ limit. Within this context, we identify only three distinct types of limits, each associated with different values of the exponential rate for the conformal dimension of the HS tower. Swampland intuition suggests that these differing rates correspond to distinct microscopic origins for the HS towers in the bulk, yielding different behaviours for the high-energy density of states below the black hole threshold. 

To test this bulk interpretation, we compute the large $N$ thermal partition function of all these 4d SCFTs in the free limit. Our analysis includes any simple gauge group and any type of matter compatible with UV completeness, thereby extending previous results in the literature. As expected, we find that each type of limit is associated with a different Hagedorn-like density of states and obtain a universal formula for the Hagedorn temperature in terms of the exponential rate of the HS tower. This correspondence holds even after restricting the theory to its flavor singlet sector, which ensures that the CFT spectrum is sparse at large $N$. Separately, we also show that the exponential rate of the HS tower in the weak-coupling limit is determined by the ratio of the central charges $a/c$ and viceversa. Altogether, we conclude that these large $N$ SCFTs fall into three universality classes, which are characterized by the exponential rate of the HS tower in the weak-coupling limit, the Hagedorn temperature(s) at the free point, and the ratio of the central charges $a/c$.

Our results for the Hagedorn temperature are of general interest for CFTs and suggest that the three types of infinite distance limits correspond to different strings becoming tensionless in the bulk. Notably, our analysis reveals that the overall free limit of holographic CFTs always corresponds to the tensionless string limit of the ten-dimensional Type IIB string. The other two types of limits, however, seem to correspond to tensionless limits of non-critical strings, and only occur in non-holographic CFTs with $a\neq c$ to leading order at large $N$. 
They are non-critical in the sense that they propagate in less than ten spacetime dimensions, but still lead to a (non-Einstein) gravitational theory with an AdS factor.
One of these non-critical strings corresponds to the one proposed in \cite{Gadde:2009dj} as the dual of $\mathcal{N}=2$ SQCD (and very recently revisited in \cite{Dei:2024frl}), while the bulk origin of the other one is yet to be found.
Our results are consistent with the Emergent String Conjecture \cite{Lee:2019wij} if applied only to theories with an Einstein gravity description at low energies, while implies that this conjecture must be modified for non-Einstein theories.

Furthermore, we extend our study to include the interplay between these higher-spin towers and the Kaluza-Klein (KK) bulk towers associated with the internal space. By plotting the convex hull of these towers --a method that has proven valuable in recent studies of the Swampland in multi-dimensional field spaces \cite{Calderon-Infante:2020dhm,Etheredge:2022opl,Etheredge:2023odp,Calderon-Infante:2023ler,Etheredge:2023usk,Castellano:2023stg,Castellano:2023jjt,Etheredge:2024tok}-- we show that the sharpened lower bound proposed in \cite{Etheredge:2022opl} remains valid in AdS backgrounds, although in a non-trivial manner (which we double-check using integrability results in the case of $\mathcal N=4$ SYM). The exponential rate of the HS tower changes in strongly curved backgrounds compared to the flat space result, yet the bound is still preserved as long as the gravitational vacuum is not scale-separated. This observation provides a new argument for the absence of scale separation between the AdS space and the internal space in the gravitational dual of holographic 4d SCFTs with a conformal manifold and an overall free limit.

\medskip

The outline of the paper is as follows. In Section \ref{sec:review}, we review previous works on infinite distance limits in conformal manifolds and the CFT Distance Conjecture. We present the classification of infinite distance limits in the conformal manifold of 4d gauge theories with simple gauge group and admitting a large $N$ limit in Section \ref{sec:types-limits}. We then propose that they correspond to tensionless limits of three different strings, two of them being non-critical. To gather evidence for this claim, we compute the Hagedorn temperature of these theories at large $N$ and at the free point in Section \ref{sec:Hagedorn}. In Section \ref{sec:convex-hulls}, we include the dependence on $N$ to enlarge the field space and study the interplay between the higher-spin and the KK towers. Using these results, we provide a new argument for the absence of AdS scale separation in those theories that are holographic. We discuss extensions of our results beyond gauge theories with simple gauge group in Section \ref{sec:beyond}, as well as the bulk interpretation of one of the non-critical strings. Finally, we give a self-contained summary of our results and conclude with general comments and future directions in Section \ref{sec:conclusions}. Several technical derivations are presented in Appendix \ref{app:single-vs-multi}, \ref{app:orthogonality-relations} and \ref{app:flavor-singlet-computations}. 

\section{Review: Infinite Distances in Conformal Manifolds} \label{sec:review}

We are interested in $d$-dimensional local\footnote{In particular, a local CFT always has a local conserved stress tensor. Through AdS/CFT, this means that the bulk theory is coupled to gravity. This is relevant for the connection to the Swampland program.} and unitary CFTs with exactly marginal couplings. These marginal couplings parametrize the conformal manifold $\mathcal{M}$, which is endowed with a natural notion of a distance measured by the Zamoldchikov metric \cite{Zamolodchikov:1986gt}:
\beq
g_{ij}(t^i)=|x-y|^{2d}\langle\mathcal{O}_i(x)\mathcal{O}_j(y)\rangle \, ,
\eeq
where $t^i$ are the marginal couplings which act as local coordinates on $\mathcal{M}$, and $\mathcal{O}_i(x)$ their associated marginal operators. This conformal manifold is holographically mapped to the moduli space of the bulk dual theory.

Of special interest are points at infinite distance on $\mathcal{M}$ (namely, any trajectory approaching such point has infinite length as measured by the Zamolodchikov metric). 
According to the Distance Conjecture in the Swampland program, one expects the existence of an infinite tower of operators saturating the unitarity bound at these infinite distance points. 
From a CFT perspective, known examples of these infinite distance points in $d>2$ are limiting points where a subsector of the CFT becomes free. 
Consider for example $d = 4$ $\mathcal N = 2$ conformal manifolds. These are typically parametrized
locally by complexified gauge couplings and exhibit “cusp” points where one or more gauge
couplings are going to zero. The existence of several such cusps is linked to the existence of a duality, namely, the same SCFT can be described by different weakly-coupled frames in different regions on $\mathcal{M}$. In practice, the nature of the alleged tower of operators provides information about the 
 microscopic interpretation of the corresponding weakly-coupled description.

It is known that all free CFTs enjoy a higher-spin (HS) symmetry and viceversa \cite{Maldacena:2011jn,Stanev:2013qra, Boulanger:2013zza, Alba:2013yda, Alba:2015upa,
Hartman:2015lfa, Li:2015itl}: that is, a free CFT contains at least one Regge
trajectory comprised of an infinite tower of HS conserved currents. These HS currents are local primary operators of increasing spin $J>2$, whose
conformal dimensions saturate the unitarity bound, 
\beq
\Delta_J \geq d-2+J \, .
\eeq
Thus, whenever a subsector of the CFT gets free, the CFT contains an infinite tower of HS operators saturating the above bound.
This is indeed the tower of operators predicted by the Distance Conjecture, as studied in \cite{Baume:2020dqd,Perlmutter:2020buo}.

\medskip

We have now all the ingredients to review the CFT Distance Conjecture of \cite{Perlmutter:2020buo}, which is a stronger refinement of the Swampland Distance Conjecture (SDC), written purely in CFT language. The conjecture reads: 
\begin{center}
    \textbf{CFT Distance Conjecture} \\
   \emph{For any local and unitary CFT in $d>2$ dimensions:}\\
    \textbf{I.} All HS points in the conformal manifold are at infinite distance. \\
    \textbf{II.}  All CFTs at infinite distance in the conformal manifold are HS points. \\
    \textbf{III.}  The conformal dimensions of the HS currents approach the unitarity \\ bound exponentially fast in the geodesic distance.
\end{center}
On one hand, it is a stronger refinement of the SDC since it specifies the nature of the tower of operators by identifying them with higher-spin operators. On the other hand, it is more general as it applies to CFTs with arbitrary central charge, even if they do not have a weakly coupled Einstein gravity dual. In that sense, the Distance Conjecture seems to apply more generally than originally envisioned, since the evidence coming from known CFTs suggests that it is a general and intrinsic property of any local unitary CFT.

The first item of the CFT Distance Conjecture has been recently proven in \cite{Baume:2023msm} using conformal bootstrap techniques, namely the evolution equations of the conformal dimensions in the conformal manifold (see e.g. \cite{Komargodski:2016auf,Bashmakov:2017rko,Sen:2017gfr,Hollands:2017chb,Behan:2017mwi,Balthazar:2022hzb}) and properties of theories with weakly broken higher-spin symmetry. If the free point contains free vectors, then the third item is also proven \cite{Baume:2020dqd,Perlmutter:2020buo}, as we will review below. The second item remains the hardest to prove, although it is supported by all known examples so far.\footnote{A potential counterexample has been proposed and investigated in \cite{Bobev:2021yya,Giambrone:2021zvp,Guarino:2021kyp,Cesaro:2021tna,Bobev:2023bxs}. It features an infinite distance limit which so far seems not to correspond to a HS point. On the other hand, the AdS bulk has a KK tower becoming light exponentially with the distance in this limit. Thus, there is no contradiction with the original Distance Conjecture, which does not require the tower to contain HS fields. }

\medskip

In this work, we will not aim to prove the conjecture (nor assume it). We are simply interested in better characterizing the physics at the free points of known conformal manifolds. We initiate a classification of the possible infinite distance limits in known CFTs with $d>2$ and the microscopic nature of the emerging towers of higher-spin operators. 

For this purpose, we are going to focus on 4d SCFTs, since these are the only known $d> 2$ CFTs with non-compact conformal manifolds. This is because all known CFTs in higher dimensions are isolated fixed points, without a conformal manifold. Moreover, even if there are examples of CFTs in $d=3$ with conformal manifolds, they are all compact (see though \cite{Bobev:2021yya,Giambrone:2021zvp,Guarino:2021kyp,Cesaro:2021tna,Bobev:2023bxs} for a potential candidate of 3d CFT with a non-compact conformal manifold which deserves further investigation). We also need to consider 4d CFTs with at least four supercharges, since there are no known examples of non-supersymmetric conformal manifolds either.\footnote{See though \cite{Giambrone:2021wsm,Eloy:2023acy,Macpherson:2024frt,Eloy:2024lwn} for holographic candidates of non-supersymmetric conformal manifolds in various dimensions. In the absence of supersymmetry, we expect that these couplings will stop being exactly marginal at finite $N$, although this deserves further investigation.} Therefore, we are only left with 4d SCFTs if we want to classify the types of known infinite distance limits in conformal manifolds. This is what we will do from now on.

As explained in \cite{Perlmutter:2020buo}, all known infinite distance limits in 4d SCFTs correspond to weak-coupling limits where a subsector of the theory becomes free. This subsector must contain free vectors, since otherwise there is no marginal operator consistent with the superconformal algebra that can move us away from the HS point \cite{Perlmutter:2020buo}. As we take the infinite distance limit given by $\tau\rightarrow i\infty$, the marginal operator is then given to leading order in perturbation theory by $\mathcal{O}_\tau=\text{Tr} (F^2)$ where $F$ is the field strength and $\tau=\frac{4\pi i}{g^2}+\frac{\theta}{2\pi}$, with $g$ being the gauge coupling. By perturbation theory, one can then compute the Zamolodchikov metric, obtaining \cite{Perlmutter:2020buo}
\beq
\label{metrictau}
ds^2=\beta^2 \frac{d\tau d\bar\tau}{(\text{Im}\tau)^2}\quad \text{as }\text{Im}\tau\rightarrow\infty\, ,\quad \beta=24\, {\text{dim}\,G_{\text{free}}}  \, ,
\eeq
where $\text{dim}\,G_{\text{free}} $ is the dimension of the gauge group getting free in the limit. The proper distance from an arbitrary point $\tau'\in\mathcal{M}$ to the HS point is $d(\tau,\tau')\approx \beta\log \text{Im}(\tau)$ as $\text{Im}(\tau)\to i\infty$. The higher-spin symmetry must get broken at perturbative level at order $\mathcal{O}(g^2)$, so that the anomalous dimension of the higher-spin tower behaves as
\beq
\gamma_J \equiv \Delta_J - (d-2+J) \sim f(J) \, g^2 \sim f(J)\exp\left(-\frac1{\beta}\,d(\tau,\tau')\right)\quad \text{as }\text{Im}(\tau)\rightarrow\infty
\eeq
in terms of the Zamolodchikov distance, as required by Point III of the CFT Distance Conjecture. Here $f(J)$ is some analytic function of the spin. 

In the cases in which the CFT admits a large $N$ limit, this HS tower is composed of single-trace operators, such that they can be mapped to a infinite tower of higher-spin fields becoming exponentially light in the dual gravity theory as,
\beq
m_J\sim \exp\left(-\alpha \, \hat d(\tau,\tau')\right)\rightarrow 0 \, .
\eeq
In order to compare with previous results in the Swampland literature, the distance $d(\tau,\tau')$ is now written in Planck units and maps to the moduli space distance for massless scalars coupled to gravity in the Einstein frame. In this convention, the exponential mass decay rate in the gravity side becomes
\beq
\alpha=\sqrt{\frac{2c}{\text{dim}\,G_{\text{free}}} }\, ,
\label{alpha}
\eeq
where $c$ is the c-type central charge of the CFT and appears in the formula due to change of units from AdS units to Planck units. This expression is also valid when more than one gauge coupling is sent to zero at the same rate. Hence, the value of the exponential rate and, therefore, how fast the tower becomes light, depends on what fraction of the theory is getting free. The value of this exponential rate will play a key role in our work.  For completeness, notice that $\alpha$ is lower bounded, taking the smallest possible value when the whole theory gets free; that is, $\alpha\geq 1/\sqrt{3} $ for $\mathcal N=2$ and $\alpha\geq 1/2$ for $\mathcal N=1$ theories \cite{Perlmutter:2020buo}.

\medskip

Before finalizing the section, let us comment on how the story changes in 2d CFTs. Interestingly, a similar story holds there but the infinite tower of operators seems to be composed by low-spin operators (rather than HS) in all known examples. Similarly, it was previously conjectured in \cite{Kontsevich:2000yf,Roggenkamp:2003qp,Acharya:2006zw,soibelman} that every infinite distance point should have vanishing scalar gap. The converse has been recently proven in \cite{Ooguri:2024ofs}, namely that a point with $\Delta_{\text{gap}}=0$ is at infinite distance on $\mathcal{M}$, which is the analogous statement of Point I above for $d=2$. Furthermore, it is shown in \cite{Ooguri:2024ofs} that a subsector of the CFT in the $\Delta_{\text{gap}}\to 0$ is described by the decompactification limit of an emergent target space in the CFT. The parameter $\alpha$ is then lower bounded as\footnote{Recall that we use $\alpha$ to denote the exponential decay rate in reduced Planck units in the bulk. This corresponds to $\alpha_{\rm AdS}$ in the notation of \cite{Ooguri:2024ofs}.}
\begin{equation}
    \alpha \geq \sqrt{\frac{2c}{3 n}} \, ,
\end{equation}
where $c$ is the central charge and $n$ the number of dimensions in the emergent target space. Notice that the form of this bound is very similar to \eqref{alpha}. From the CFT viewpoint, infinite distance limits in 2d conformal manifolds are then naturally classified by the number of target space dimensions that are being decompactified, in relation to the value of the central charge.

Arguably, the equivalent to the overall free limit in $d=2$ would be to take the limit that maximizes $n$. Without supersymmetry, this yields $\alpha=\sqrt{2/3}$ which violates the sharpen bound of \cite{Etheredge:2022opl} and probably only arise  in highly non-holographic theories. With minimal supersymmetry one gets $\alpha =1$, which saturates the sharpen bound and suggests a critical emergent string limit.\footnote{This seems to be indeed the case for the large volume limit in AdS$_3\times S^3 \times K3$ or $T^4$ associated to sending ${\rm Vol}_{T^4,K3}\rightarrow \infty$. Using that ${\rm Vol} \sim g_s^2$ along the flat direction, and the supergravity expectations $T_{D1} \sim g_{s}^{-1/2}$ and $M_{\rm KK}\sim M_s ({\rm Vol})^{-1/4}$, one gets that: $\sqrt{T_{D1}} \sim M_{\rm KK} \sim \rm Vol^{-1/8}$ as $\rm Vol \to \infty$. Hence, the string and the KK modes decay at the same rate, which is indeed a signature of an emergent string limit.}

\section{Types of Free Limits in 4d SCFTs and Non-Critical Strings} \label{sec:types-limits}

Consider 4d local and unitary SCFTs with exactly marginal couplings. As reviewed above, all known infinite distance limits in these conformal manifolds correspond to weak-coupling limits in which one or several gauge couplings go to zero. At these points, a subsector of the CFT gets free and there is a tower of HS operators saturating the unitarity bound. As we approach the free point, the HS anomalous dimension decays exponentially with the Zamoldchikov distance. From the dual bulk perspective, these HS operators map to an infinite tower of HS fields whose mass behaves exponentially with the moduli space distance, and the exponential rate is given by \eqref{alpha}. Our goal is to initiate a classification of these infinite distance limits and determine the microscopic description of the tower of HS fields. From the gravity side, the intuition gathered from string theory examples is that the concrete value of the exponential mass decay rate is in one-to-one correspondence with the microscopic nature of the tower. Therefore, we will start by computing the possible values of this exponential rate \eqref{alpha}.

As a first step in this classification, we are going to restrict in this work to the case in which the 4d SCFT is Lagrangian and only contains one simple gauge factor, so that there is only one gauge coupling and everything gets free at the infinite distance limit. We will see that this mini-landscape of SCFTs is already very rich and will bring interesting results both for the Swampland and CFT communities. These 4d SCFTs with simple gauge group have been fully classified in \cite{Bhardwaj:2013qia} for $\mathcal N=2$ and \cite{Razamat:2020pra} for $\mathcal N =1$, and we summarize the results in Table \ref{table:SCFTs}, where we include only theories admitting large $N$ limit. Using these classifications, the concrete numerical value that $\alpha$ takes for each of the CFTs was already computed in \cite{Perlmutter:2020buo}. Interestingly, as noticed already in \cite{Perlmutter:2020buo}, only three possible values of $\alpha$ emerged in the large $N$ limit: 
\begin{equation} \label{eq:three-alphas}
    \alpha=\frac1{\sqrt{2}},\sqrt{\frac7{12}},\sqrt{\frac23} \, .
\end{equation}
This is the observation which motivated this paper, and sets the beginning of our new results leading to three types of free points and, consequently, three types of tensionless strings in this mini-landscape of SCFTs.

\setlength{\arrayrulewidth}{0.25mm} 
\renewcommand{\arraystretch}{1.3} 
\begin{table}[h]
\centering
    \addtolength{\leftskip} {-2cm}
    \addtolength{\rightskip}{-2cm}
\begin{tabular}{| c | c c c c c c c | c c |}
    \multicolumn{10}{c}{\boldmath $\alpha = 1/\sqrt{2}$} \\
    \hline
     SUSY \& Group & $n_{Ad}$ & $n_{A}$ & $n_{\bar A}$ & $n_{S}$ & $n_{\bar S}$ & $n_{F}$ & $n_{\bar F}$ & $c$ & $a$ \\
    \hline
    $\mathcal N =4 \; \; SU(N)$ & $3$ & $0$ & $0$ & $0$ & $0$ & $0$& $0$ & $\frac{N^2-1}{4}$ & $\frac{N^2-1}{4}$ \\
    \hline
    $\mathcal N =2 \; \; SU(N)$ & $1$ & $2$ & $2$ & $0$ & $0$ & $4$& $4$ & $\frac{3 N^2+3 N-2}{12}$ & $\frac{6 N^2+3 N-5}{24}$ \\
    \hline
    $\mathcal N =2 \; \; SU(N)$ & $1$ & $1$ & $1$ & $1$ & $1$ & $0$& $0$ & $\frac{3 N^2-2}{12}$ & $\frac{6 N^2-5}{24}$ \\
    \hline
    $\mathcal N =1 \; \; SU(N)$ & $2$ & $1$ & $1$ & $0$ & $0$ & $2$& $2$ & $\frac{6 N^2+3 N-5}{24}$ & $\frac{12 N^2+3 N-11}{48}$ \\
    \hline
    $\mathcal N =4 \; \; USp(2N)$ & $-$ & $0$ & $-$ & $3$ & $-$ & $0$& $-$ & $\frac{N (2 N+1)}{4}$ & $\frac{N (2 N+1)}{4}$ \\
    \hline
    $\mathcal N =2 \; \; USp(2N)$ & $-$ & $2$ & $-$ & $1$ & $-$ & $8$& $-$ & $\frac{6 N^2+9 N-1}{12}$ & $\frac{12 N^2+12 N-1}{24}$ \\
    \hline    
    $\mathcal N =1 \; \; USp(2N)$ & $-$ & $3$ & $-$ & $0$ & $-$ & $12$& $-$ & $\frac{4N^2+8N-1}{8}$ & $\frac{8 N^2+10 N-1}{16}$ \\
    \hline
    $\mathcal N =4 \; \; SO(N)$ & $-$ & $3$ & $-$ & $0$ & $-$ & $0$& $-$ & $\frac{N (N-1)}{8}$ & $\frac{N (N-1)}{8}$ \\
    \hline
\end{tabular}
\\ \medskip
\begin{tabular}{| c | c c c c c c c | c c |}
    \multicolumn{10}{c}{\boldmath $\alpha = \sqrt{7/12}$} \\
    \hline
    SUSY \& Group & $n_{Ad}$ & $n_{A}$ & $n_{\bar A}$ & $n_{S}$ & $n_{\bar S}$ & $n_{F}$ & $n_{\bar F}$ & $c$ & $a$ \\
    \hline
    $\mathcal N =2 \; \; SU(N)$ & $1$ & $1$ & $1$ & $0$ & $0$ & $N+2$ & $N+2$ & $\frac{7 N^2+3 N-4}{24}  $ & $\frac{13 N^2+3 N-10}{48}  $ \\
    \hline
    $\mathcal N =2 \; \; SU(N)$ & $1$ & $0$ & $0$ & $1$ & $1$ & $N-2$ & $N-2$ & $\frac{7 N^2-3 N-4}{24}  $ & $\frac{13 N^2-3 N-10}{48}  $ \\
    \hline
    $\mathcal N =1 \; \; SU(N)$ & $2$ & $0$ & $0$ & $0$ & $0$ & $N$ & $N$ & $\frac{7 N^2-5}{24}  $ & $\frac{13 N^2-11}{48}  $  \\
    \hline
    $\mathcal N =1 \; \; SU(N)$ & $1$ & $0$ & $1$ & $1$ & $0$ & $N-4$ & $N+4$ & $\frac{7 N^2-4}{24}  $ & $\frac{13 N^2-10}{48}  $ \\
    \hline
    $\mathcal N =1 \; \; USp(2N)$ & $-$ & $1$ & $-$ & $1$ & $-$ & $2N+6$ & $-$ & $\frac{14 N^2+15 N-1}{24}  $ & $\frac{26 N^2+21 N-1}{48}  $ \\
    \hline    
    $\mathcal N =1 \; \; USp(2N)$ & $-$ & $2$ & $-$ & $0$ & $-$ & $2N+10$ & $-$ & $\frac{14 N^2+21 N-2}{24}  $ & $\frac{26 N^2+27 N-2}{48}  $ \\
    \hline
    $\mathcal N =1 \; \; SO(N)$ & $-$ & $0$ & $-$ & $2$ & $-$ & $N-10$ & $-$ & $\frac{7 N^2-21 N-4}{48}  $ & $\frac{13 N^2-27 N-4}{96}  $ \\
    \hline
    $\mathcal N =1 \; \; SO(N)$ & $-$ & $1$ & $-$ & $1$ & $-$ & $N-6$ & $-$ & $\frac{7 N^2-15 N-2}{48}  $ & $\frac{13 N^2-21 N-2}{96}  $ \\
    \hline
\end{tabular}
\\ \medskip
\begin{tabular}{| c | c c c c c c c | c c |}
    \multicolumn{10}{c}{\boldmath $\alpha = \sqrt{2/3}$ }  \\
    \hline
    SUSY \& Group & $n_{Ad}$ & $n_{A}$ & $n_{\bar A}$ & $n_{S}$ & $n_{\bar S}$ & $n_{F}$ & $n_{\bar F}$  & $c$ & $a$ \\
    \hline
    $\mathcal N =2 \; \; SU(N)$ & $1$ & $0$ & $0$ & $0$ & $0$ & $2N$& $2N$ & $\frac{2 N^2-1}{6}  $ & $\frac{7 N^2-5}{24}  $ \\
    \hline
    $\mathcal N =1 \; \; SU(N)$ & $0$ & $0$ & $1$ & $1$ & $0$ & $2N-4$& $2N+4$ & $\frac{8 N^2-3}{24} $ & $\frac{14 N^2-9}{48}  $ \\
    \hline
    $\mathcal N =2 \; \; USp(2N)$ & $-$ & $0$ & $-$ & $1$ & $-$ & $4N+4$& $-$ & $\frac{N(4 N+3)}{6}  $ & $\frac{N(14 N+9)}{24}  $ \\
    \hline    
    $\mathcal N =2 \; \; SO(N)$ & $-$ & $1$ & $-$ & $0$ & $-$ & $2N-4$& $-$ & $\frac{ N (2 N-3)}{12}$ & $\frac{N (7 N-9)}{48} $ \\
    \hline
    $\mathcal N =1 \; \; SO(N)$ & $-$ & $0$ & $-$ & $1$ & $-$ & $2N-8$& $-$ & $\frac{4 N^2-9 N-1}{24}  $ & $\frac{7 N^2-12 N-1}{48}  $ \\
    \hline
\end{tabular}
\medskip
\caption{Mini-lanscape of large $N$ CFTs with simple gauge group. The theories are grouped in different tables by their shared value for $\alpha$, yielding three universality classes. For each theory, we indicate the amount of supersymmetry, the gauge group, the number of $\mathcal N=1$ chiral multiplets in the representation $R$ --denoted as $n_R$-- and the central charges $c$ and $a$. Dashes indicate that the representation does not exist for this gauge group. For more details on the available representations for $SU(N)$, $USp(2N)$ and $SO(N)$ groups, see sections \ref{sec:SU-case}, \ref{sec:USp-case} and \ref{sec:SO-case}, respectively.}
\label{table:SCFTs}
\end{table}
\renewcommand{\arraystretch}{1} 

\medskip

To show this, let us first re-write \eqref{alpha} in a more convenient way in terms of the two central charges of the CFT, namely $c$ and $a$. These central charges appear in the conformal anomaly (see e.g. \cite{Myers:2010tj}) as follows
\begin{equation}
\begin{split}
    \left< T^{\mu}_{\ \mu} \right> &= \frac{c}{16 \pi^2} W^2 - \frac{a}{16 \pi^2} G + \frac{a^\prime}{16 \pi^2} \nabla^2 R \, , \\
    W^2 &= R_{\mu \nu \rho \sigma} R^{\mu \nu \rho \sigma} -2 R_{\mu \nu} R^{\mu \nu} + \frac{1}{3} R^2 \, , \\
    G &= R_{\mu \nu \rho \sigma} R^{\mu \nu \rho \sigma} -4 R_{\mu \nu} R^{\mu \nu} + R^2 \, ,
\end{split}
\end{equation}
where $a^{\prime}$ is a scheme-dependent coefficient that we will not discuss further.  For Lagrangian SCFTs, they can be written in terms of the number of vector and hyper/chiral multiplets as
\begin{equation} \label{eq:central-charges}
\begin{array}{lll}
\mathcal N=2 : \ & \ c = \dfrac{n_v}{6} + \dfrac{n_h}{12} \, , \ & \ a = \dfrac{5n_v}{24} + \dfrac{n_h}{24}  \, , \\[10pt] 
\mathcal N=1 : \ & \ c = \dfrac{N_v}{8} + \dfrac{N_c}{24} \, , \ & \ a = \dfrac{3 N_v}{16} + \dfrac{N_c}{48}   \, .
\end{array}
\end{equation}
Using that $\text{dim}\,G=n_v$ and $\text{dim}\,G=N_v$, we can use this to write $\alpha$ as
\begin{equation} \label{alpha-nv-nh}
  \alpha = \sqrt{\frac{1}{3} + \frac{1}{6} \frac{n_h}{n_v}}  \quad \text{for }\mathcal N=2\ , \quad
  \alpha=\sqrt{\frac{1}{4} + \frac{1}{12} \frac{N_c}{N_v}}\quad \text{for }\mathcal N=1.
\end{equation}
Similarly, by inverting \eqref{eq:central-charges} and replacing the result we obtain 
\beq
\label{aac}
\alpha= \frac{1}{\sqrt{2}} \frac{1}{\sqrt{2\frac{a}{c}-1}} \, ,
\eeq
yielding the same result for $\alpha(a/c)$ regardless of the level of supersymmetry.
We then see that the value of $\alpha$ is completely determined by the ratio $a/c$. Notice that \eqref{aac} is not only valid for SCFTs with a simple gauge factor, but for any SCFT as long as we explore a limit in which the whole theory becomes free.

This way of writing the exponential rate $\alpha$ allows to translate the unitarity constraints on the ratio $a/c$ \cite{Hofman:2008ar} to bounds on $\alpha$. We have
\begin{equation} \label{eq:bounds}
\begin{array}{l l l l}
\mathcal N=2 : & \ \dfrac{a}{c} \leq \dfrac{5}{4} & \, \Longrightarrow & \, \alpha \geq \dfrac{1}{\sqrt{3}}  \, , \\[10pt]
\mathcal N=1 : & \ \dfrac{a}{c} \leq \dfrac{3}{2} & \, \Longrightarrow & \, \alpha \geq \dfrac{1}{2}  \, , \\[10pt]
\mathcal N=0 : & \ \dfrac{a}{c} \leq \dfrac{31}{18} & \, \Longrightarrow & \, \alpha \geq \dfrac{3}{2\sqrt{11}}  \, .
\end{array}
\end{equation}
We naturally recover the results of \cite{Perlmutter:2020buo} for the supersymmetric theories since these upper bounds on $a/c$ are saturated by a theory of free vectors (multiplets).\footnote{The ratio $a/c$ also satisfies lower bounds \cite{Hofman:2008ar}. For supersymmetric theories one has $a/c \geq 1/2$, yielding the trivial bound $\alpha < \infty$. Non-supersymmetric theories satisfy the weaker bound $a/c \geq 1/3$, which seems to be in conflict with $\alpha$ being real. However, there is no contradiction since \eqref{alpha} assumes a gauge theory, for which $a/c \geq 1/2$ is guaranteed.} Since the latter are not conformal at the interacting level, one can promote the bounds on $\alpha$ to strict inequalities as in \cite{Perlmutter:2020buo}. Notice that the $\mathcal{N}=2$ result automatically guarantees satisfying the sharpened bound for the Distance Conjecture proposed in \cite{Etheredge:2022opl}, even for finite $N$! This is not the case with less supersymmetry. Nevertheless, we have used the classification of $\mathcal{N}=1$ theories with simple gauge group and conformal manifold in \cite{Razamat:2020pra} to check that, remarkably, all of them satisfy the sharpened bound, regardless of having a large $N$ limit. Even though the gravitational bulk dual will not be weakly coupled in general, it is interesting that we find no violation of this bound for general SCFTs.

\medskip

But let us come back to the main focus of this paper and summarize the results obtained so far: Supersymmetric gauge theories with simple gauge group, conformal manifold, and large $N$ limit (summarized in Table \ref{table:SCFTs}), can be divided in three universality classes according to the type of their overall free limit. This type is determined by the value of the exponential rate $\alpha$ of the higher-spin tower, which is also associated to the value of $a/c$:
\vspace{-7.5pt}
\renewcommand{\arraystretch}{2} 
\begin{equation} \label{list-lambda}
  \begin{array}{llll}
    \text{Type 1: } & \alpha = \dfrac{1}{\sqrt{2}} \ & \ \text{and } & \dfrac{a}{c} = 1 \, , \\
    \text{Type 2: } & \alpha = \sqrt{\dfrac{7}{12}}  \ & \ \text{and } & \dfrac{a}{c} = \dfrac{13}{14}  \, , \\
	\text{Type 3: } & \alpha = \sqrt{\dfrac{2}{3}}  \ & \ \text{and } & \dfrac{a}{c} = \dfrac{7}{8}  \, .
  \end{array}
\end{equation}
\renewcommand{\arraystretch}{1} 
We have denoted the three values as three different \emph{types} of limits to emphasize that they will correspond to three different types of tensionless strings in the bulk, as we will argue in the following.

\medskip

The intuition that each value of $\alpha$ should correspond to a different microscopic nature of the bulk modes comes from recent research on the Swampland program.  There has been plethora of works analysing in detail the asymptotic towers of states at infinite distance in moduli spaces of Minkowksi compactifications and deriving the exponential mass decay rate of the tower. One of the lessons is that the exponential rate is tied to the microscopic nature of the tower, which in all known examples of flat space corresponds to a Kaluza-Klein tower or to a critical perturbative string (these only two options for the tower of states are referred to as the Emergent String Conjecture \cite{Lee:2019wij}). It is therefore reasonable to expect the same in AdS compactifications, in the sense that each value should correspond to a different microscopic nature of the bulk tower of states. Since all these infinite distance limits in the conformal manifold contain higher-spin fields, they are not going to correspond to pure decompactification limits in which we have only Kaluza-Klein towers, but must originate from some extended object in the bulk. Thereby the expectation that they correspond to different strings becoming tensionless at the weak-coupling point.

The first piece of evidence in this direction comes from the above re-writing of the exponential rate in terms of the ratio $a/c$. Notice that if the bulk dual has an Einstein gravity description at low energies, then the CFT must admit a large $c$ limit in which the ratio  $a/c$ is 1 to leading order \cite{Henningson:1998gx,Nojiri:1998dh,Nojiri:1999mh}. This automatically implies that the weak-coupling limit of all holographic CFTs (i.e. with Einstein gravity dual) must be of Type 1 above, and the higher-spin modes have always $\alpha =  1/\sqrt{2}$. This suggests that the bulk description of the higher-spin modes of Type 1 corresponds to the perturbative critical string in ten dimensions. Indeed, this set of Type 1 limits includes $\mathcal{N}=4$ SYM in the $g_{\text{YM}}\to 0$ limit, which is known to correspond to the tensionless and weak-coupling limit of the Type IIB string in the bulk. This check extends to all the $\mathcal N \geq 2$ theories with $\alpha =  1/\sqrt{2}$ in Table \ref{table:SCFTs}, which also have known Type IIB duals \cite{Ennes:2000fu}. In Section \ref{sec:matching}, we explain how the value $\alpha =  1/\sqrt{2}$ can be connected to the supergravity result for the perturbative Type IIB string, further confirming our expectation. 

On the other hand, the limits of Type 2 and 3 occur in non-holographic CFTs, namely whose bulk dual theories do not admit Einstein gravity as the low energy limit. Since they all have a large $N$ limit (i.e. gravity is weakly coupled in the bulk), we expect thay they fail to have a large gap, in the sense that there should be low lying higher-spin fields (spin bigger than 2) with mass of order the AdS scale. This is indeed confirmed by examples.
For instance, an example of a theory with a Type 3 limit (i.e. with $\alpha=\sqrt{2/3}$) is $\mathcal{N}=2$ $SU(N)$ SQCD with $N_f=2N$, which corresponds to the first row in the third group of Table \ref{table:SCFTs}. The properties of its bulk dual were discussed in \cite{Gadde:2009dj}, where they indeed  argued for the presence of low lying higher-spin BPS operators  by computing the superconformal index. For the bulk, this implies the presence of higher-spin modes that cannot be decoupled from the AdS scale, yielding non-supressed higher derivative corrections over Einstein gravity. This explains why the supergravity approximation is not valid at low energies. Moreover, they also argued that the dual string theory is not ten dimensional, but propose a dual sub-critical string background in eight dimensions. Following their proposal, the Type 3 limit then corresponds to the weak-coupling limit of this non-critical bulk string, as we explain in more detail in Section \ref{sec:beyond}. According to our proposal, all the other SCFTs exhibiting a Type 3 limit should then also contain the same non-critical closed string.

\medskip

To sum up, the weak-coupling limits of 4d large $N$ SCFTs with simple gauge factor can be divided in three types according to the value of the exponential rate $\alpha$. Holographic theories with an Einstein gravity dual exhibit the first type of limit, which corresponds to the tensionless limit of the critical perturbative Type IIB string; while the other two types of limits arise in non-holographic theories and seem to correspond to tensionless limits of two different non-critical bulk strings. 
To gather evidence for this, in Section \ref{sec:Hagedorn} we compute the large $N$ thermal partition function in the free limit for all these theories, in order to get information about the density of states in the infinite distance limit. Recall that if our expectation is correct and they correspond to three different types of bulk strings, they should exhibit a different density of states. Indeed, we obtain that each type exhibits a different Hagedorn density of states, with a Hagedorn temperature, $T_H$, purely determined by the value of $\alpha$ (or equivalently, of the ratio $a/c$). Furthermore, we show that this remains true for the Hagedorn temperature of the flavor singlet spectrum of the theory (both if considering the flavor group at the free point, $T_{H}^{f.s.}$, or that of the interacting theory, $T_H^{f.s.i.}$).

\renewcommand{\arraystretch}{1.5} 
\begin{table}[h]
\centering 
\begin{tabular}{| c | c c c c c | c c |}
    \hline
    \multirow{2}*{\text{Limits}} & \multirow{2}*{$\alpha$} & \multirow{2}*{$a/c$} & \multirow{2}*{$T_H$} & \multirow{2}*{$T_H^{f.s.}$} & \multirow{2}*{$T_H^{f.s.i.}$} & $(\mathcal{N}=2)$ & ($\mathcal{N}=1$)  \\
     &  &  & & &  & $n_h/n_v$ & $N_c/N_v$ \\
    \hline
    \text{Type 1} & $\dfrac{1}{\sqrt{2}}$ & $1$ & $\approx 0.380$ & $\approx 0.380$ & $\approx 0.380$ & $1$ & $3$ \\[5pt]
    \hline
    \text{Type 2} &  $\sqrt{\dfrac{7}{12}}$ & $\dfrac{13}{14}$ & $\approx 0.414$ & $\approx 0.406$ & $\approx 0.400$ & $\dfrac{3}{2}$ & $4$ \\[5pt]
    \hline
    \text{Type 3} & $\sqrt{\dfrac{2}{3}}$ & $\dfrac{7}{8}$ & $\approx 0.471$ & $\approx 0.452$ & $\approx 0.439$ & $2$ & $5$ \\[5pt]
    \hline
\end{tabular}
\medskip
\caption{Three types of infinite distance limits in the conformal manifold of 4d gauge theories with simple gauge factor and a large $N$ limit (summarized in Table \ref{table:SCFTs}). The types of limits are characterized by the values for the exponential rate $\alpha$ of the HS tower, the ratio of central charges $a/c$, and the Hagedorn temperature $T_H$. We also include the Hagedorn temperature if restricting to the flavor singlets of the free theory ($T_H^{f.s.}$) or of the interacting theory ($T_H^{f.s.i.}$), although the latter only applies to $\mathcal N=2$ theories as explained in Section \ref{sec:flavor-singlets}. For $\mathcal N=2$ and $\mathcal N=1$ theories, we also show the ratios $n_h/n_v$ and $N_c/N_v$, respectively.}
\label{table:summary}
\end{table}
\renewcommand{\arraystretch}{1} 

We summarize the relevant information about these three types of weak-coupling limits in Table \ref{table:summary}. As the reader can notice, the Hagedorn temperature increases as $\alpha$ increases, which is consistent with the intuition that towers with a denser spectrum decay at a slower rate \cite{Castellano:2023jjt, Castellano:2023stg}.

\section{Hagedorn Temperature in the Free Limits} \label{sec:Hagedorn}

In this section, we will study the thermal partition function of the gauge theories of Section \ref{sec:types-limits} on the 3-sphere. The goal is to study the Hagedorn-like behaviour of the density of states in the free limit, and whether it depends on the value of the exponential rate $\alpha$ characterizing the three types of limits of Table \ref{table:summary}. As usual, we organize the spectrum of the theory in eigenstates of the Hamiltonian with energy $E_i$ such that the thermal partition function is given by
\begin{equation} \label{partion-function}
  Z(x) = \sum_{states} x^{E_i} = \int_0^\infty \rho(E) \, x^E dE \, .
\end{equation}
To ease notation, we have introduced $x \equiv e^{-\beta} = e^{-1/T}$ with $\beta$ the length of the thermal circle and $T$ the temperature. In the last equality we have also introduced the density of states $\rho(E)$ with energy $E$.

We are especially interested on whether the spectrum exhibits at some point an exponential Hagedorn-growth of states of the form\footnote{More generally, one could also include a polynomial in front of the exponential law in \eqref{Hagedorn-density}. We are omitting it since it will play no role in our considerations. \label{footnote1}}
\begin{equation} \label{Hagedorn-density}
  \rho(E) \sim e^{E/T_H} \, .
\end{equation} 
This can be detected by a divergence on the partition function $Z(x)$ for $x>x_H$, i.e., for temperatures higher than the so-called Hagedorn temperature $T_H$ \cite{Hagedorn:1965st}. This exponential growth of the density of states is considered to be a stringy signature of the spectrum. Indeed, the excitation modes of all the strings we know have this feature at tree level in $g_s$ (see e.g. \cite{Huang:1970iq,Fubini:1969qb,Atick:1988si,Polchinski:1998rq,Green:1987sp}). 

As can be seen from \eqref{Hagedorn-density}, the Hagedorn temperature $T_H$ controls the exponential density of states. Assuming that this exponential degeneracy is related to the excitation modes of a string, it seems reasonable to use the Hagedorn temperature to tell different strings apart. In other words, one expects that different strings (in the sense of different spectrum of oscillator modes) will have different Hagedorn temperatures. Thus, following our reasoning above that different values of $\alpha$ correspond to different tensionless strings, we expect that the Hagedorn temperature of all the above theories should only depend on $\alpha$.\footnote{Let us clarify that the exponential degeneracy of states that we find is not due to the HS currents that become conserved in the free limit with exponential rate $\alpha$, but to other operators that do not saturate the unitarity bound. This is not in contradiction with our logic, since these modes are also expected to be part of the spectrum of excitations of the string that is becoming tensionless. In fact, their conformal dimensions also fall to their free value with the same exponential rate $\alpha$ as the HS conserved currents.} This is what we will check in this section.

\medskip

This Hagedorn behavior has already been discussed for certain large $N$ gauge theories at weak-coupling in the literature. For instance, it was observed in $SU(N)$ gauge theories with matter in the adjoint representation in \cite{Sundborg:1999ue,Aharony:2003sx}.\footnote{The distiction between $U(N)$ and $SU(N)$ is irrelevant for these purposes.} This was also studied for $SU(N)$ theories with large number of fields in the fundamental representation at large $N$ in \cite{Schnitzer:2004qt} and for $\mathcal N=2$ circular quivers in \cite{Larsen:2007bm}. In this paper, we extend these results to any theory with simple gauge group and fields transforming in any representation with at most two indices. This includes all the theories with conformal manifold introduced in Section \ref{sec:types-limits} (and summarized in Table \ref{table:SCFTs}). Even though this is not enough evidence by itself, we believe that finding a Hagedorn behavior of the spectrum is a nice check to our emergent strings proposal in AdS. Furthermore, we will remarkably show that the Hagedorn temperature in the free limit only depends on $\alpha$ (and therefore on the ratio of the central charges $a/c$) as expected from the discussion above.

Before going on, one comment is in order. The partition function varies non-trivially as we move in the conformal manifold. This is in contrast to the supersymmetric index, that only has contributions from BPS-protected operators and does not depend on marginal couplings (see e.g. \cite{Kinney:2005ej,Romelsberger:2005eg,Romelsberger:2007ec,Dolan:2008qi,Gadde:2009dj,Rastelli:2016tbz,Gadde:2020yah}). Even though we will not discuss the supersymmetric index further in this paper, the techniques we will employ can also be used for computing it. We will focus on the partition function of the theory at the free point, which encodes the spectrum (including non-BPS states) at the infinite distance limit, that we expect to be dominated by the tensionless string in the bulk.\footnote{Unlike in flat space, where the Hagedorn temperature is proportional to the tension of the string and goes to zero in the infinite distance limit, in AdS/CFT one finds a finite Hagedorn temperature. This will allow us to use the result at the free point to compare different theories and tell different strings apart, without having to compute the first perturbative corrections, which are quite involved \cite{Aharony:2003sx}.}

\subsection{Thermal partition functions as matrix models} \label{sec:partition-function-intro}

As shown in \cite{Aharony:2003sx} (see also \cite{Sundborg:1999ue}), the thermal partition function of a free theory with gauge group $G$ can be written as the matrix integral
\begin{equation} \label{eq:matrix-integral}
  Z(x) = \int_{G} d\mu(u) \, \exp \left\{ \sum_{n=1}^{\infty} \frac{1}{n} \left( \sum_{R} z_R(x,n) \, \chi_R(u^n) \right) \right\} \, ,
\end{equation}
with $d\mu(u)$ the $G$ invariant measure. As made explicit by the notation, this can be written as an integral over the maximal torus of the group $G$, that we parametrize with the complex variables
\begin{equation}
  u_i \, , \quad i=1,\ldots, \text{rank}(G) \,  \quad \text{with} \quad |u_i| = 1 \, .
\end{equation}

For convenience, we have defined the functions
\begin{equation} \label{z-functions}
  z_R (x,n) = z_B^R(x^n) + (-1)^{n+1} z_F^R(x^n) \, .
\end{equation}
In the RHS we have the single-particle partition functions for bosons and fermions transforming in the representation $R$ of the gauge group $G$. These encapsulate the contribution of each letter/field and all its derivatives (taking into account their free equations of motion). In four dimensions, each real scalar, Weyl fermion and vector field contribute to the single-particle partition function with
\begin{equation} \label{single-particles}
  z_S(x) = \frac{x^2 + x}{(1-x)^3} \, , \quad z_F(x) = \frac{4 x^{\frac{3}{2}}}{(1-x)^3} \, , \quad z_V(x) = \frac{6 x^2 - 2 x^3}{(1-x)^3} \,  \, .
\end{equation}

Inserting these single-particle partition functions into the plethystic exponential in \eqref{eq:matrix-integral}, we obtain the partition function that counts all the possible combinations of letters in the theory. Notice that each single-particle partition function comes with the character of its representation, $\chi_R(u)$. This is equivalent to introducing chemical potentials that keep track of the charge of each state under the gauge group $G$. Finally, the integral over the gauge group takes care of projecting into gauge singlets.

Even if our results will be valid both for $\mathcal N=2$ and $\mathcal N=1$ theories, we will always write the spectrum in the language of $\mathcal N =1$ for convenience.
It will be then convenient to encode the contribution from vector and chiral multiplets in the functions
\begin{equation}
\begin{split} \label{susy-single-particles}
	z_v(x,n) &= z_V(x^n) + (-1)^{n+1} z_F(x^n) \, , \\
	z_{c}(x,n) &= 2 z_S(x^n) + (-1)^{n+1} z_F(x^n) \, ,
\end{split}
\end{equation}
where we have taken into account that these multiplets contain one vector and one Weyl fermion, and one complex scalar and one Weyl fermion, respectively.

To give an example, a chiral multiplet transforming in a complex representation $R$ of the gauge group $G$ will produce
\begin{equation}
\begin{split}
	z_{R}(x,n) &= \frac{1}{2} \, z_{c}(x,n) + \text{other contributions} \, , \\
	z_{\bar R}(x,n) &= \frac{1}{2}  \, z_{c}(x,n) + \text{other contributions} \, .
\end{split}
\end{equation}
Crucially, here we have taken into account that a complex field transforming in a complex representation $R$ contributes with two types of letters: half of them come from the field itself and transform in $R$, while the other half come from its complex conjugate and transform in $\bar R$. This is just the usual statement that, in a CPT symmetric theory, each particle comes with its antiparticle. As a consequence, as long as we only care about the dependence on the temperature and not in some other variable such as chemical potentials, we can set
\begin{equation} \label{CPT-consequence}
  z_{R}(x,n) = z_{\bar R}(x,n) \, .
\end{equation}

\medskip

Before going on, let us point out that there is a similar matrix integral formula for the computation of the supersymmetric index of gauge theories. It takes the form of \eqref{eq:matrix-integral}, but replacing the single-particle partition functions by similar objects that encode the contribution from BPS protected operators (see e.g. \cite{Kinney:2005ej,Romelsberger:2005eg,Romelsberger:2007ec,Dolan:2008qi,Gadde:2009dj,Rastelli:2016tbz,Gadde:2020yah}). Thus, even though we are not explicitly considering it in this paper, our results can be easily generalised to determine the supersymmetric index of these theories. 

\subsection{Theories with $SU(N)$ gauge group} \label{sec:SU-case}

In this section we focus on theories with $SU(N)$ gauge group. First, we will introduce the representations that will be allowed and their characters. Then, we will consider the large $N$ limit and compute the thermal partition function. Finally, we will find the condition that determines the Hagedorn temperature and, upon restriction to theories with vanishing one-loop beta function, we will show that it can be written as a function of only $\alpha$ (defined in \eqref{aac}).

\subsubsection*{Matter content and the integrand}
For an UV-complete $SU(N)$ gauge theory, the available representations (and their complex conjugates) are: adjoint ($Ad$), two-index symmetric ($S$, $\bar S$), two-index antisymmetric ($A$, $\bar A$) and fundamental ($F$, $\bar F$). Their characters read (see e.g. \cite{Dolan:2008qi}):
\begin{equation} \label{eq:characters-SU}
\begin{split}
	\chi_F(u) &= \sum_{i=1}^{N} u_i \equiv p_N(u) \, , \\
	\chi_{Ad}(u) &= \sum_{1\leq i,j \leq N} u_i u_j^{-1} -1  = p_N(u) p_N(u^{-1}) - 1 \, , \\
	\chi_S(u) &= \sum_{1\leq i \leq j \leq N} u_i u_j = \frac{1}{2} p_N(u)^2 + \frac{1}{2} p_N(u^2) \, , \\
	\chi_A(u) &= \sum_{1\leq i < j \leq N} u_i u_j = \frac{1}{2} p_N(u)^2 - \frac{1}{2} p_N(u^2) \, , \\
	\chi_{\bar R}(u) &=  \chi_{R}(u^{-1}) \, \quad \forall R \, .
\end{split}
\end{equation}
For simplicity, we write them in terms of the $N$ eigenvalues of the matrix, $u_i$, that satisfy $|u_i|=1$ and $\prod_{i=1}^{N} u_i =1$. The maximal torus over which the integral in \eqref{eq:matrix-integral} is performed correspond to the $N-1$ first eigenvalues, replacing the last one by solving the last constraint. From now on this replacement should be understood implicitly. We have also introduced the symmetric polynomial $p_N(u)$ in the first line for convenience.

Given these representations, in the integrand in \eqref{eq:matrix-integral} we find
\begin{equation} 
\begin{split}
	  \sum_{R} z_R(x,n) \, \chi_R(u^n) &= z_{Ad}(x,n) \Big[ p_N(u^{n}) p_N(u^{-n}) - 1 \Big] \\
	  & + \frac{1}{2} z_{S}(x,n) \Big[ p_N(u^{n})^2 + p_N(u^{2n}) + p_N(u^{-n})^2 + p_N(u^{-2n}) \Big] \\
	  & + \frac{1}{2} z_{A}(x,n) \Big[ p_N(u^{n})^2 - p_N(u^{2n}) + p_N(u^{-n})^2 - p_N(u^{-2n}) \Big] \\
	  & + z_{F}(x,n) \, p_N(u^{n}) + z_{\bar F}(x,n) \, p_N(u^{-n}) \, . \\
\end{split}
\end{equation}
Notice that we have used \eqref{CPT-consequence} for all the complex representations except for the fundamental. The reason is that in latter sections we will be interested in introducing chemical potentials to keep track of the flavor charge of the different states. To do this we need to allow for $z_{F} \neq z_{\bar F}$.

The integrand in \eqref{eq:matrix-integral} can be further simplified due to the sum over $n$, which allows us to recast the terms with a $p_N(u^{2n})$ factor as extra contributions to the $p_N(u^{n})$ terms for $n$ odd. For example, we have
\begin{equation}
\begin{split}
	\sum_{n=1}^{\infty} \frac{1}{n} \left( \sum_{R} z_R(x,n) \, \chi_R(u^n) \right) &\supset \sum_{n=1}^{\infty} \frac{1}{2n} \Big[ z_{S}(x,n) - z_{A}(x,n) \Big] p_N(u^{2n}) \\
	& = \sum_{n \in 2\mathbb{Z}_{>0}} \frac{1}{n} \Big[ z_{S}\left( x,\frac{n}{2} \right) - z_{A}\left( x,\frac{n}{2} \right) \Big] p_N(u^{n}) \, .
\end{split}
\end{equation}

Taking this into account, we end up with
\begin{equation} \label{eq:integral-simplified}
\begin{split}
	  \sum_{n=1}^{\infty} \frac{1}{n} \left( \sum_{R} z_R \, \chi_R(u^n) \right) = \sum_{n=1}^{\infty} \frac{1}{n} &\bigg( z_{Ad} \Big[ p_N(u^{n}) p_N(u^{-n}) - 1 \Big] \\
	  & + z_{AS} \Big[ p_N(u^{n})^2 + p_N(u^{-n})^2 \Big] \\
	  & + \tilde z_{F} \, p_N(u^{n}) + \tilde z_{\bar F} \, p_N(u^{-n}) \bigg) \, , 
\end{split}
\end{equation}
where we have defined the effective single-particle partition functions
\begin{equation} \label{z-effectives}
	z_{AS} = \frac{1}{2} (z_A + z_S) \, , \qquad 
	\tilde z_{F} = z_{F} + z_{\delta F} \, , 
    \qquad \tilde z_{\bar F} = z_{\bar F} + z_{\delta F} \, .
\end{equation}
To ease notation, we have omitted the $(x,n)$ dependence in front of the various single-particle partition functions. We will do this from now own whenever it is possible. For this, we have also introduced the function
\begin{equation}
    z_{\delta F}(x,n) = 
    \begin{cases} 
    0 & \text{if } n \text{ odd} \, , \\
    z_{S}\left( x,\frac{n}{2} \right) - z_{A}\left( x,\frac{n}{2} \right) & \text{if } n \text{ even} \, .
    \end{cases}
\end{equation}

\subsubsection*{The integration measure and the effective action}

The $SU(N)$ invariant integral can be explicitly written by replacing (see e.g. \cite{Dolan:2008qi})
\begin{equation}
  \int_{SU(N)} d\mu(u) \to  \frac{1}{N!} \int \prod_{j=1}^{N-1} \frac{du_j}{2\pi i u_j} \, \Delta(u)\Delta(u^{-1}) \, ,
\end{equation}
where $\Delta(u)$ is the Vandermonde determinant
\begin{equation} \label{Vandermonde}
  \Delta (u) = \prod_{1 \leq i < j \leq N} \left( u_i - u_j \right) \, .
\end{equation}
By rewriting the integration measure as\footnote{This expression can be derived by starting with the RHS and using the identity $\log(1-x)=-\sum_{n=1}^{\infty} x^n/n$.}
\begin{equation} \label{measure-SU}
  \Delta(u)\Delta(u^{-1}) = \exp \left( - \sum_{n=1}^{\infty} \frac{1}{n} \sum_{i \neq j} u_i^{n} u_j^{-n} \right) \, ,
\end{equation}
we bring the matrix integral in \eqref{eq:matrix-integral} to the convenient form
\begin{equation} \label{integral-eff-action}
\begin{gathered}
  Z(x) = \frac{1}{N!} \int \prod_{j=1}^{N-1} \frac{du_j}{2\pi i u_j} \ e^{- S_{\text{eff}}} \, \\
  S_{\text{eff}} = \sum_{n=1}^{\infty} \frac{1}{n} \left( \sum_{i \neq j} u_i^{n} u_j^{-n} - \sum_{R} z_R \, \chi_R(u^n) \right)  \, .	
\end{gathered}
\end{equation}
In the second line we have introduced the effective action $S_{\text{eff}}$ which, upon using \eqref{eq:integral-simplified}, takes the form
\begin{equation}
\begin{split}
  S_{\text{eff}} = \sum_{n=1}^{\infty} \frac{1}{n} \bigg( & \sum_{i \neq j} u_i^{n} u_j^{-n} - z_{Ad}\, p_N(u^{n}) p_N(u^{-n}) \\
	  & - z_{AS} \, \Big[ p_N(u^{n})^2 + p_N(u^{-n})^2 \Big] \\
	  & - \tilde z_{F} \, p_N(u^{n}) - \tilde z_{\bar F} \, p_N(u^{-n})  + z_{Ad} \bigg) \, .
	  \end{split}
\end{equation}

\subsubsection*{Large $N$ limit partition function}

In what follows, we consider the large $N$ limit of \eqref{integral-eff-action}. For this, we first parametrize the eigenvalues as
\begin{equation} \label{angles}
  u_i = e^{i \, \theta_i} \quad i=1,\ldots, N \,  \quad \text{with} \quad -\pi < \theta_i \leq \pi \, , \quad \sum_{i=1}^{N} \theta_i = 0 \, .
\end{equation}
Following \cite{Aharony:2003sx}, we then introduce a density of eigenvalues $\rho (\theta)$ normalized as
\begin{equation}
  \int_{-\pi}^{\pi} d\theta \, \rho(\theta) = N \, .
\end{equation}
For any function of the eigenvalues, we then have
\begin{equation}
  \sum_{i=1}^{N} f(\theta_i) \quad \to \quad \int_{-\pi}^{\pi} d\theta \, \rho(\theta) f(\theta) \, .
\end{equation}

With this definition, we find that the effective action is naturally written in terms of the Fourier modes of $\rho(\theta)$. We have
\begin{equation} \label{largeN-replacements}
\begin{gathered}
  p_N (u^n) = \sum_{i=1}^{N} e^{i n \theta_i} \quad \to \quad \rho_n \equiv \int_{-\pi}^{\pi} d\theta \, \rho(\theta) \, e^{i n \theta_i} \, , \\	
    \sum_{i \neq j} u_i^{n} u_j^{-n} \quad \to \quad \int_{\theta_1 \neq \theta_2} d\theta_1 d\theta_2 \, \rho(\theta_1) \rho(\theta_2) e^{i n(\theta_i - \theta_j)}  \simeq\rho_n \, \rho_{-n} \, .
\end{gathered}
\end{equation}
In the last step we have ignored the (lack of) contribution from $\theta_1 = \theta_2$ at large $N$ since it is a subspace of zero measure in the integral. Otherwise, this contribution can be reabsorbed in the definition of the $d[\rho]$ measure to be introduced below.  Plugging these replacements, we get
\begin{equation}
  S_{\text{eff}} = \sum_{n=1}^{\infty} \frac{1}{n} \bigg( (1 - z_{Ad}) \, \rho_n \bar\rho_n  - z_{AS} \, ( \rho_n^2 + \bar\rho_n^2 ) - \tilde z_{F} \, \rho_n + \tilde z_{\bar F} \, \bar\rho_n  + z_{Ad} \bigg) \, ,
\end{equation}
where we have used that, since $\rho(\theta)$ is real, its Fourier modes satisfy
\begin{equation}
  \rho_{-n} = \bar\rho_n \, .
\end{equation}

In this language, the integral over the eigenvalues becomes a functional integral over all the possible densities of eigenvalues $\rho(\theta)$. This is,\footnote{Strictly speaking, given the normalization of $\rho(\theta)$, we should integrate in the range $-N \leq \text{Re}\,\rho_n \leq N $ and $ -N \leq \text{Im}\,\rho_n \leq N$ and then take $N\to\infty$. Nevertheless, this is irrelevant for sufficiently low temperatures. For theories with large number of fields in the (anti)fundamental, this leads to the critical temperature $T_c < T_H$ introduced in \cite{Schnitzer:2004qt}, as discussed below.}
\begin{equation} \label{replacement-integral-SU}
  \frac{1}{N!} \int \prod_{j=1}^{N-1} \frac{du_j}{2\pi i u_j} = \frac{1}{(2\pi)^{N-1} \, N!} \int_{-\pi}^{\pi} \prod_{j=1}^{N-1} d\theta_j \quad \to \quad \int d[\rho] = \prod_{n=1}^{\infty} \frac{1}{n \, \pi} \int d^2\rho_n \, .
\end{equation}
In the last step, we have fixed the integration measure for each of the Fourier mode by requiring the $SU(N)$ invariant measure to be normalized to one, as it should be. This is,
\begin{equation}
  \int d[\rho] \exp \left\{ -\sum_{n=1}^{\infty} \frac{1}{n} \, \rho_n \bar\rho_n \right\} = \prod_{n=1}^{\infty} \frac{1}{n \, \pi} \int d^2\rho_n  \exp \left( -\frac{1}{n} \, \rho_n \bar\rho_n \right) = 1 \, .
\end{equation}

All in all, in the large $N$ limit the partition function becomes
\begin{equation} \label{SU-Z-integral}
  Z(x) \simeq \prod_{n=1}^{\infty} \frac{1}{n \, \pi} \int d^2\rho_n \, \exp \left\{ - \frac{1}{n} \bigg( (1 - z_{Ad}) \, \rho_n \bar\rho_n  - z_{AS} \, ( \rho_n^2 + \bar\rho_n^2 ) - \tilde z_{F} \, \rho_n + \tilde z_{\bar F} \, \bar\rho_n  + z_{Ad} \bigg) \right\} \, .
\end{equation}
Upon performing the integral, this yields
\begin{equation} \label{SU-partition-function}
\begin{split}
  Z(x) \simeq & \exp \left\{ \sum_{n=1}^{\infty} \frac{1}{n} \left( \frac{ \left( 1 -z_{Ad} \right) \tilde{z}_{F} \tilde{z}_{\bar F} + z_{AS} \left( \tilde{z}_{F}^{2} + \tilde{z}_{\bar F}^{2} \right) }{\left( 1-z_{Ad} \right)^{2} -  4 z_{AS}^{2} } -z_{Ad} \right) \right\} \\ 
  & \prod_{n=1}^{\infty} \frac{1}{\sqrt{\left( 1-z_{Ad} \right)^{2} - 4z_{AS}^{2} }} \, .
\end{split}
\end{equation}
As a check, let us note that for a theory with only fields in the adjoint, i.e. only $z_{Ad}\neq 0$, we recover the result in \cite{Aharony:2003sx}.\footnote{We have an extra $\exp \left( - \sum_{n=1}^{\infty} \frac{1}{n} z_{Ad} \right) $ coming from considering $SU(N)$ instead of $U(N)$.} Additionally, in Appendix \ref{app:orthogonality-relations} we recover our result for the case $\tilde z_{F} = \tilde z_{\bar F}=0$ using a different method that exploits the orthogonality properties of the polynomial $p_N(u)$ (see e.g. \cite{Dolan:2007rq,Dolan:2008qi}), which serves as an important double-check of our computations.

Before going on, let us emphasize that this result is valid in the limit $N\rightarrow \infty$, so it should be understood as the leading contribution for large $N$. As discussed in \cite{Aharony:2003sx}, only the planar theory (i.e. for $N=\infty$) features the Hagedorn divergence in the partition function. For finite $N$, $Z(x)$ remains finite for any $x$ since at high temperatures it must recover the thermodynamics of a field theory with finitely many fields. For the spectrum, this means that the exponential density of states cannot hold for arbitrary high energies. Indeed, there is a maximum energy of order $N^2$ for which this exponential growth stops.\footnote{From the gauge theory point of view, this is due to trace relations. The trace of order $N^2$ fields can break down into smaller traces. Thus, these two states are the same and only one of them should be included in the partition function.} Nevertheless, it holds for energies in the range $1 \ll E \ll N^2$. The divergence of the strict large $N$ partition function captures correctly the exponential growth in this regime. Our goal is not to find Hagedorn-like thermodynamics, but rather to detect this exponential degeneracy of states. Taking $N\to\infty$ is a good way of extending it to arbitrary high energies so that it is easier to detect it as a divergence in the partition function.

Moreover, certain issues may arise when taking the strict limit $N\rightarrow \infty$ if the number of chiral multiplets in the (anti-)fundamental representation grows with $N$ (which occurs for some of the gauge theories under consideration). First of all, strictly speaking, the partition function in \eqref{SU-partition-function} blows up for any $x$ as $N\to\infty$ due to the infinitely large flavor symmetry group, so one may consider restricting to the flavor singlets. This will be discussed in detail in Section \ref{sec:flavor-singlets}. Until then, we will take \eqref{SU-partition-function} as the leading term for $N\gg 1$ but finite. Secondly, 
it was argued in \cite{Schnitzer:2004qt} that the thermodynamics of $SU(N)$ gauge theories with large number of fields in the fundamental representation at large $N$ do not show a Hagedorn phase transition. The reason is the presence of a previous phase transition at a temperature $T_c < T_H$. Despite this, the exponential behavior of the density of states that we are interested in is still controlled by $T_H$. In fact, fields in the (anti)fundamental representations do not contribute to this exponential degeneracy of states, as they do not appear in the denominator of \eqref{SU-partition-function} and thus play no role in the Hagedorn divergence. 
This can also be argued by counting gauge invariant operators in the CFT;\footnote{For instance, consider an $SU(N)$ gauge theory with fields in the adjoint. As explained in \cite{Aharony:2003sx}, its exponential degeneracy of states comes from the possibility of choosing different fields when building a very large trace of the schematic form $\text{Tr}\left( \phi_1 \cdots \phi_n \right)$. By contrast, fields in the (anti)fundamental representations alone cannot form large arrays of operators; they can only be the first or the last one. Fields in the adjoint representation have two color indices, which allow them to link together while leaving some free index to continue the chain. On the other hand, fields in the (anti)fundamental have only one color index. Thus, they cut the chain whenever they link to another field. As a consequence, adding fields in the (anti)fundamental representations does not modify the exponential degeneracy of states. They serve as end points of a large chain of fields in the adjoint, which gives new possibilities to construct a gauge invariant operator in addition of taking the trace. However, this does not lead to an exponential but to a polynomial enhancement in the number of states at high energies.} they can only lead to a polynomial enhancement in the number of states at high energies, which forces a non-sparse spectrum when $N\rightarrow \infty$ as further discussed in Section \ref{sec:flavor-singlets}. Therefore, since we are only interested in signaling the exponential degeneracy of states controlled by $T_H$ (and not on the full thermodynamics), it is sufficient for us to consider the divergence of the partition function in \eqref{SU-partition-function}.

\subsubsection*{Hagedorn temperature for SCFTs}

The Hagedorn temperature $T_H$ is reached for the minimum $x$ such that the large $N$ partition function blows up. From \eqref{SU-partition-function}, and taking into account the definitions in \eqref{z-effectives}, we see that this happens when\footnote{The partition function in \eqref{SU-partition-function} diverges for $(1-z_{Ad})^2 = (z_S + z_A)^2 $ leading to $z_{Ad} \pm (z_S + z_A) = 1$. Taking into account that $z_{Ad}$, $z_{S}$ and $z_{A}$ are monotonically increasing for $0\leq x <1$, we pick the plus sign since it leads to the smallest temperature.}
\begin{equation} \label{SU-Hagedorn-condition}
  z_{Ad}(x_H,n) + z_S(x_H,n) + z_A(x_H,n) = 1 \, .
\end{equation}

One might worry that the partition function in \eqref{partion-function} also counts descendants and multi-trace operators, and not only single-trace primary operators; being the latter what would map to single particle exictation modes of the string.\footnote{Via state-operator correspondence, each state of the CFT on the 3-sphere maps to an operator in the CFT on flat space. AdS/CFT then maps each operator to a state in the AdS side. To compare with the spectrum of excitations of a string, one would like to only consider the minimum energy state of each single particle state, which in AdS/CFT are dual to single-trace primary operators.} However, this will not affect the results for the Hagedorn temperature since only single-trace primary operators can be responsible for the exponential degeneracy at high energies, so that the full partition function  \eqref{partion-function} diverges at the same Hagedorn temperature than the single trace partition function. This is explained in more detail in Appendix \ref{app:single-vs-multi}.

The condition \eqref{SU-Hagedorn-condition} can be written explicitly in terms of the matter content of the theory. Following the discussion in Section \ref{sec:partition-function-intro}, we have
\begin{equation} \label{z-matter-content}
	z_{Ad} = z_v + n_{Ad} \, z_{c} \, , \quad
	z_{S} = \frac{1}{2} \left( n_{S} + n_{\bar S}  \right) z_{c} \, , \quad
	z_{A} = \frac{1}{2} \left( n_{A} + n_{\bar A}  \right) z_{c} \, ,
\end{equation}
with $n_R$ denoting the number of $\mathcal N =1$ chiral multiplets transforming in the representation $R$ of the gauge group. Combining the two equations above, the Hagedorn condition reads
\begin{equation} \label{Hagedorn-SU}
  z_v(x_H,n) + \left\{ n_{Ad} + \frac{1}{2} \left( n_{S} + n_{\bar S} +  n_{A} + n_{\bar A} \right) \right\} z_{c}(x_H,n) =1 \, .
\end{equation}
This equation holds for any $\mathcal N \geq 1$ $SU(N)$ gauge theory with matter in the representations introduced above (written in $\mathcal N = 1$ language for convenience). Our next goal is to show that, only imposing the vanishing of the beta function at one loop (which is a minimal requirement for having a conformal manifold), this condition can be written in terms of $\alpha$ only. 

Recall equation \eqref{alpha} for the parameter $\alpha$, as well as \eqref{eq:central-charges} for the central charge $c$ in terms of the number of vector and chiral degrees of freedom. The latter are given by
\begin{equation}
\begin{split}
	N_{v} &= \text{dim} G = N^2 - 1 \, , \\
	N_{c} &= \sum_R n_R  \,\text{dim}(R) \\
	&= n_{Ad} \left( N^2 - 1 \right) + \left( n_{S} + n_{\bar S} \right) \frac{N (N+1)}{2} + \left( n_{A} + n_{\bar A} \right) \frac{N (N-1)}{2} + \left( n_{F} + n_{\bar F} \right) N \, .\end{split}
\end{equation}
Taking $n_F = n_F^{(0)} + n_{F}^{(1)} N$ and $n_{\bar F} = n_{\bar F}^{(0)} + n_{\bar F}^{(1)} N$,  this yields
\begin{equation}
	  12 \,\alpha^{2} - 3 = n_{Ad} + \frac{1}{2} \left( n_{S} + n_{\bar S} +  n_{A} + n_{\bar A} \right) + n_{F}^{(1)} + n_{\bar F}^{(1)} \, ,
\end{equation}
at leading order in the large $N$ expansion.
From here we can see that, in general, it is not possible to rewrite the combination of parameters appearing in \eqref{Hagedorn-SU} in terms of only $\alpha$.

Next, we impose the vanishing of the beta function at one loop. This gives (see e.g. \cite{Razamat:2020pra})
\begin{equation}
  n_{Ad} \, N + \left( n_F + n_{\bar F} \right) \frac{1}{2} + \left( n_S + n_{\bar S} \right)\frac{N+2}{2} + \left( n_A + n_{\bar A} \right) \frac{N-2}{2} = 3N \, ,
\end{equation}
which at large $N$ yields
\begin{equation} \label{beta=0}
   n_{F}^{(1)} + n_{\bar F}^{(1)} = 6 - 2\left( n_{Ad} + \frac{1}{2} \left( n_{A} + n_{\bar A} +  n_{S} + n_{\bar S} \right) \right)  \, .
\end{equation}
With this, the previous equation for $\alpha$ can be written as
\begin{equation} \label{alpha-vs-matter}
	  3 \left( 3 - 4 \alpha^2 \right) = n_{Ad} + \frac{1}{2} \left( n_{S} + n_{\bar S} +  n_{A} + n_{\bar A} \right) = 3 - \frac{n_{F}^{(1)} + n_{\bar F}^{(1)}}{2}  \, .
\end{equation}

As advertised, this allows us to write the Hagedorn condition only in terms of $\alpha$ as
\begin{equation} \label{Hagedorn-alpha}
  \boxed{ z_v(x_H,n) + 3 \left( 3 - 4 \alpha^2 \right) z_{c}(x_H,n) =1 }
\end{equation}
Given the relation in \eqref{aac}, this condition can also be written in terms of the ratio $a/c$ as
\begin{equation}
      z_v(x_H,n) + 3 \left( 3 + \frac{2}{1-2a/c} \right) z_{c}(x_H,n) =1 \, .
\end{equation}
Equivalently, using \eqref{alpha-vs-matter}, we can also write it in terms of the total number of flavors at large $N$:
\begin{equation}
  z_v(x_H,n) + \left( 3 - \frac{n_{F}^{(1)} + n_{\bar F}^{(1)}}{2} \right) z_{c}(x_H,n) =1 \, .
\end{equation}
Using \eqref{susy-single-particles} one can show that the smallest $x_H$, and thus the smallest temperature, satisfying this equation is found for $n=1$. For the theories in Table \ref{table:SCFTs}, we thus find the numerical values for the Hagedorn temperature $T_H$ reported in Table \ref{table:summary}.

Let us stress that, to get this result, we only needed to impose the vanishing of the one-loop beta function, so our result is more general than expected. For $4d$ $\mathcal N =1 $ theories, this is necessary but not enough to guarantee the presence of a conformal manifold. For example, the theory with $n_{F}=n_{\bar F}=3 N$ has vanishing beta function but has no conformal manifold \cite{Razamat:2020pra}.

\subsection{Theories with $USp(2N)$ gauge group} \label{sec:USp-case}

In this section we focus on theories with $USp(2N)$ gauge group. We will follow a similar structure to that of the previous section, focusing only on the aspects that are different.

\subsubsection*{Matter content and the integrand}
For an UV-complete $USp(2N)$ gauge theory, the available representations are: two-index symmetric ($S$), two-index antisymmetric ($A$) and fundamental ($F$). Using again the symmetric polynomial $p_N(u)$, their characters read (see e.g. \cite{Dolan:2008qi}):\footnote{The character of the antisymmetric can be obtained as $\chi_A(u) = \chi_F(u)^2 - \chi_S(u) - 1$.}
\begin{equation} \label{eq:characters-USp}
\begin{split}
	\chi_F(u) &= p_N(u) + p_N(u^{-1}) \, , \\
	\chi_S(u) &= p_N(u)p_N(u^{-1})+ \frac{1}{2} p_N(u)^2 + \frac{1}{2} p_N(u^2) + \frac{1}{2} p_N(u^{-1})^2 + \frac{1}{2} p_N(u^{-2})  \, , \\
	\chi_A(u) &= p_N(u)p_N(u^{-1}) + \frac{1}{2} p_N(u)^2 - \frac{1}{2} p_N(u^2) + \frac{1}{2} p_N(u^{-1})^2 - \frac{1}{2} p_N(u^{-2}) - 1 \, .
\end{split}
\end{equation}
Before plugging this into the integrand of \eqref{eq:matrix-integral}, let us discuss the integration measure.

\subsubsection*{The integration measure and the effective action}

The $USp(2N)$ invariant integral can be explicitly written by replacing (see e.g. \cite{Dolan:2008qi})
\begin{equation} 
  \int_{USp(2N)} d\mu(u) \to  \frac{(-1)^{N}}{2^{N} N!} \int \prod_{j=1}^{N} \frac{du_j}{2\pi i u_j} \, \prod_{j=1}^{N} \left( u_j - u_j^{-1} \right)^{2} \Delta\left( u+u^{-1} \right)^{2} \, ,
\end{equation}
where $\Delta(u)$ is again the Vandermonde determinant defined in \eqref{Vandermonde}. This integration measure can be rewritten as
\begin{equation} \label{measure-USp}
\begin{gathered}
  (-1)^{N}\prod_{j=1}^{N} \left( u_j - u_j^{-1} \right)^{2} \Delta\left( u+u^{-1} \right)^{2} =\\= \exp \Bigg\{ - \sum_{n=1}^{\infty} \frac{1}{n} \bigg( \sum_{i \neq j} u_i^{n} u_j^{-n} + \frac{1}{2} p_N(u^{n})^2 + \frac{1}{2} p_N(u^{2n}) + \frac{1}{2} p_N(u^{-n})^2 + \frac{1}{2} p_N(u^{-2n}) \bigg) \Bigg\} \, ,
\end{gathered}
\end{equation}
thus bringing the matrix integral in \eqref{eq:matrix-integral} to
\begin{equation} \label{integral-eff-action}
\begin{gathered}
  Z(x) = \frac{1}{2^{N} N!} \int \prod_{j=1}^{N} \frac{du_j}{2\pi i u_j} \ e^{- S_{\text{eff}}} \, , \\
  S_{\text{eff}} = \sum_{n=1}^{\infty} \frac{1}{n} \Bigg( \sum_{i \neq j} u_i^{n} u_j^{-n} + \frac{1}{2} p_N(u^{n})^2 + \frac{1}{2} p_N(u^{2n}) + \frac{1}{2} p_N(u^{-n})^2 + \frac{1}{2} p_N(u^{-2n}) - \sum_{R} z_R \, \chi_R(u^n) \Bigg)  \, .	
\end{gathered}
\end{equation}
Plugging the characters in \eqref{eq:characters-USp} and performing similar simplifications to those used for the $SU(N)$ case, the effective action reads
\begin{equation}
\begin{gathered}
  S_{\text{eff}} = \sum_{n=1}^{\infty} \frac{1}{n} \bigg( \sum_{i \neq j} u_i^{n} u_j^{-n} - \left( z_S + z_A \right)\, p_N(u^{n}) p_N(u^{-n}) \\
	   + \frac{1}{2} \left( 1-z_S-z_A \right) \, \Big[ p_N(u^{n})^2 + p_N(u^{-n})^2 \Big] - \tilde z_{F} \Big[ p_N(u^{n}) + p_N(u^{-n}) \Big] + z_{A} \bigg) \, .
\end{gathered}
\end{equation}
As for the $SU(N)$ case, we have defined the effective single-particle partition function
\begin{equation} \label{eff-zF-USp}
  \tilde z_F = z_F + z_{\delta F} \, ,
\end{equation}
but with the function $z_{\delta F}$ satisfying
\begin{equation}
    z_{\delta F}(x,n) = 
    \begin{cases} 
    0 & \text{if } n \text{ odd} \, , \\
    z_{S}\left( x,\frac{n}{2} \right) - z_{A}\left( x,\frac{n}{2} \right) - 1 & \text{if } n \text{ even} \, .
    \end{cases}
\end{equation}

\subsubsection*{Large $N$ limit partition function}

Let us now consider the large $N$ limit. As before, we first parametrize the eigenvalues as
\begin{equation} 
  u_i = e^{i \, \theta_i} \quad i=1,\ldots, N \,  \quad \text{with} \quad -\pi < \theta_i \leq \pi  \, 
\end{equation}
and introduce a density of eigenvalues $\rho (\theta)$ with the same properties as the one introduced for $SU(N)$. Doing the replacements in \eqref{largeN-replacements}, the effective action reads
\begin{equation}
  S_{\text{eff}} = \sum_{n=1}^{\infty} \frac{1}{n} \bigg( (1 - z_S - z_A) \left[ \rho_n \bar\rho_n + \frac{1}{2} \rho_n^2 + \frac{1}{2} \bar\rho_n^2 \right] - \tilde z_{F} \bigg[ \rho_n + \bar\rho_n  \bigg]  + z_{A} \bigg) \, .
\end{equation}

For the integral over the eigenvalues we replace
\begin{equation} \label{replacement-integral-USp}
  \frac{1}{2^N N!} \int \prod_{j=1}^{N} \frac{du_j}{2\pi i u_j} = \frac{1}{(4\pi)^{N} \, N!} \int_{-\pi}^{\pi} \prod_{j=1}^{N} d\theta_j \quad \to \quad \int d[\rho] = \prod_{n=1}^{\infty} \sqrt{\frac{2}{n \, \pi}} \int_{-\infty}^{\infty} d\rho_n \, .
\end{equation}
As for the previous case, the integration measure for each $\rho_n$ has been set by requiring the $USp(2N)$ invariant measure to be unit normalized.
In contrast with the $SU(N)$ case, we see that the integrals must be performed over real-valued $\rho_n$, otherwise the result is divergent. This amounts to restricting the density of eigenvalues to be even, i.e.,
\begin{equation} \label{even-rho}
  \rho(\theta) = \rho(-\theta) \quad \rightarrow \quad \rho_n \in \mathbb R \, .
\end{equation}
This extra restriction can be understood in the following way: For general gauge group $G$, the integral over the maximal torus in \eqref{eq:matrix-integral} is modded out by the group encoding the different ways in which it can be embedded into $G$, i.e., the Weyl group. This group is larger for $USp(2N)$ than for $SU(N)$, as made evident by comparing the suppressions in front of the integrals in the LHS of equations \eqref{replacement-integral-SU} and \eqref{replacement-integral-USp}. At large $N$, this is implemented by restricting to even $\rho(\theta)$ as above.

All in all, in the large $N$ limit the partition function becomes
\begin{equation}
  Z(x) \simeq \prod_{n=1}^{\infty} \sqrt{\frac{2}{n \, \pi}} \int_{-\infty}^{\infty} d\rho_n \, \exp \left\{ - \frac{1}{n} \bigg( 2(1 - z_S - z_A) \rho_n^2 - 2\tilde z_{F} \rho_n + z_{A} \bigg) \right\} \, ,
\end{equation}
where we have already imposed $\rho_n = \bar\rho_n$ in the integrand. This yields
\begin{equation} \label{USp-partition-function}
\begin{split}
  Z(x) \simeq & \exp \left\{ \sum_{n=1}^{\infty} \frac{1}{n} \left( \frac{ \tilde z_F^2 }{ 2\left( 1-z_S-z_A \right) } -z_{A} \right) \right\} \\ 
  & \prod_{n=1}^{\infty} \frac{1}{\sqrt{1-z_{S}-z_{A}}} \, .
\end{split}
\end{equation}

\subsubsection*{Hagedorn temperature for SCFTs}

The partition function \eqref{USp-partition-function} diverges at a Hagedorn temperature satisfying the following condition
\begin{equation} \label{USp-Hagedorn-condition}
  z_{S}(x_H,n) + z_{A}(x_H,n) = 1 \, .
\end{equation}
Following the discussion in Section \ref{sec:partition-function-intro}, and taking into account that in this case the vector transforms in the two-index symmetric representation of $USp(2N)$, we have
\begin{equation} \label{z-matter-content-USp}
	z_{S} =  z_v + n_{S} \, z_{c} \, , \qquad
	z_{A} = n_{A} \, z_{c} \, ,
\end{equation}
with $n_R$ again denoting the number of $\mathcal N =1$ chiral multiplets transforming in the representation $R$. With this, the condition above simplifies to
\begin{equation}
    z_v(x_H,n) + \left\{ n_S + n_A \right\} z_{c}(x_H,n) =1 \, .
\end{equation}
As for $SU(N)$, our next goal is to write this condition in terms of only $\alpha$ by imposing the vanishing of the one-loop beta function.

In this case, the number of vector and chiral degrees of freedom are given by
\begin{equation}
\begin{split}
	N_{v} &= \text{dim}\, (G) = N(2N+1) \, , \\
	N_{c} &= \sum_R n_r  \,\text{dim}(R) = n_{S}\, N(2N+1) + n_A \left( N(2N-1) -1 \right)+ n_F \, 2 N \, .
\end{split}
\end{equation}
Thus, using \eqref{alpha} and \eqref{eq:central-charges}, and taking $n_F = n_F^{(0)} + n_{F}^{(1)} N$, at large $N$ we find
\begin{equation}
	  12 \,\alpha^{2} - 3 = n_S + n_A + n_F^{(1)} \, .
\end{equation}
For a $USp(2N)$ gauge theory, the vanishing of the one-loop beta function requires
\begin{equation}
   n_F \, \frac{1}{2} + n_S (N+1) + n_A (N-1) = 3(N+1) \, .
\end{equation}
At large $N$, this condition yields
\begin{equation} \label{beta=0-USp}
  n_{F}^{(1)} = 6 - 2 \left( n_S + n_A \right) \, .
\end{equation}
Plugging this into the previous expression for $\alpha$, we get
\begin{equation}
	  3 \left( 3 - 4 \alpha^2 \right) = n_S + n_A \, ,
\end{equation}
which finally allows us to write
\begin{equation}
  z_v(x_H,n) + 3 \left( 3 - 4 \alpha^2 \right) z_{c}(x_H,n) =1 \, .
\end{equation}

This shows that the Hagedorn temperature for $USp(2N)$ gauge theories with vanishing one-loop beta function is a function of $\alpha$. Moreover, this Hagedorn condition is identical to the one we had for the $SU(N)$ case in \eqref{Hagedorn-alpha}. The numerical values for $T_H$ are shown in Table \ref{table:summary}.

\subsection{Theories with $SO(2N)$ gauge group} \label{sec:SO-case}

In this section we focus on theories with $SO(2N)$ gauge group. Since we are interested in the large $N$ limit, the final result will be the same if considering $SO(2N+1)$ gauge group. We will follow a similar structure to that of the previous sections.

\subsubsection*{Matter content and the integrand}
For an UV-complete $SO(2N)$ gauge theory, the available representations are: two-index symmetric ($S$), two-index antisymmetric ($A$) and fundamental (or vector) ($F$). Using once more the symmetric polynomial $p_N(u)$, their characters read (see e.g. \cite{Dolan:2008qi}):
\begin{equation} \label{eq:characters-SO}
\begin{split}
	\chi_F(u) &= p_N(u) + p_N(u^{-1}) \, , \\
	\chi_S(u) &= p_N(u)p_N(u^{-1})+ \frac{1}{2} p_N(u)^2 + \frac{1}{2} p_N(u^2) + \frac{1}{2} p_N(u^{-1})^2 + \frac{1}{2} p_N(u^{-2}) -1 \, , \\
	\chi_A(u) &= p_N(u)p_N(u^{-1}) + \frac{1}{2} p_N(u)^2 - \frac{1}{2} p_N(u^2) + \frac{1}{2} p_N(u^{-1})^2 - \frac{1}{2} p_N(u^{-2}) \, .
\end{split}
\end{equation}

\subsubsection*{The integration measure and the effective action}

The $SO(2N)$ invariant integral can be explicitly written by replacing (see e.g. \cite{Dolan:2008qi})
\begin{equation}
  \int_{SO(2N)} d\mu(u) \to  \frac{1}{2^{N-1} N!} \int \prod_{j=1}^{N} \frac{du_j}{2\pi i u_j} \, \Delta\left( u+u^{-1} \right)^{2} \, ,
\end{equation}
where we again find the Vandermonde determinant defined in \eqref{Vandermonde}. As in the previous cases, we can rewrite the integration measure as
\begin{equation} \label{measure-SO}
  \Delta\left( u+u^{-1} \right)^{2} = \exp \Bigg\{ - \sum_{n=1}^{\infty} \frac{1}{n} \bigg( \sum_{i \neq j} u_i^{n} u_j^{-n} + \frac{1}{2} p_N(u^{n})^2 - \frac{1}{2} p_N(u^{2n}) + \frac{1}{2} p_N(u^{-n})^2 - \frac{1}{2} p_N(u^{-2n}) \bigg) \Bigg\} \, ,
\end{equation}
which allows us to bring the matrix integral in \eqref{eq:matrix-integral} to
\begin{equation} \label{integral-eff-action}
\begin{gathered}
  Z(x) = \frac{1}{2^{N-1} N!} \int \prod_{j=1}^{N} \frac{du_j}{2\pi i u_j} \ e^{- S_{\text{eff}}} \, , \\
  S_{\text{eff}} = \sum_{n=1}^{\infty} \frac{1}{n} \Bigg( \sum_{i \neq j} u_i^{n} u_j^{-n} + \frac{1}{2} p_N(u^{n})^2 - \frac{1}{2} p_N(u^{2n}) + \frac{1}{2} p_N(u^{-n})^2 - \frac{1}{2} p_N(u^{-2n}) - \sum_{R} z_R\, \chi_R(u^n) \Bigg)  \, .	
\end{gathered}
\end{equation}
Plugging the characters in \eqref{eq:characters-USp} and performing similar simplifications to those used for the $SU(N)$ case, the effective action reads
\begin{equation}
\begin{gathered}
  S_{\text{eff}} = \sum_{n=1}^{\infty} \frac{1}{n} \bigg( \sum_{i \neq j} u_i^{n} u_j^{-n} - \left( z_S + z_A \right)\, p_N(u^{n}) p_N(u^{-n}) \\
	   + \frac{1}{2} \left( 1-z_S-z_A \right) \, \Big[ p_N(u^{n})^2 + p_N(u^{-n})^2 \Big] - \tilde z_{F} \Big[ p_N(u^{n}) + p_N(u^{-n}) \Big] + z_{S} \bigg) \, .
\end{gathered}
\end{equation}
Once more, we have defined
\begin{equation} \label{eff-zF-SO}
  \tilde z_F = z_F + z_{\delta F} \, ,
\end{equation}
but in this case the function $z_{\delta F}$ satisfies
\begin{equation}
    z_{\delta F}(x,n) = 
    \begin{cases} 
    0 & \text{if } n \text{ odd} \, , \\
    z_{S}\left( x,\frac{n}{2} \right) - z_{A}\left( x,\frac{n}{2} \right) + 1 & \text{if } n \text{ even} \, .
    \end{cases}
\end{equation}

\subsubsection*{Large $N$ limit partition function}

Let us now consider the large $N$ limit. Once more, we first parametrize the eigenvalues as
\begin{equation}
  u_i = e^{i \, \theta_i} \quad i=1,\ldots, N \,  \quad \text{with} \quad -\pi < \theta_i \leq \pi  \, 
\end{equation}
and introduce the density of eigenvalues $\rho (\theta)$. Doing the replacements in \eqref{largeN-replacements}, the effective action reads
\begin{equation}
  S_{\text{eff}} = \sum_{n=1}^{\infty} \frac{1}{n} \bigg( (1 - z_S - z_A) \left[ \rho_n \bar\rho_n + \frac{1}{2} \rho_n^2 + \frac{1}{2} \bar\rho_n^2 \right] - \tilde z_{F} \bigg[ \rho_n + \bar\rho_n  \bigg]  + z_{S} \bigg) \, .
\end{equation}

For the integral over the eigenvalues we replace
\begin{equation} \label{replacement-integral-SO}
  \frac{1}{2^{N-1} N!} \int \prod_{j=1}^{N} \frac{du_j}{2\pi i u_j} = \frac{2}{(4\pi)^{N} \, N!} \int_{-\pi}^{\pi} \prod_{j=1}^{N} d\theta_j \quad \to \quad \int d[\rho] = \prod_{n=1}^{\infty} \sqrt{\frac{2}{n \, \pi}} \int_{-\infty}^{\infty} d\rho_n \, .
\end{equation}
The integration measure for each $\rho_n$ has been set again by requiring the $SO(2N)$ invariant measure to be normalized to one. As already happened for $USp(2N)$, the integral is performed over real-valued $\rho_n$. This is, the density of eigenvalues is taken to be even. The reason is the same one as for $USp(2N)$ (see the discussion below equation \eqref{even-rho}). 

All in all, in the large $N$ limit the partition function becomes
\begin{equation}
  Z(x) \simeq \prod_{n=1}^{\infty} \sqrt{\frac{2}{n \, \pi}} \int_{-\infty}^{\infty} d\rho_n \, \exp \left\{ - \frac{1}{n} \bigg( 2(1 - z_S - z_A) \rho_n^2 - 2\tilde z_{F} \rho_n + z_{S} \bigg) \right\} \, ,
\end{equation}
where we have already imposed $\rho_n = \bar\rho_n$ in the integrand. This yields
\begin{equation} \label{SO-partition-function}
\begin{split}
  Z(x) \simeq & \exp \left\{ \sum_{n=1}^{\infty} \frac{1}{n} \left( \frac{ \tilde z_F^2 }{ 2\left( 1-z_S-z_A \right) } -z_{S} \right) \right\} \\ 
  & \prod_{n=1}^{\infty} \frac{1}{\sqrt{1-z_{S}-z_{A}}} \, .
\end{split}
\end{equation}

\subsubsection*{Hagedorn temperature for SCFTs}

From \eqref{SO-partition-function}, the Hagedorn temperature is given by the same condition as for $USp(2N)$
\begin{equation} \label{SO-Hagedorn-condition}
  z_{S}(x_H,n) + z_{A}(x_H,n) = 1 \, .
\end{equation}
Following the discussion in Section \ref{sec:partition-function-intro}, and taking into account that in this case the vector transforms in the two-index antisymmetric representation of $USp(2N)$, we have
\begin{equation} \label{z-matter-content-USp}
	z_{S} =  n_{S} \, z_{c} \, , \qquad  z_{A} = z_v + n_{A} \, z_{c} \, ,
\end{equation}
with $n_R$ again denoting the number of $\mathcal N =1$ chiral multiplets transforming in the representation $R$. With this, the condition above takes the same form as for $USp(2N)$:
\begin{equation}
    z_v(x_H,n) + \left\{ n_S + n_A \right\} z_{c}(x_H,n) =1 \, .
\end{equation}
As for two previous cases, our next goal is to write this condition in terms of only $\alpha$ by imposing the vanishing of the one-loop beta function.

In this case, the number of vector and chiral degrees of freedom are given by
\begin{equation}
\begin{split}
	N_{v} &= \text{dim}\, (G) = \frac{2N(2N-1)}{2} \, , \\
	N_{c} &= \sum_R n_r  \,\text{dim}(R) = n_{S} \, \frac{2N(2N+1)-2}{2} + n_A \, \frac{2N(2N-1)}{2}+ n_F \, 2N \, .
\end{split}
\end{equation}
Thus, using \eqref{alpha} and \eqref{eq:central-charges}, and taking $n_F = n_F^{(0)} + n_{F}^{(1)} 2N$, at large $N$ we find
\begin{equation}
	  12 \,\alpha^{2} - 3 = n_S + n_A + 2 n_F^{(1)} \, .
\end{equation}
For a $SO(2N)$ gauge theory, the vanishing of the one-loop beta function requires
\begin{equation}
   n_F + n_S (2N+2) + n_A (2N-2) = 3(2N-2) \, ,
\end{equation}
which at large $N$ reduces to
\begin{equation} \label{beta=0-SO}
  n_{F}^{(1)} = 3 - \left( n_S + n_A \right) \, .
\end{equation}
Plugging this into the previous equation for $\alpha$ we end up with
\begin{equation}
	  3 \left( 3 - 4 \alpha^2 \right) = n_S + n_A \, .
\end{equation}
Plugging this into the Hagedorn condition above, we finally find
\begin{equation}
  z_v(x_H,n) + 3 \left( 3 - 4 \alpha^2 \right) z_{c}(x_H,n) =1 \, .
\end{equation}

This shows that the Hagedorn temperature for $SO(2N)$ gauge theories with vanishing one-loop beta function is a function of $\alpha$. Moreover, this Hagedorn condition is identical to the one we had for the $SU(N)$ and $USp(2N)$ cases. With this, we see that the Hagedorn temperature is only a function of $\alpha$, regardless of the gauge group.  The numerical values for $T_H$ are shown in Table \ref{table:summary}.

\subsection{Non-sparse spectrum and restriction to flavor singlets} \label{sec:flavor-singlets}

As pointed out before, we found that $Z(x)\to\infty$ as $N\to\infty$ when the number of chiral multiplets in the (anti-)fundamental representation grows with $N$. This is apparent in equations \eqref{SU-partition-function}, \eqref{USp-partition-function} and \eqref{SO-partition-function}. Even though the matter in the (anti-)fundamental do not contribute to the piece controlling the Hagedorn divergence, they do appear in the numerator of the exponential term. Given the definitions for $\tilde z_F$ and $\tilde z_{\bar F}$ in equations \eqref{z-effectives}, \eqref{eff-zF-USp} and \eqref{eff-zF-SO}, and writing down
\begin{equation} \label{no-flavor-potentials}
\begin{array}{ll}
	z_F = z_{\bar F}  = \frac{1}{2} \left( n_F + n_{\bar F} \right) z_c  & \text{ for } SU(N) \, , \\ \\
	z_F = n_F\, z_c  & \text{ for } USp(2N) \text{ and } SO(2N) \, ,
\end{array}
\end{equation}
we see that this term blows up in the large $N$ limit when $n_F$ or $n_{\bar F}$ grow with $N$. In this section we discuss two possible takes on this issue and the implications for our results.

\medskip

Notice that this divergence is of different nature to the Hagedorn one. It happens for any temperature! It is signaling that the spectrum of the theory is not sparse in the large $N$ limit. If the number of chiral multiplets in the (anti-)fundamental representation grows with $N$, so it does the number of operators at fixed conformal dimension. This can also be understood as a consequence of the flavor symmetry of the free theory becoming infinitely large. On the other hand, for any fixed $N$, this multiplicity of operators is not responsible for the exponential degeneracy giving raise to the Hagedorn behavior. Thus, one option is to take this as a feature and not a bug, and think of the results in equations \eqref{SU-partition-function}, \eqref{USp-partition-function} and \eqref{SO-partition-function} as the leading order approximation when $N\gg 1$ but finite. As discussed at the beginning of this section, this should not be taken as a good approximation to the finite $N$ partition function, which cannot diverge at any finite temperature, but rather as a good approximation to the exponential density of states in its regime of validity.

From the point of view of the bulk, it may sound strange to have a diverging partition function in the large $N$ limit. In known holographic examples, this limit (with fixed 't Hooft coupling) corresponds to decoupling the Planck scale from the string and the AdS scales. It is indeed counterintuitive that decoupling gravity could lead to a diverging partition function. However, notice that this intuition relies on having a low-energy Einstein gravity description in the bulk. From the CFT, this divergence precisely happens for theories with $a \neq c$ at large $N$, which are not holographic! This also fits with the spectrum of these theories not being sparse in the large $N$ limit, which has been argued to be required for the CFT to be holographic in this sense (see e.g. \cite{Alday:2019qrf}). 

\medskip

Despite all these considerations, one may want to avoid this divergence, which can be done by restricting the theory to its flavor singlet sector. In fact, it was proposed in \cite{Gadde:2009dj} that the flavor-singlet sector of a gauge theory in the Veneziano limit is dual to a purely closed string theory.
Hence, from the bulk perspective, we expect that by restricting to the flavor singlet of the theory we can neglect the open strings (which will also become light in the weak-coupling limit) and focus only on the closed string spectrum.
It is, therefore, interesting to check whether, even after restricting to the flavor singlets, the effective closed string background still exhibits a Hagedorn temperature that depends only on $\alpha$.  In this section, we explore this and demonstrate that this is indeed the case, so that our conclusions remain unchanged in a non-trivial way.
We will also discuss the limitations when it comes to extending our results to small but non-zero coupling.

\subsubsection*{Theories with $SU(N)$ gauge group}

Let us start with the $SU(N)$ case. We will first restrict to the singlets of the flavour group at the free point, postponing the discussion of considering the flavour group of the interacting theory to the end of this Section. At the free point, a theory with  $n_F$ and $n_{\bar F}$ chiral multiplets in the fundamental and anti-fundamental representations enjoys a $SU(n_F) \times SU(n_{\bar F})$ flavor symmetry.\footnote{We could also consider $U(n_F) \times U(n_{\bar F})$ and our results would be unchanged.} Our goal is to restrict to the singlet sector of this symmetry. The procedure is the same that we applied when projecting onto the singlet sector of the gauge symmetry.

First, we introduce chemical potentials to keep track of the flavor charge of the different states. For concreteness, let us take the chiral multiplets in the fundamental and anti-fundamental of the gauge group to transform in the fundamental of $SU(n_F)$ and anti-fundamental of $SU(n_{\bar F})$, respectively. We then have
\begin{equation} \label{SU-flavor-replacement}
\begin{split}
  z_{F} &= \frac{1}{2} z_c \big( \chi_F(v^n) + \chi_F(w^n) \big) = \frac{1}{2} z_c  \big( p_N(v^n) + p_N(w^n) \big) \, , \\
  z_{\bar F} &= \frac{1}{2} z_c \big( \chi_{\bar F}(v^{n}) + \chi_{\bar F}(w^n) \big) = \frac{1}{2} z_c  \big( p_N(v^{-n}) + p_N(w^{-n}) \big) \, ,
\end{split}
\end{equation}
where $v$ and $w$ are the eigenvalues of the $SU(n_F)$ and $SU(n_{\bar F})$ matrices, respectively. As a check, notice these expressions reduce to those in \eqref{no-flavor-potentials} when setting all the eigenvalues to one, i.e., when turning off the chemical potentials for the flavor charges. 

To project onto the singlet sector, we plug this into the partition function in \eqref{SU-partition-function} and perform the integral over the $SU(n_F) \times SU(n_{\bar F})$ flavor symmetry group. In the $n_{F},n_{\bar F}\to \infty$ limit, we can use the same techniques as in Section \ref{sec:SU-case}. The computations and the resulting partition function can be found in Appendix \ref{app:SU-singlets}. At the end of the day, we find the following Hagedorn conditon:
\begin{equation}
  z_{Ad}(x_H,n) + z_{S}(x_H,n) + z_{A}(x_H,n) + \frac{1}{2} z_c(x_H,n)^{2} = 1 \, .
\end{equation}

As in Section \ref{sec:SU-case}, we impose the vanishing of the one loop beta function and rewrite this condition only in terms of $\alpha$ as
\begin{equation} \label{Hagedorn-flavor-singlets}
  \boxed{ z_v(x_H,n) + 3 \left( 3 - 4 \alpha^2 \right) z_{c}(x_H,n) + \frac{1}{2} z_c(x_H,n)^{2} = 1 }
\end{equation}
We see that restricting to the flavor singlet sector amounts to adding the third term in the LHS, which is independent of any parameter of the theory. Thus, the Hagedorn temperature is still a function of only $\alpha$. Let us however clarify that this new term should not be added in the $\alpha=1/\sqrt{2}$ case (Type 1 limit), for which $n_{F}^{(1)}+n_{\bar F}^{(1)}=0$, but only in the other two types of limits for which we have a large number of flavors at large $N$ (see Table \ref{table:SCFTs}). The numerical results for the Hagedorn temperature obtained upon resolving \eqref{Hagedorn-flavor-singlets} are shown in Table \ref{table:summary} denoted as $T_H^{f.s.}$.

Let us also stress that we assumed both $n_{F}$ and $n_{\bar F}$ to blow up at large $N$ to derive this result. There are theories with vanishing one-loop beta functions for which only $n_{F}^{(1)}$ or $n_{\bar F}^{(1)}$ are non-vanishing \cite{Razamat:2020pra}. In these cases we would restrict to the singlets of the $SU(n_F)$ or $SU(n_{\bar F})$ flavor group, and the result would differ from equation \eqref{Hagedorn-flavor-singlets}. On the other hand, this does not happen for theories with a conformal manifold, for which we observe that $n_{F}^{(1)}=n_{\bar F}^{(1)}$ (see Table \ref{table:SCFTs}). All in all, after restricting to the singlet sector of the flavor symmetry group of the free theory, our main conclusion does no longer hold for all the theories with vanishing one-loop beta function, but it does for theories that have a conformal manifold.

\subsubsection*{Theories with $USp(2N)$ or $SO(2N)$ gauge group}

A similar computation can be carried out for theories with $USp(2N)$ and $SO(2N)$ gauge groups. In this case we have $n_F$ chiral multiplets transforming in the fundamental of both the gauge group and the $SU(n_F)$ flavor symmetry. This corresponds to
\begin{equation} \label{z-flavor-USp-SO}
  z_{F} = \frac{1}{2} z_c \big( \chi_F(v^n) + \chi_{\bar F}(v^n) \big) = \frac{1}{2} z_c  \big( p_N(v^n) + p_N(v^{-n}) \big) \, , 
\end{equation}
which reproduces \eqref{no-flavor-potentials} when turning off the chemical potentials for the flavor symmetry.

We restrict to the flavor singlet sector by plugging this into equations \eqref{USp-partition-function} and \eqref{SO-partition-function} and integrating over the $SU(n_F)$ flavor symmetry group in the $n_F \to \infty$ limit. The details of this computation can be found in Appendix \ref{app:USp-SO-singlets}. For both cases, the Hagedorn condition reads
\begin{equation}
  z_S(x_H,n) + z_A(x_H,n) + \frac{1}{2} z_c(x_H,n)^{2} = 1 \, .
\end{equation}

Upon imposing the vanishing of the one-loop beta function as in Section \ref{sec:USp-case} and \ref{sec:SO-case}, this condition reduces to
\begin{equation}
  z_v(x_H,n) + 3 \left( 3 - 4 \alpha^2 \right) z_{c}(x_H,n) + \frac{1}{2} z_c(x_H,n)^{2} = 1 \, .
\end{equation}
Again, restricting to the flavor singlet only adds a parameter-independent term to the Hagedorn condition. As in the $SU(N)$ case, this extra term should not be added for the $\alpha = 1/\sqrt{2}$ case, for which to $n_{F}^{(1)}=0$.

Rather non-trivially, we find exactly the same result as for the $SU(N)$ case. It is remarkable that the new term in the Hagedorn condition coincides between the three possible gauge groups. As we will see next, this is for instance not the case if we consider the flavor symmetry that is preserved by $\mathcal N=2$ interactions.

\subsubsection*{Considering the flavor group beyond the free theory}

So far, we have considered the free theory and restricted to its flavor singlet sector. At zero coupling, this is consistent since the operator product expansion of operators in the flavor singlet sector is closed thanks to the global flavor symmetry. Nevertheless, part of this symmetry is generically broken for any small but finite coupling and this restriction is no longer consistent. This raises the question of how to solve the issue of having a non-sparse spectrum at large $N$ at finite coupling, and whether the results above are a good proxy for the Hagedorn temperature for small but non-zero coupling.

One could take the perspective of restricting the theory to the flavor singlet sector of the interacting theory, so that the results can be more easily extended to finite coupling. In fact, this seems more natural from a bulk perspective, if we assume that the typical holographic brane construction still works for those theories with $a\neq c$ at leading order in the large $N$ expansion. In that case, the flavor group comes from the worldvolume gauge sector of the branes. Therefore, restricting to the flavor singlet of the interacting theory seems to correspond to neglecting the open string sector and considering only the closed strings (as in \cite{Gadde:2009dj}). Nevertheless, this approach can, at best, work for $\mathcal N =2$ theories, but will not solve the issue for $\mathcal N =1$ theories. The reason is that the flavor symmetry group at a generic point of the conformal manifold is too small. The most extreme case of this happens for the theory in second to last entry of Table \ref{table:SCFTs}, whose flavor symmetry group is empty at a generic point in the conformal manifold (c.f. first entry of table 18 in \cite{Razamat:2020pra}). Restricting to the singlet sector of the flavor symmetry cannot solve the problem of the divergence of the partition function because there is no flavor symmetry at all. From the bulk perspective, this might occur e.g. due to a Higgsing of the gauge group of the stack of branes, which gives mass to the open string gauge fields  in a way consistent with this little supersymmetry. Hence, if the goal is to focus on the closed string degrees of freedom, restricting to the flavour singlets of the interacting theory does not seem to be useful (or enough) for $\mathcal{N}=1$ theories. 

In any case, we can at least check it for $\mathcal N =2$ theories, where the situation is different and this exercise can indeed provide us with information about the purely closed string background. For instance, $\mathcal N=2$ $SU(N)$ gauge theories enjoy a $SU(n_F)$\footnote{Recall that $\mathcal N=2$ $SU(N)$ theories satisfy $n_F=n_{\bar F}$ in $\mathcal N =1$ language.} flavor symmetry at any point of the conformal manifold. It is then natural to restrict to the flavor singlet sector by plugging\footnote{One can check that this recovers equation \eqref{no-flavor-potentials} for $n_F=n_{\bar F}$ when turning off the chemical potentials.}
\begin{equation} \label{N=2-flavor-replacement}
  z_{F} = z_c \, \chi_F(v^n) = z_c \, p_N(v^n) \, , \qquad
  z_{\bar F} = z_c \, \chi_{\bar F}(v^{n})=  z_c \, p_N(v^{-n}) \, ,
\end{equation}
into \eqref{SU-partition-function} and integrating over the $SU(n_F)$ flavor group for $n_F \to \infty$ using the techniques in Section \ref{sec:SU-case}. The resulting Hagedorn condition is (see Appendix \ref{app:N=2-singlets} for the details on the computation)
\begin{equation} \label{SU-N=2-prescription}
  \boxed{ z_v(x_H,n) + 3 \left( 3 - 4 \alpha^2 \right) z_{c}(x_H,n) + z_c(x_H,n)^{2} = 1 }
\end{equation}
This reproduces the functional form of the result in \cite{Gadde:2009dj} for $\mathcal N=2$ $SU(N)$ SQCD with $N_f=2N$. In that paper, they compute the supersymmetry index by restricting to the flavor singlets of the interacting theory, which is a very similar computation to our partition function, although conceptually different.  The supersymmetric index only counts BPS operators and is independent of the marginal coupling (see e.g. \cite{Kinney:2005ej,Romelsberger:2005eg,Romelsberger:2007ec,Dolan:2008qi,Gadde:2009dj,Rastelli:2016tbz,Gadde:2020yah}), so it is an object of the interacting theory. In fact, it was also observed in \cite{Gadde:2009dj} that for exactly zero coupling the supersymmetric index can change drastically due to long multiplets splitting into BPS ones. We do not expect, though, this drastic change to occur for the full partition function that we are computing,  since there should be some continuity in the density of states of the entire spectrum when taking the free limit.
Hence, we expect that our partition function at the free limit still provides a good approximation for the Hagedorn temperature of the closed string background at weak-coupling.

Similarly, we can compute the result that one would get for $\mathcal N=2$ $USp(2N)$ and $SO(2N)$ theories if following the same procedure. $\mathcal N=2$ $USp(2N)$ theories enjoy a $SO(n_F)$ flavor symmetry at any point of the conformal manifold, while for $\mathcal N=2$ $SO(2N)$ the flavor group includes a $USp(n_F)$ factor \cite{Argyres:1995fw}.\footnote{The fundamental representation of $USp(2N)$ is pseudo-real. Thus, it is possible to introduce $\mathcal N =2$ half-hypermultiplets (i.e. $\mathcal N=1$ chiral multiplets) transforming under it (see e.g. \cite{Bhardwaj:2013qia}). In $\mathcal N=1$ language this means that $n_F$ can be any non-negative integer. On the other hand, the fundamental (or vector) representation of $SO(2N)$ is strictly real. Thus, $\mathcal N=2$ requires full-hypermultiplets which in $\mathcal N=1$ language translate to $n_F$ being even.} In both cases, we can turn on a chemical potential for this flavor group by replacing\footnote{Notice that we are taking $n_F$ to be even even for the case of $USp(2N)$ gauge group, for which this is not required by $\mathcal N=2$ supersymmetry. This is irrelevant since we are interested in taking the $n_F \to \infty$ limit. Also notice that this expression recovers the corresponding formula in \eqref{no-flavor-potentials} when turning off the chemical potentials.}
\begin{equation} \label{SO-USp-N=2-prescription}
  z_{F} = z_c \big( \chi_F(v^n) + \chi_{\bar F}(v^n) \big) = z_c \big( p_{n_F/2}(v^n) + p_{n_F/2} (v^{-n}) \big) \, 
\end{equation}
in their large $N$ partition functions \eqref{USp-partition-function} and \eqref{SO-partition-function}. We then restrict to the flavor singlet sector by integrating over their $SO(n_F)$ and $USp(n_F)$ flavor groups for $n_F \to \infty$ using the techniques in sections \ref{sec:USp-case} and \ref{sec:SO-case}. The Hagedorn condition for the flavor singlets in both cases reads (see Appendix \ref{app:N=2-singlets-USp-SO} for the details on the computation)
\begin{equation} \label{USp-SO-N=2-prescription}
  z_v(x_H,n) + 3 \left( 3 - 4 \alpha^2 \right) z_{c}(x_H,n) + z_c(x_H,n)^{2} = 1 \, ,
\end{equation}
which coincides with the one found for $\mathcal N=2$ $SU(N)$ gauge theories above. The numerical results for the Hagedorn temperature obtained upon resolving the above equation are shown in Table \ref{table:summary} denoted as $T_H^{f.s.i.}$. 

Hence, we find that this $\mathcal N=2$ prescription gives the same results for the three different gauge groups. Thus, the Hagedorn temperature is again controlled by only $\alpha$, as befits our proposal. Let us recall, though, that this prescription is not valid in general for $\mathcal N=1$ theories, so in any case we cannot give a general result for the Hagedorn temperature if restricting to the flavor singlets of the interacting theory. It would be interesting to understand better how to generalize our results beyond the free point and whether one can find a universal prescription to restrict to the closed string background at finite value of the gauge coupling that is valid for all gauge theories. 

\section{Convex Hulls and Scale Separation in AdS} \label{sec:convex-hulls}

In this section, we will compare our CFT results with the expectations from the string theory bulk duals and the Swampland program. We will first explain how the result for the exponential mass decay rate of the higher-spin fields is compatible with the expectation of the critical perturbative string in the AdS gravity dual. Then we will plot the convex hull of the towers of states (as typically done in the Swampland program \cite{Calderon-Infante:2020dhm,Etheredge:2022opl,Etheredge:2023odp,Calderon-Infante:2023ler,Etheredge:2023usk,Castellano:2023stg,Castellano:2023jjt,Etheredge:2024tok}) and use this to provide a new argument for the absence of scale separation in the Einstein gravity AdS$_5$ duals to 4d SCFTs exhibiting overall weak-coupling limits.

\subsection{Matching results for the perturbative critical string} \label{sec:matching}

Consider a string theory compactification whose low energy limit is given by Einstein gravity in a weakly curved background. From the gravity side, one can show that the exponential mass decay rate for the string oscillator modes of the critical perturbative string satisfies \cite{Etheredge:2022opl}
\beq
\label{asugra}
\left|\vec\nabla \log m_{\text str}\right| = \frac1{\sqrt{D-2}} \, ,
\eeq
where $D$ is the space-time dimension of the gravity theory. The 4d SCFTs studied in this paper are dual to AdS$_5\times X$ with $X$ being some internal compact space. Hence, $D=5$. However, a perceptive reader might have noticed that our CFT result for the exponential rate of the higher-spin modes in the Einstein gravity case (namely, the Type 1 limit with $a=c$ to leading order) is
\beq
\alpha=\frac1{\sqrt{2}}\neq \frac1{\sqrt{3}} \, ,
\eeq
so it does not reproduce \eqref{asugra}. What is going on? In the following, we explain how both results are compatible with each other, since it is important to take into account two aspects:
\begin{itemize}
\item The quantization of the string in a highly curved AdS background differs from the quantization in flat space, which will change the value of the exponential rate.
\item The value $\alpha=1/\sqrt{2}$ corresponds to the exponential mass decay rate in the direction of the string coupling (the conformal manifold) while \eqref{asugra} is the gradient with respect to all the scalar fields parametrizing the string mass (even if the scalars are not massless).
\end{itemize}

For simplicity, let us exemplify how to take into account the above aspects by using the well-known example of AdS$_5\times$S$^5$ dual to $\mathcal N=4$ SYM (although our results apply more generally). According to the holographic dictionary, the Yang-Mills marginal coupling $g_{\text{YM}}$ in the CFT is dual to the bulk string dilaton $\Phi$ --which remains as a massless scalar in the bulk theory-- as follows,
\beq
g_{\text{YM}}^2=g_s=\exp\Phi \, .
\eeq
Hence, the weak coupling limit $g_{\text{YM}}\rightarrow 0$ maps to the weak string coupling limit $g_s\rightarrow 0$. However, the string dilaton is not the only scalar of relevance to compute \eqref{asugra}: we also have the radion $R_{S^5}$ parametrizing the radius of the $S^5$.\footnote{The Type IIB axion field, which is dual to the Yang-Mills theta angle, will not be relevant for our computations in the $g_s \to 0$ limit. For this reason, we will ignore it all along.} This radion gets stabilized in the gravity theory by the curvature of the 5-sphere and the $F_5$ flux, such that the vev is set to
\beq
\label{radius}
R_{S^5} M_{Pl,5}\sim R_{AdS} M_{Pl,5}\sim N^{2/3} \, ,
\eeq
where $N$ is the units of $F_5$ flux on the $S^5$ and $R_{AdS}$ the AdS length scale. The $N\to \infty$ limit then explores the large radius limit for the five-sphere, i.e., $ R_{S^5} \to \infty$. From the CFT perspective, $N$ sets the rank of the gauge group and the value of the central charge $c \sim N^2$. The 't Hooft coupling is given by $\lambda=g_{\text{YM}}^2N$. 

We can now distinguish two different regimes depending on the value of $g_{\text{YM}}$ and $N$. Taking $g_{\text{YM}}$ small and $N$ large, we can have $\lambda$ being large or small depending on the relative growth of both parameters. For large values of the 't Hooft coupling, i.e. $\lambda\gg 1$, we can trust the supergravity bulk description, since the $S^5$ radius (and consequently, the AdS length scale) remains large in string units. Indeed, in this regime we have
\begin{equation}
    (R_{AdS} M_s)^4 \sim \lambda \gg 1 \, .
\end{equation}
However, for $\lambda\ll 1$, the AdS length becomes small in string units and the supergravity description breaks down due to significant $\alpha'$ corrections (even if the string coupling is small). In the latter case, the theory is better described by the CFT in the perturbative regime. The mass of the string modes will be different in these two regimes, as we explain in the following.

\medskip

\noindent $\bullet$ \textbf{Weakly-curved supergravity regime:} $\lambda\gg 1$.\\
The string tension is given by
\begin{equation} \label{string-tension}
  T_s \sim e^{\Phi/2} M_{Pl,10}^2 \sim e^{\Phi/2}  R^{-10/3} M_{Pl,5}^2  \, ,
\end{equation}
where we have used that $ M_{Pl,5}^3 \sim  R_{S^5}^5 M_{Pl,10}^8$ and $ R$ is the radion given in 10d Planck units $R=R_{S^5} M_{Pl,10}$. Since the string propagates in a weakly-curved background (i.e., $\lambda\gg 1$), the tower of string excitation modes have a mass scale of order
$ M_s \sim \sqrt{T_s} $. Recall that this mass (in lower dimensional Planck units) is a function of both the radius and the string dilaton. Thus, at a given point of the field space we can compute the following gradient vector
\beq
\vec\zeta_{\rm str}\equiv \vec\nabla \log m_{\text str}=\left( \nabla_\Phi \log m_{\text str},\nabla_R \log m_{\text str}\right) \, ,
\eeq
which encodes the variation of the string mass as a function of the scalars. This vector is known as the \emph{scalar charge-to-mass ratio} in the Swampland literature \cite{Calderon-Infante:2020dhm,Etheredge:2022opl,Etheredge:2023odp,Calderon-Infante:2023ler,Etheredge:2023usk,Castellano:2023stg,Castellano:2023jjt,Etheredge:2024tok}. The dot product (necessary to compute the length of the vector) is defined with respect to the field space metric (i.e., the field metric appearing in the kinetic term of the scalars in Einstein frame). Equivalently, we can compute the derivatives  with respect to the canonically normalised scalar fields such that the dot product becomes cartesian (i.e., $g_{ij}=\delta_{ij}$). In terms of the canonically normalized fields $\hat R$ and $\hat \Phi$, the string mass behaves as
\begin{equation} \label{string-scale}
  M_s \sim \sqrt{T_s} \sim M_{Pl,5}\,e^{- \frac{1}{2\sqrt{2}} \hat\Phi - \frac{\sqrt{30}}{12} \hat R} \, ,
\end{equation}
yielding
\begin{equation} \label{string-vector}
   \vec\zeta_{\rm str}^{\rm{\, sugra}}=  \vec\nabla \log m_{\rm str}^{\rm sugra} = \left( \frac{1}{2\sqrt{2}} , \frac{\sqrt{30}}{12} \right) \, .
\end{equation}
Here we have used that $  R =  e^{\frac{\sqrt{30}}{20} \, \hat R}  $
and $ \hat\Phi = - \frac{1}{\sqrt{2}} \,\Phi $, as can be read from the bulk action.\footnote{Taking the following ansatz for the 10d metric in the Einstein frame
\begin{equation}
  ds_{10}^2 = e^{-\frac{\sqrt{30}}{6} \hat{R}} \, ds_{5}^{2} + e^{\frac{\sqrt{30}}{10}\hat{R}}  \, ds_{S^5}^{2} \, ,
\end{equation}
one gets the relevant part of the 5d effective action written in terms of the canonically normalized fields
\begin{equation}
  S_{5d} = M_{Pl,5}^{3} \int d^{5}x \sqrt{-g_{5}} \left[ \frac{1}{2} \mathcal{R} - \frac{1}{2} (\partial \hat \Phi)^{2} - \frac{1}{2} (\partial \hat R)^{2} - V(\hat R) \right] \, .
\end{equation}
with $ \hat\Phi = - \frac{1}{\sqrt{2}} \,\Phi $ and $  R = \, e^{\frac{\sqrt{30}}{20} \, \hat R}  $ in 10d Planck units. \label{footnote-compactification}} The length of this vector  is 
\beq
|\vec\nabla \log m_{\rm str}^{\rm sugra}|=\frac{1}{\sqrt{3}},
\eeq
as expected for a string in five dimensions (see \eqref{asugra}).

\medskip

\noindent $\bullet$ \textbf{Field theory perturbative regime:} $\lambda\ll 1$.\\
For small $\lambda$, the gravitational background becomes strongly curved with an AdS scale of order the string scale, so that the above supergravity formulae are no longer valid. However, this is precisely the regime in which the gauge theory in the field theory side becomes perturbative, and our previous results about the HS operators in the free limit apply. The anomalous dimension of the tower of higher-spin operators scale as
\begin{equation} \label{anomalous-dimension}
	\gamma \sim Ng_{\text{YM}}^2 \sim \lambda \, ,
\end{equation}
where we have restored the dependence with $N$. The mass of the dual higher-spin fields will scale as $(M_{s} R_{\text{AdS}})^2\sim \gamma$. In order to compute the scalar charge-to-mass ratio vector, we need to know the metric in this space of parameters that maps to the field metric in the bulk dual. For $g_{\text{YM}}$ is easy, since it corresponds to the Zamolodchikov metric as explained in Section \ref{sec:review}; while for $N$ is not so obvious. Using \eqref{metrictau} and \eqref{alpha} we know that the \emph{canonically normalized coupling} $\hat g$ that directly maps to the canonically normalised bulk string dilaton is $\hat g=-\alpha^{-1}\log g_{\text{YM}}$ with $\alpha$ given in \eqref{alpha}. Using for the moment  the following parametrization
\beq
\label{distance-N}
N=\exp( \alpha_N \hat N),
\eeq
we find that the mass of the string excitation modes (in 5d Planck units) reads
\begin{equation} \label{string-scale-CFT}
\begin{split}
	& M_s \sim \frac{\sqrt{\lambda}}{N^{2/3}} \sim g_{\text{YM}} N^{-1/6} \sim e^{-\frac{1}{\sqrt 2} \hat g - \frac{1}{6} \alpha_N \hat N} \, ,
\end{split}
\end{equation}
where we have used that $R_{AdS} M_{Pl,5}\sim N^{2/3}$ and replaced $\alpha=1/{\sqrt{2}}$ since $a=c$. This yields the following scalar charge-to-mass ratio
\begin{equation} \label{HS-vectors}
	   \vec\zeta_{\rm str}^{\rm{\, CFT}}=\vec\nabla \log m_{\rm str}^{\rm CFT} = \left( \frac{1}{\sqrt 2} , \frac{\alpha_N}{6} \right) \, .
\end{equation}
Now we can see that the exponential mass decay rate of the tower when moving in the conformal manifold obtained in \eqref{alpha} corresponds precisely to the the projection of $\vec\nabla \log m_{\rm str}^{\rm CFT} $ on the dilaton direction (i.e. the first component):
\beq
\alpha=\vec\nabla \log m_{\rm str}\cdot(1,0)=\frac1{\sqrt{2}} \, .
\eeq

Fixing the coefficient $\alpha_N$ purely from the CFT is not trivial, as it is not related to any exactly marginal coupling. Moving in this direction requires breaking the conformal symmetry, triggering an RG flow. In this work, we will simply fix it by matching the result with the bulk, in such a way that it reproduces the canonically normalized radion. Using \eqref{radius} and the field metric for the radion we obtain that $\alpha_N=\sqrt{6/5}$. This implies that the length of the vector above becomes $\sqrt{8/15}$ in the small 't Hooft coupling limit. Interestingly, this is precisely the value that would correspond to a KK tower (or equivalently, to winding modes of a T-dual theory) associated to five extra dimensions.\footnote{When compactifying Einstein gravity in $D+n$ down to $D$ dimensions, the KK modes satisfy $|\vec \nabla \log m_{\rm KK}| = \sqrt{\frac{D-2+n}{n(d-2)}}$ \cite{Etheredge:2022opl}. For $D=5$ and $n=5$ this indeed yields $\sqrt{8/15}$. \label{footnote-KK} } Thus, as $\lambda$ decreases, the scaling of the string oscillator modes shifts from the original supergravity result of a perturbative string \eqref{string-scale} to the value expected for a tower of string winding modes. The intuition behind is that, as $\lambda$ gets small, the AdS scale becomes of order string scale; so the string propagates in a dimension of the same size than its string length. Hence, the mass of the oscillator string modes scales formally in the same way as that of a tower of wrapping modes of a string feeling five extra AdS-size dimensions.  

Finally, let us point out that these results are valid for any of the holograhpic SCFTs studied in this paper, and not only for  $\mathcal N=4$ SYM. The formulae for the weakly-curved supergravity regime are valid for any ten-dimensional string theory on AdS$_5\times X_5$, with $X_5$ some internal manifold.\footnote{To see this, notice that focusing on the breathing mode of the $X_5$ (which controlls its overall volume) allows us to just replace $S^5 \to X_5$ in Footnote \ref{footnote-compactification}, thus obtaining the same $5d$ effective action. The rest of the computations proceed in exactly the same way, again replacing $S^5 \to X_5$. \label{footnote-X5}} Similarly, the results in the field theory perturbative regime are valid for any of the SCFTs in our mini-landscape with $\alpha=1/\sqrt{2}$, or more generally, for any 4d SCFT exhibiting a Type 1 limit. Hence, all the previous equations remain unchanged and the results for the charge-to-mass ratio vectors of the towers are universal for all holographic 4d SCFTs with simple gauge group. For the non-holographic ones, one can still easily generalize the field theory result \eqref{HS-vectors} by simply changing the value of $\alpha$ as necessary, but the gravity result will not necessarily hold anymore.

\subsection{Convex Hull Distance Conjecture in $\mathcal N=4$ SYM} \label{sec:convex-hull-N=4}

As it will get clear momentarily, it becomes very useful to plot the convex hull of the scalar charge-to-mass ratio vectors of the different towers of states. This allows us to test several refinements of the Distance Conjecture along any direction of the field space.

\medskip

In the previous subsection, we have computed the scalar charge-to-mass ratio vector of the tower of string modes in two different regimes, when $\lambda$ is very large or very small. However, this not the only tower of states becoming light at large radius and weak coupling: there is also the Kaluza-Klein tower of states of the $S^5$. These KK modes are BPS and can be equivalently computed from the supergravity or the field theory description, since the expression for their mass does not change as we move in the field space. 

From the supergravity perspective, the KK mass is given by
\begin{equation}\label{KKbulk}
  M_{\text{KK}} \sim R_{S^5}^{-1} \sim M_{Pl,5}\,e^{-\sqrt{\frac{8}{15}} \hat R} \, ,
\end{equation}
yielding the following scalar charge-to-mass ratio
\begin{equation} \label{KK-vector}
  \vec{\zeta}_{\text{KK}} \equiv \vec \nabla \log M_{\text{KK}} = \left( 0 , \sqrt{\frac{8}{15}} \right) \, .
\end{equation}
The length of this vector is $\sqrt{8/15}$, as expected from a KK tower decompactifying five dimensions (see Footnote \ref{footnote-KK}).

From the CFT perspective, the KK modes are dual to BPS operators whose conformal dimension does not depend on either $g_\text{YM}$ or $N$. As a consequence, the mass of the dual fields in the bulk behave as
\begin{equation}
  M_{\text{KK}}R_{\text{AdS}} \sim \mathcal O(1) \, 
\end{equation}
for any value of $\lambda$. This reproduces \eqref{KKbulk} since there is no separation of scales in AdS$_5\times$S$^5$, namely $R_{S^5}\sim R_{AdS}$. With this, we can aim to compute the scalar charge-to-mass ratio vector purely from the CFT. Using \eqref{radius}, we find
that the KK mass in 5d Planck units reads
\begin{equation}
  M_{\text{KK}} \sim N^{-2/3} \sim e^{- \frac{2}{3} \alpha_N \hat N} \, ,
\end{equation}
where in the last step we have plugged the parametrization \eqref{distance-N}. From here we read off the scalar charge-to-mass ratio
\begin{equation} \label{KK-vector-CFT}
  \vec \zeta_{\text{KK}} = \left( 0 , \frac{2 \alpha_N}{3} \right) \, .
\end{equation}
As a double-check, this result agrees with \eqref{KK-vector} for $\alpha_N = \sqrt{6/5}$, which is precisely the coefficient obtained from matching the sugra-CFT results for the string modes. Hence, matching the result for the KK modes is an equivalent way to obtain $\alpha_N$. This matching is legit since it can be done in the weakly-curved regime of the bulk, where both \eqref{KK-vector} and \eqref{KK-vector-CFT} can be trusted.

\medskip

We now have all the ingredients to plot the convex hull of the towers. Consider the field space spanned by the canonically normalized radion and dilaton. We will focus on the first quadrant, where the radius is large and string coupling is small. Notice that these vacua exhibit Type IIB S-duality, so that it suffices to focus on the region in which $g_s$ is small, as the large $g_s$ regime will mirror the results in the S-dual theory. On the contrary, the limit in which the radius of the five-sphere becomes small cannot be explored by a family of AdS vacua of varying $N$, since the potential becomes a runaway when the flux $N\rightarrow 0$.\footnote{For $N=0$ the potential receives no contribution from the $F_5$ flux and is only sourced by the $S^5$ curvature, which indeed gives a runaway potential without any local minima.}

\begin{figure}[htb]
\begin{center}
\includegraphics[scale=.55]{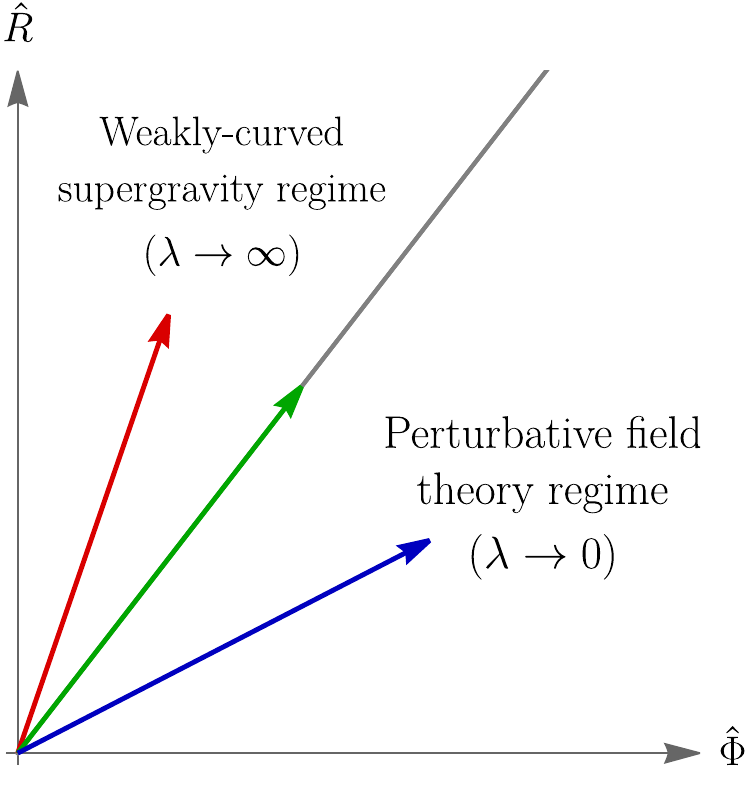}
\caption{\small Regions in the $(\hat R, \hat \Phi)$ separated by the 't Hooft line (gray line). The arrows represent different directions in this space leading to $\lambda\rightarrow\infty$ (red), $\lambda\rightarrow 0$ (blue) and $\lambda\rightarrow const.$ (green).} 
\label{fig:regions}
\end{center}
\end{figure}

Since we are using flat coordinates in the field space, i.e. the canonically normalized fields $(\hat R, \hat \Phi)$, asymptotic geodesics approaching an infinite distance limit ($\hat R\rightarrow\infty, \hat \Phi\rightarrow\infty$) correspond to straight lines with different slopes. This set of infinite distance trajectories is divided into two regions based on whether $\hat R$ or $\hat\Phi$ grows faster, which in turn determines whether the 't Hooft coupling $\lambda$ goes to zero or diverges in the limit. These two regions are separated by the 't Hooft line defined by keeping $\lambda$ constant in the limit, and correspond to the weakly coupled supergravity regime and to the perturbative field theory regime. We will also refer to these regions as two different duality regimes, since they are better described by one or the other of the above dual theories. In Figure \ref{fig:regions} we plot these two regions of the field space in the orthonormal flat coordinates, as well as  the tangent vector of some asymptotic trajectories illustrating the three options with $\lambda\rightarrow\infty$ (red), $\lambda\rightarrow 0$ (blue) and $\lambda\rightarrow const.$ (green) in the limit. 

\begin{figure}[htb]
\begin{center}
\includegraphics[scale=.55]{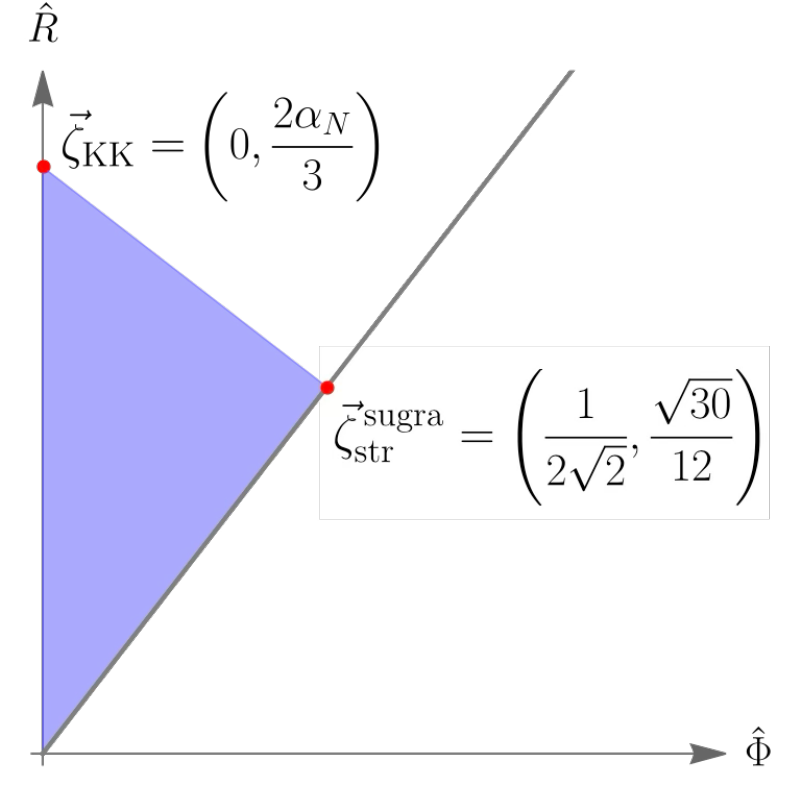} \quad
\includegraphics[scale=.55]{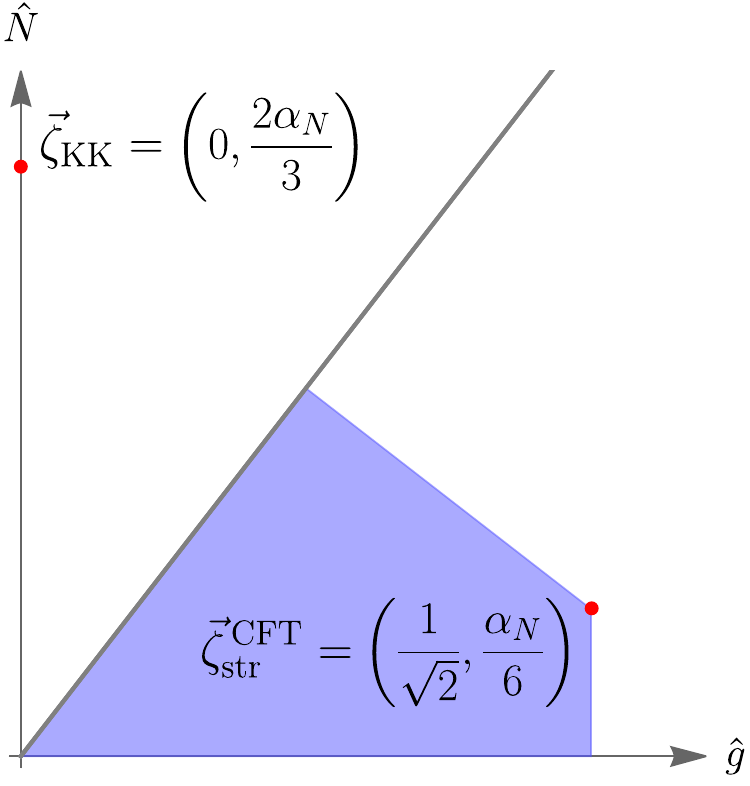}
 \caption{\small Convex hulls of the asymptotic towers (i.e. the KK tower and the string oscillator modes) depending on the duality regime. Left: weakly-curved supergravity regime ($\lambda \to \infty$), using \eqref{KK-vector} and \eqref{string-vector}. Right: perturbative field theory regime ($\lambda \to 0$), using \eqref{KK-vector-CFT} and \eqref{HS-vectors}. The convex hull is only shown in its regime of validity. The position of the KK tower does not change, while the string tower changes depending on the duality regime. Holographic duality fixes $\alpha_N = \sqrt{6/5}$.} 
\label{fig:ch-local}
\end{center}
\end{figure}

Given a particular geodesic approaching an infinite distance limit, we can now plot the scalar charge-to mass ratio vectors of the towers in the tangent plane at a given point of the field space. The convex hull of these vectors forms a \emph{frame simplex} (see \cite{Etheredge:2024tok}) which remains rigid under variations of the infinite distance trajectory as long as we stay within the same duality regime. In Figure \ref{fig:ch-local} we plot the convex hull of the scalar charge-to-mass ratio vectors of the asymptotic towers of states for the two duality regimes: $\lambda\rightarrow\infty$ (better described by supergravity, left figure) and $\lambda\rightarrow 0$ (better described by the boundary field theory, right figure).\footnote{We are using the prescription in \cite{Calderon-Infante:2020dhm} to build the convex hull when focusing on a subspace of directions. Apart from the scalar charge-to-mass ratios (red dots in Figure \ref{fig:ch-local}), we also include their projections onto the boundary of the allowed region (not shown in the figure) and the origin to build the convex hull.} The convex hull of the left figure is the typical one associated with a perturbative string theory regime (the geometry of such convex hulls was classified in \cite{Etheredge:2024tok} for flat space compactifications). This paper provides the first study of the convex hull in a strongly curved background.

Note that the location of the KK modes remains the same regardless of the value of $\lambda$. However, the location of the string modes differs when considering trajectories with $\lambda\rightarrow\infty$ or $\lambda\rightarrow 0$. 
This variation arises because the quantization of the string changes in strongly versus weakly curved backgrounds, affecting the mass dependence of the oscillator modes on the radius and the dilaton, as explained in the previous subsection. Interestingly, this change ensures that the convex hull continues along the straight line joining the KK modes.

We emphasize that this field space is not a moduli space since the radion is massive. The moduli space (which maps to the conformal manifold) is one-dimensional and limited to the horizontal axis, i.e. the string coupling. Therefore, to take the infinite distance decompactification limit $\hat R\rightarrow\infty$ one must consider a discrete family of AdS vacua with different $N$, and take the limit $N\rightarrow\infty$ within this family. This complicates defining a proper notion of distance purely from the CFT (see \cite{Lust:2019zwm,Kehagias:2019akr,DeBiasio:2020xkv,DeBiasio:2022omq,Velazquez:2022eco,DeBiasio:2022nsd,DeBiasio:2022zuh,Shiu:2022oti,Farakos:2023nms,Li:2023gtt,Tringas:2023vzn,Basile:2023rvm,Shiu:2023bay,Palti:2024voy,Mohseni:2024njl,Debusschere:2024rmi} for proposals from the AdS perspective). For our purposes, we do not need to define a notion of distance in this discrete family of vacua, since we are simply interested in the variation of the masses as a local function of the fields. This variation of the masses (encoded in the scalar charge-to-mass ratios) depends only on the kinetic field metric and is independent of whether the scalars are stabilized by a potential. Therefore, we can compare the exponential rates of the towers with results from the literature on the Distance Conjecture, which simply uses this kinetic field metric.

\medskip

Using results from AdS/CFT integrability in the planar limit  (i.e. $N\to\infty$ with the 't Hooft coupling $\lambda = N g_{\text{YM}}^2$ fixed) we can also plot the convex hull for any value of $\lambda$ and understand how to interpolate between the two images in Figure \ref{fig:ch-local} in a continuous way. The string excitation modes are dual to non-protected higher-spin operators whose conformal dimensions depend non-trivially on both $g_\text{YM}$ and $N$. 
As reviewed e.g. in \cite{Dorigoni:2015dha}, we have
\begin{equation} \label{integrability}
  \Delta_J - J = 2 \Gamma_{\text{cusp}}(\lambda) \log J \, + \, \cdots \, , \quad \text{for }J \to \infty \, ,
\end{equation}
where $J$ is the spin of the higher-spin operator with conformal dimension $\Delta_J$. The function $\Gamma_{\text{cusp}}(\lambda)$ is the famous cusp anomalous dimension, which can be computed for any value of $\lambda$ by solving the so-called Beisert-Eden-Staudacher (BES) equation. There is no analytic expression, but one can find the results and numerical plots in \cite{Gromov:2023hzc}.
It is also possible to obtain systematic expansions of this quantity for $\lambda \to 0 , \infty$:
\begin{equation}
\begin{split}
	&\Gamma_{\text{cusp}}(\lambda) \approx \frac{\lambda}{4 \pi^2} \, , \quad \lambda \to 0 \, , \\
	&\Gamma_{\text{cusp}}(\lambda) \approx \frac{\sqrt{\lambda}}{2 \pi} \, , \quad \lambda \to \infty \, .
\end{split}
\end{equation}
Using the AdS/CFT dictionary, the mass of the dual higher-spin fields read
\begin{equation} \label{eq:string-states-CFT}
  (M_{s} R_{\text{AdS}})^2 = (\Delta_J + J - 2)(\Delta_J - J - d + 2) \approx 4 \, \Gamma_{\text{cusp}}(\lambda) \, J \log J \, ,
\end{equation}
where, for consistency, we are only keeping the leading order in $J\to \infty$. The string mass in 5d Planck units will then be given by
\begin{equation}
\label{Ms1}
  M_{s} \sim \frac{\sqrt{\Gamma_{\text{cusp}}(\lambda)}}{N^{2/3}} \ .
\end{equation}
For $\lambda \to 0 , \infty$ we then find
\begin{equation} \label{string-scale-CFT2}
\begin{split}
	& M_s \sim \frac{\sqrt{\lambda}}{N^{2/3}} \sim g_{\text{YM}} N^{-1/6} \sim e^{-\frac{1}{\sqrt 2} \hat g - \frac{1}{6} \alpha_N \hat N} \, , \quad \text{for }\lambda \to 0 \, , \\
	& M_s \sim \frac{\lambda^{1/4}}{N^{2/3}} \sim g_{\text{YM}}^{1/2} N^{-5/12} \sim e^{-\frac{1}{2\sqrt 2} \hat g - \frac{5}{12} \alpha_N \hat N} \, , \quad \text{for }\lambda \to \infty \, . 	
\end{split}
\end{equation}
We can see that the scalar charge-to-mass ratio of these string states depends on the value of $\lambda$, such that we have
\begin{align}
	&\label{HSl0} \vec\zeta_{\rm str}(\lambda) = \left( \frac{1}{\sqrt 2} , \frac{\alpha_N}{6} \right) \, , \quad \text{for }\lambda \to 0 \, , \\
	& \label{HSli}\vec\zeta_{\rm str}(\lambda) = \left( \frac{1}{2\sqrt 2} , \frac{5\alpha_N}{12} \right) \, , \quad \text{for }\lambda \to \infty \, .
\end{align}
Recall that the $\lambda \to 0,\infty$ limits in the CFT correspond to the strongly and weakly-curved regime for the string in the bulk.  Indeed, the $\lambda \to \infty$ scalar charge-to-mass ratio above recovers the bulk supergravity result $\vec\zeta^{\, \rm sugra}_{\rm str}$ in \eqref{string-vector} for $\alpha_N = \sqrt{6/5}$; while the $\lambda \to 0$ result reproduces the perturbative field theory result $\vec\zeta^{\, \rm CFT}_{\rm str}$ in \eqref{HS-vectors}. 

\medskip

But what happens in the planar limit in which $\lambda$ remains constant as $N\rightarrow\infty$ and $g_{\text{YM}}\rightarrow 0$?
Expressing the 't Hooft coupling in terms of the canonically normalized parameters $\hat g$ and $\hat N$, we see that taking $N\to \infty$ with $\lambda$ fixed corresponds to the trajectory
\begin{equation} \label{tHooft-line}
  \hat g = \frac{\alpha_N}{ \sqrt 2} \hat N \, ,
\end{equation}
which we denoted above as the 't Hooft line.
This geodesic is characterized by the unit tangent vector 
\begin{equation}
  \hat T_{\lambda} = \frac{1}{\sqrt{\alpha_{N}^2 +2}} \left( \alpha_{N} ,  \sqrt 2 \right) \, ,
\end{equation}
which was represented as a green vector in Figure \ref{fig:regions}. All infinite distance geodesics that keep $\lambda$ constant in the limit have this same tangent vector (i.e. the same asymptotic direction), while the concrete value of $\lambda$ acts as the impact parameter to this trajectory. 

Since $\Gamma_{\text{cusp}}(\lambda)$ only depends on $\lambda$, the exponential rate of the tower along a direction with tangent vector $ \hat T_{\lambda}$ will remain constant and independent on the concrete value of $\lambda$. More explicitly, using \eqref{Ms1}, we can decompose
\beq
\label{orth}
\vec\zeta_{\rm str}(\lambda) \equiv \vec \nabla\log M_s =-\frac23(0,\alpha_N)+\vec\nabla\log \sqrt{\Gamma(\lambda)} \, ,
\eeq
where $\vec\nabla \log \sqrt{\Gamma(\lambda)}=\partial_\lambda\log \sqrt{\Gamma(\lambda)} \, \lambda \, (-\sqrt{2},\alpha_N)$. Hence, one gets that $\vec\nabla\log \sqrt{\Gamma(\lambda)} \cdot  \hat T_{\lambda}=0$, so that the exponential rate along $ \hat T_{\lambda}$ reads
\begin{equation} \label{t Hooft-limit}
  \alpha_\lambda = \hat T_{\lambda} \cdot \vec\zeta_{\rm str}(\lambda) = \frac{2\sqrt{2} \, \alpha_N}{3 \sqrt{\alpha_N^2 + 2}} \, ,
\end{equation}
which is indeed independent of $\lambda$. In other words, the vector $\vec\zeta_{\rm str}(\lambda)$  always lies in the same line, which is orthogonal to the 't Hooft line. As $\lambda$ decreases, it interpolates continuously between the supergravity result for large 't Hooft coupling \eqref{HSli} and the perturbative field theory result \eqref{HSl0} for small 't Hooft coupling, following the straight line connecting them. Therefore, the scalar charge-to-mass ratio vector \emph{jumps} from one result to the other as we change the direction of the asymptotic trajectory, while \emph{slides} continuously\footnote{This sliding is continuous since the function $\Gamma(\lambda)$ is continuous \cite{Gromov:2023hzc}. It would be interesting to study the sliding in more detail by using the results of \cite{Gromov:2023hzc} and check for instance whether it is also monotonous.} from one to the other as a function of the impact parameter $\lambda$. 
This behaviour is similar to the one observed for the sliding of the non-BPS KK tower of Type I' found in \cite{Etheredge:2023odp}. The latter is a completely unrelated setup, but it is interesting to notice that, whenever the scalar charge-to-mass ratio vector of a tower of states changes as we move in field space, it always seems to do so in a way perpendicular to the asymptotic direction. In this example, we can be very explicit about this \emph{sliding}, thanks to the results from integrability in $\mathcal{N}=4$ SYM.

Furthermore, it turns out that not only $\vec\zeta_{\rm str}(\lambda)$ lies in the same line for any $\lambda$. Given
\begin{equation}
  \hat T_\lambda \cdot \vec\zeta_{\text{KK}} = \frac{2\sqrt{2} \, \alpha_N}{3 \sqrt{\alpha_N^2 + 2}} = \hat T_\lambda \cdot \vec\zeta_{\rm str}(\lambda) \, ,
\end{equation}
we find that $\vec\zeta_{\text{KK}}$ lies in the same line as $\vec\zeta_{\rm str}(\lambda)$, regardless of the value of $\alpha_N$. This means that in the 't Hooft limit both the string excitations and the KK modes are becoming massless at the same rate. This identifies this limit as the usual emergent string limits studied in flat space vacua \cite{Lee:2019wij}.

\medskip

\begin{figure}[htb]
\begin{center}
\includegraphics[scale=.55]{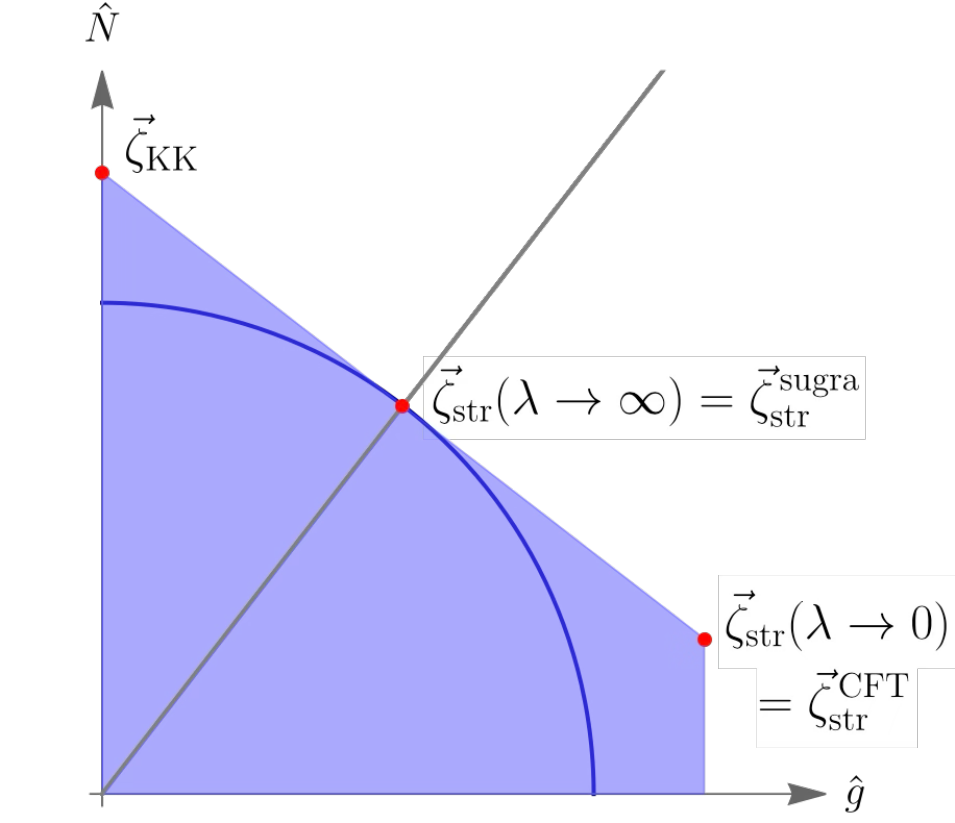}
\caption{\small Global convex hull for $\alpha_N=\sqrt{6/5}$, i.e., the value predicted by the bulk. The circular blue line represents the Sharpened SDC bound for $D=5$, $\alpha \geq 1/\sqrt{3}$. The bound is satisfied for any direction and saturated along the 't Hooft line (gray straight line). When necessary, we indicate the regime of validity of the vector in brackets (c.f. Figure \ref{fig:regions}).} 
\label{fig:ch-global}
\end{center}
\end{figure}

At this moment, we know the asymptotic scalar charge-to-mass ratio vectors (and their convex hull) of the towers of states at any point of the field space parametrized by $(\hat\Phi, \hat R)$ (or equivalently by $(\hat g, \hat N)$). We remark that these convex hulls are plotted locally at a given point of the field space, and characterize the different duality frames. As a final step, we can aim to \emph{glue} the different convex hulls that are valid at the different regions of the field space, to build a global figure that captures the mass decay rate of the lightest tower of states for any asymptotic direction. By gluing the left and right-hand side of Figure \ref{fig:ch-local} for $\alpha_N=\sqrt{6/5}$, we obtain Figure \ref{fig:ch-global}. There are two key properties of this global convex hull that we would like to remark, and that we discuss in detail now.

First, this global convex hull is continuous; the two local convex hulls glue nicely together without producing any discontinuity. We note that this continuity holds for any $\alpha_N$. In fact, it is guaranteed by one of the properties of the vector $\vec\zeta_{\rm str}(\lambda)$ shown above, namely that it always lies in a straight line that is perpendicular to the 't Hooft line. This feature was derived by using the structure in \eqref{integrability} for the conformal dimension of the higher-spin operators in the 't Hooft limit (see argument around eq. \eqref{orth}). One can generalize this argument to show that it holds for any $\Delta_J = N^a f(\lambda)$ with $f(\lambda)$ a continuous function and $a$ independent of $\lambda$,\footnote{We get $\vec\zeta_{\rm str}(\lambda\rightarrow \infty) =(\frac{a}2-\frac23)(0,\alpha_N)+\vec\nabla\log \sqrt{f(\lambda)}$ with $\vec\nabla\log \sqrt{f(\lambda)} \cdot  \hat T_{\lambda}=0$ so that the exponential rate along $ \hat T_{\lambda} $ is independent of $\lambda$.} which is precisely what one expects from a well-behaved 't Hooft limit. Therefore, we conclude that the continuity of the global convex hull along the line of constant $\lambda$ is guaranteed by having a bona fide 't Hooft limit.

Secondly, the convex hull includes a ball of radius $\alpha_0=|\vec\zeta_{\rm str}(\lambda\rightarrow \infty)|$. Note that if the global convex hull of the towers includes a ball of radius $\alpha_0$, it implies that there is always a tower of states becoming light with exponential rate $\alpha\geq \alpha_0$ along any infinite distance geodesic. This is known as the Convex Hull Distance Conjecture \cite{Calderon-Infante:2020dhm}. Recent evidence from flat space compactifications \cite{Etheredge:2022opl,Etheredge:2023odp,Castellano:2023stg,Castellano:2023jjt} suggests that the minimal value for this exponential rate is $\alpha_0=1/\sqrt{D-2}$, as proposed in \cite{Etheredge:2022opl}. This lower bound for the exponential rate is known as the Sharpened Distance Conjecture, and it is saturated by the tower of string oscillator modes in a weakly curved background (see \eqref{asugra}). A priori, there is no obvious reason to expect that this minimal value of $\alpha$ should also hold in AdS compactifications, since the quantization of the string changes in strongly curved backgrounds. Nevertheless, in Figure \ref{fig:ch-global} we can see that the Convex Hull Distance Conjecture is still satisfied with $\alpha_0=1/\sqrt{D-2}$, implying that the exponential rate of the different towers is always bigger or equal than $1/\sqrt{3}$, in agreement with the sharpened bound of the Distance Conjecture. This is satisfied in a very non-trivial way, as the mass of the higher-spin fields changes in a way that follows the straight line touching (but without cutting) the ball of radius $1/\sqrt{3}$ that connects with the KK modes. Unlike in flat space, though, the bound is not saturated whenever the leading tower is given by string oscillator modes, but only when the string propagates in a weakly curved background (and it yields the leading tower). Only in that case does one obtain the usual emergent string limit in which the string tower and the KK modes become massless at the same rate. Otherwise, as we move into the strongly curved regime of small $\lambda$, the string modes decay at a faster rate than the KK tower, and the resulting theory is a free field theory in four dimensions. As observed in other examples \cite{Etheredge:2022opl,Etheredge:2023odp,Calderon-Infante:2023ler,Castellano:2023stg,Castellano:2023jjt,Etheredge:2024tok}, the convex hulls of the towers can be used as a new tool to reveal the different duality frames of a given theory. In this case, it clearly shows the two duality frames: a perturbative string theory above the 't Hooft line and a conformal field theory below the 't Hooft line, related by the holographic duality.

\medskip

Before finalizing this section, let us remark that the convex hull of all the other theories with Type 1 limits in our mini-landscape are expected to look exactly the same as the one in Figure \ref{fig:ch-global}. Following our proposal, their bulk duals are described by a critical string background of the form $AdS_5 \times X_5$, with $X_5$ some five-dimensional internal space. Thus, the scalar charge-to-mass ratios of the string excitation and KK modes for $\lambda\to\infty$ will take the form in \eqref{string-vector} and \eqref{KK-vector} (see footnote \ref{footnote-X5}). Similarly, the CFT result in \eqref{HS-vectors} will be unchanged. We also expect that the KK modes are dual to BPS operators in the CFT, which will satisfy \eqref{KK-vector-CFT} and allow us to fix $\alpha_N$ by matching with \eqref{KK-vector}. In this way, we end up with the same local convex hulls for both $\lambda \to \infty$ and $\lambda \to 0$. Even though in this case we do not have the integrability results for the HS operators, we still expect a bona fide 't Hooft limit, so that $\vec\zeta_{\rm str}(\lambda)$ will again lie in a straight line orthogonal to the 't Hooft line for finite $\lambda$. Hence, the picture is exactly the same.

\subsection{New argument against AdS scale separation in 4d SCFTs} \label{sec:scale-separation}

In the previous subsection, we have seen how the two convex hulls of the towers of states for $\lambda\rightarrow 0$ and $\lambda\rightarrow \infty$ nicely glue together along the 't Hooft line, in such a way that the sharpened lower bound for the Distance Conjecture is satisfied. This nice fit between the convex hulls shows that nothing strange happens in the 't Hooft limit (as expected from having a well-defined large $N$ limit at constant 't Hooft coupling), and we can continuously follow the mass of the towers as continuous functions in the field space. Notice that this continuous interpolation occurs for any value of $\alpha_N$, which simply sets the normalization  of the vertical axis. However, it would be crucially violated if the AdS vacuum would exhibit scale separation between the AdS length and the size of the extra dimensions, as we explain in the following.

\medskip

When plotting in the same figure the scalar charge-to-mass vectors associated to the higher-spin fields for $\lambda\ll 1$ and the KK modes, it was essential to use that $R_{S^5} \sim R_{AdS}$. Recall that the higher-spin fields (from a CFT perspective) are sensitive to the value of $R_{AdS}$, while the KK modes will depend on $R_{S^5}$. Only when these two scales are comparable, we obtain that the convex hull of these two towers is a straight line. Furthermore, only when fixing $\alpha_N$ to the value predicted by the bulk, this straight line precisely touches the ball of radius $1/\sqrt{3}$, as in Figure \ref{fig:ch-global}. For these reasons, one can expect that, when allowing for scale separation between the AdS and the $S^5$, both the continuity of the convex hull and the sharpened bound for the Distance Conjecture will not hold anymore. Let us see this more explicitly.

Suppose we start with a holographic $4d$ SCFT with a conformal manifold and an asymptotic limit in which the whole theory becomes free. The bulk should then be described by a critical string background of the form $AdS_5 \times X_5$ with $X_5$ some five-dimensional internal space. Of course, the only consistent backgrounds of this type that are currently known are not scale separated, but the goal here is to find a new argument as of why this has to be the case. Setting the 5d Planck scale $M_{Pl,5}=1$, let us consider the following parametrization
\beq
\label{beta}
R_{X_5}^{-1}\sim R^{-\beta}_{AdS} \sim N^{-2\beta/3}\, ,
\eeq
where we allow for a possible scale separation if $\beta\neq 1$. 

From the supergravity perspective, the scalar charge-to-mass ratio vectors of the towers remain unchanged, as they only depend on $R_{X_5}$. We thus have
\begin{equation}
\begin{split}
    \vec\zeta_{\rm KK} &= \left(0, \sqrt{\frac{8}{15}} \right) =\left(0, \frac{2 \beta}{3} \alpha_N \right) \, ,\\
    \vec\zeta_{\rm str}^{\, \rm sugra} &=  \left( \frac{1}{2\sqrt{2}} , \frac{\sqrt{30}}{12} \right) = \left( \frac{1}{2\sqrt 2} , \frac{5 \beta}{12} \alpha_N \right) \, ,
\end{split}
\end{equation}
where in the RHS we have used \eqref{beta} and \eqref{distance-N} to further write the $\vec \zeta$-vectors in CFT variables. Matching the two sides of this equation, we can fix the normalization factor to 
\begin{equation} \label{alphaN-scale-separation}
  \alpha_N = \frac{1}{\beta} \sqrt{\frac{6}{5}} \, .
\end{equation}

From the field theory side, we will not assume anything about the KK modes but only use the result for the HS tower as $\lambda\rightarrow 0$. Recall that our starting point is a holographic 4d SCFT with a non-compact conformal manifold exhibiting a limit in which the whole theory becomes free. As shown in Section \ref{sec:types-limits}, this infinite distance limit is of Type 1 in our classification, and necessarily yields a HS tower with
\begin{equation} 
	\vec\zeta_{\rm str} ^{\, \rm CFT} = \left( \frac{1}{\sqrt 2} , \frac{\alpha_N}{6} \right) = \left( \frac{1}{\sqrt 2} , \frac{1}{\beta} \frac{1}{\sqrt{30}} \right) \, 
\end{equation}
as in \eqref{HS-vectors}. In the last equality we have plugged \eqref{alphaN-scale-separation}.

\begin{figure}[htb]
\begin{center}
\includegraphics[scale=.55]{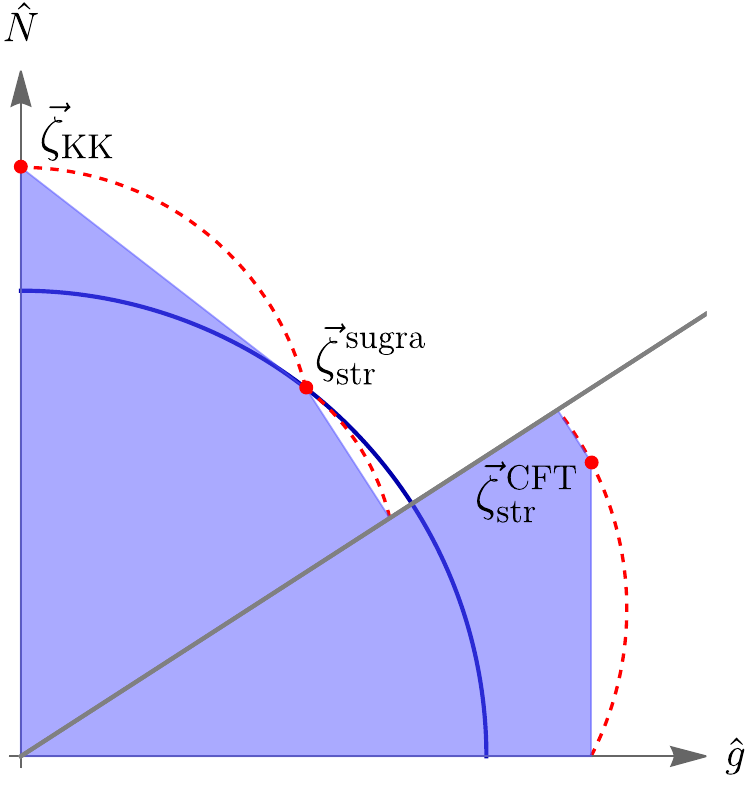} \quad
\includegraphics[scale=.55]{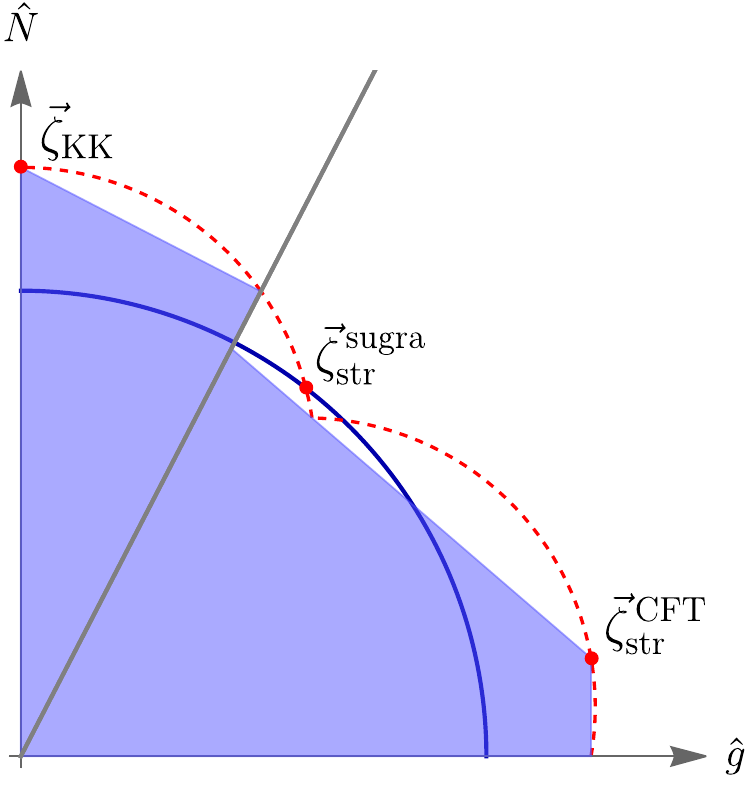}
\caption{\small Global convex hulls for $\beta=0.5$ (left) and $\beta=1.5$ (right). In both cases the convex hull is discontinuous and cuts the ball of radius $1/\sqrt{3}$ (circular blue line), leading to a violation of the Sharpened Distance Conjecture. To see more clearly how the convex hulls are built when restricting to their regime of validity, we also plot the exponential decay rate of the leading tower as function of the direction (red dashed line).} 
\label{fig:ch-separation-scales}
\end{center}
\end{figure}

The global convex hull obtained by gluing together the two regimes of large and small 't Hooft coupling is shown in Figure \ref{fig:ch-separation-scales} for an example with large extra dimensions (left figure) and small extra dimensions (right figure).
To guide the eye of the reader, we have also included the value of the exponential rate of the leading tower along each trajectory, represented by the red dotted curves. As we can see in the figure, the 't Hooft line depends on $\beta$ through \eqref{alphaN-scale-separation}. It is then clear that, for $\beta \neq 1$, the convex hull cuts the ball of radius $1/\sqrt{3}$, thus violating the Sharpened Distance Conjecture. Notice that violating this bound means that the leading tower of states would have an exponential rate smaller than $1/\sqrt{3}$ along some asymptotic trajectories, decaying at a slower rate than the perturbative critical string in a weakly curved background. To date, this has never occurred in top-down string examples.

In addition to this, we observe that the convex hulls in the perturbative field theory (i.e., $\lambda\to 0$) and supergravity (i.e., $\lambda\to\infty$) regions cannot glue continuously along the 't Hooft line anymore. This is because the line connecting the supergravity and perturbative field theory limits of the vector $\vec\zeta_{\rm str}$ (i.e. the line along which the sliding of the string tower should occur when varying $\lambda$) is no longer orthogonal to the 't Hooft line. Following the discussion at the end of Section \ref{sec:convex-hull-N=4}, we then conclude that this putative SCFT with $\beta \neq 1$ could not have a bona fide 't Hooft limit.\footnote{It can be confusing that the radial plot for $\alpha$ in the RHS of Figure \ref{fig:ch-separation-scales} is continuous, even if the convex hull is not. The reason for this is that the vector $\vec\zeta_{\rm str}^{\, \rm sugra}$ never gives the leading tower. Thus, the radial plot for $\alpha$ is blind to how this vector changes from this value to its $\lambda \to 0$ value.} 


Before going on, let us comment on the regime of validity of the supergravity approximation for the tower of higher-spin modes in this setup. On physical grounds, we expect our formula to hold whenever $T_s R_{AdS}^2 \gg 1$, i.e., when the string propagates in a weakly-curved AdS.
Following this logic, we should trust the scalar charge-to-mass ratio in \eqref{string-vector} only above the line of fixed $T_s R_{AdS}^2$. However, for $\beta \neq 1$, this line does not coincide with the 't Hooft line, which determines the regime of validity of the CFT prediction. In fact, for the most relevant case of $\beta <1$ (i.e. small extra dimensions), the region of validity for both \eqref{string-vector} and \eqref{HS-vectors} have some overlap, thus leading to a further inconsistency in this AdS/CFT model. This is nothing but another way of seeing that forcing separation of scales as we did leads to a sick model.
Limiting the regime of validity of the supergravity prediction to $\lambda \ll 1$  or shortening the regime of validity of the CFT to $T_s R_{AdS}^2 \ll 1$ does not help either, since it always results in a discontinuous convex hull and a violation of the Sharpened Distance Conjecture.

\medskip

With this we conclude that: 
\begin{center}
\emph{The bulk dual of 4d holographic SCFTs exhibiting a conformal manifold with an overall free limit cannot be scale separated unless the Sharpened Distance Conjecture is violated in the bulk or, equivalently, there is no bona fide 't Hooft limit in the boundary.}
\end{center}

This absence of scale separation is in agreement with the expectations from \cite{Polchinski:2009ch,Alday:2019qrf}. The only known 4d CFTs with conformal manifolds have at least 8 supercharges, and it is expected that this level of supersymmetry does not allow for scale separation in the bulk. The reason is that these theories have a continuous R-symmetry, and the conformal dimension of BPS charged operators is such that it would map (in case they are populated by physical states) to the mass of a non-scale separated KK tower. In fact, due to supersymmetry, we know that the Type 1 overall free limit that we have considered must lead to a theory with free vectors \cite{Perlmutter:2020buo} and should therefore have a sort of Lagrangian description in the weak-coupling limit. This facilitates to construct explicitly the BPS operators using the fields of the Lagrangian, which, upon ignoring possible subtleties, lead to operators with conformal dimension of order one and, therefore, to masses of order AdS.
Our approach, though, provides a different angle to signal the inconsistency of having scale separation in this class of holographic 4d SCFTs, without the need of constructing (or having information about) the BPS operators on the CFT side that map to the bulk KK modes. Our argument reveals that having a scale-separated AdS dual would imply a violation of the sharpened bound of the Distance conjecture and would seem incompatible with a well-defined 't Hooft limit in these theories.

On a different note, we remark that the absence of scale separation can also be related to a generalization of the Distance Conjecture in AdS spaces using a distance in the space of metric configurations. Here, though, we are not using such AdS distance but the usual field space distance associated with the kinetic term of the scalar fields. In that sense, this is the first time that a connection between results from the usual Distance Conjecture can be related to the absence of scale separation in AdS.

\section{Beyond CFTs with Simple Gauge Factors and Interpolating Models} \label{sec:beyond}

As a first step in the classification of infinite distance limits in 4d conformal manifolds, we have focused on SCFTs with a simple gauge group, such that there is a single gauge coupling that can be sent to zero. This mini-landscape of SCFTs is already rich enough to yield three types of free limits, which we argue are associated to three different tensionless strings in the bulk. The next step would be to consider more general SCFTs with several factors for the gauge group. This mimics the studies in the flat space compactifications, for which one first classifies single field limits and then move on to multi-dimensional field limits. We expect that, in analogy, our results can be used in the future as a sort of building blocks for the more general classification.

Even if this more general classification is left for future work, we can still use our results to extract some lessons about certain cases with more than one gauge factor. For instance, the result for the exponential rate $\alpha$ in terms of $a/c$ in \eqref{aac} holds generally as long as we take the limit in which the whole CFT becomes free (if it exists). In that sense, we expect that most of our results apply equally to SCFTs with more than one gauge factor, as long as all gauge couplings go to zero in that limit.
Hence, it would be interesting to extend our computation for the Hagedorn temperature to those cases. This generalization is conceptually doable, but challenging on the technical level. Such a computation was performed for $\mathcal N=2$ $SU(N)$ circular quivers in \cite{Larsen:2007bm}. These theories satisfy $a=c$ at large $N$ and thus feature a Type 1 limit with $\alpha = 1/\sqrt{2}$ in the overall free limit. The result is that the Hagedorn temperature coincide with that of $\mathcal N=4$ SYM (and thus of all the theories with Type 1 limit in our mini-landscape), which nicely fit with our results. Also befitting our proposal, these theories are known to have a Type IIB bulk dual \cite{Kachru:1998ys,Lawrence:1998ja}.

On the other hand, we do not expect our classification to capture limits in which there is a subsector of the theory that remains interacting (so not everything gets free in the limit). In those cases, we can obtain other possible values for the exponential rate of the tower of operators, which might have a different microscopic interpretation in the bulk. It would be interesting to further explore these partial decoupling limits in the future, as they can be used to connect different SCFTs with an interpolating model. For instance, $\mathcal N=2$ $SU(N)$ SQCD with $N_f=2N$ hypermultiplets in the fundamental representation (dubbed $\mathcal N=2$ SCQCD) can be obtained from taking the limit of the $SU(N)\times SU(N)$ circular quiver in which only one of the two gauge couplings go to zero \cite{Gadde:2009dj}. Following this limit is also useful to better understand the bulk dual of $\mathcal N=2$ SCQCD by starting from a holographic theory, as done in \cite{Gadde:2009dj}. Interestingly, we can use these results to also better understand the bulk description of the non-critical strings associated to the free limits in non-holographic theories discussed in this paper. In the rest of this section, we review the results of \cite{Gadde:2009dj} in our language, since we find them very illuminating to elucidate the microscopic description of the tower. In particular, it provides further support to identifying the higher-spin tower of operators in the Type 3 limit (with $\alpha=\sqrt{2/3}$) with a non-critical string in the bulk.

\medskip

Consider the $\mathcal N=2$ circular quiver theory with gauge group $SU(N)\times SU(N)$. This gauge theory is dual to the orbifold $AdS_5\times S^5/Z_2$. We have two marginal couplings associated to the two gauge couplings. There are two obvious types of infinite distance limits: sending both gauge couplings to zero, or only one goes to zero while the other remains constant.\footnote{As discussed in \cite{Aharony:2003sx,Baume:2020dqd}, the class-S description of this SCFT reveals another inequivalent weak-coupling limit at infinite distance: sending one of the gauge couplings to infinity while the other remains constant. We will not consider it further as it is nor relevant for the purposes of this work. The limit in which both gauge couplings are sent to infinity while keeping their ratio constant is also at infinite distance but equivalent to sending both gauge couplings to zero with constant ratio.} In both cases we will get a tower of higher-spin operators saturating the unitarity bound, but the exponential rate will be different. Recall from \eqref{alpha} that the exponential rate of the dual tower of higher-spin fields reads
\begin{equation} \label{eq:quiver-alpha}
  \alpha = \sqrt{\frac{2 c}{\text{dim}\,G_{\text{free}}}} = \sqrt{\frac{N^2}{ n_{v}^{\text{free}}}}  \, .
\end{equation}
Therefore, the two options\footnote{More generally, we get a HS tower with $\vec\nabla\log m=(1 ,0)$ and another one with  $\vec\nabla\log m=(0,1)$. Hence, the exponential rate is given by $\alpha=\vec\nabla\log m\cdot \hat t$ with $\hat t$ being the normalized tangent vector along the asymptotic trajectory within the two dimensional space spanned by $g_1$ and $g_2$. For $\hat t=(1,0)$ or $\hat t=(1,0)$ one gets $\alpha=1$, while $\alpha=1/\sqrt{2}$ for $\hat t=(1,1)/\sqrt{2}$. }  are:
\begin{itemize}
\item Both $g_1$ and $g_2$ go to zero with fixed ratio. We have $\alpha=1/\sqrt{2}.$
\item Only $g_1$ go to zero and $g_2$ remains constant (or viceversa). We have $\alpha=1$.
\end{itemize}
In the first limit, everything gets free and it corresponds to a Type 1 limit according to our notation in Section \ref{sec:types-limits}. The second option is a new type of limit, though. From the bulk perspective, they map to different limits on the string coupling and the NS axion. The holographic dictionary reads
\beq
\frac1{2\pi g_s}=\frac1{g_1^2}+\frac1{g_2^2}
\,  , \quad \frac{g_2^2}{g_1^2}=\frac{\beta}{1-\beta} \, ,
\eeq
with $\beta=\int_{S^2} B_2$ and $S^2$ being the blow-up of the $A_1$ orbifold singularity. The first limit simply maps to the weak-coupling limit $g_s\rightarrow 0$ as expected, while the second limit corresponds to sending both $g_s\rightarrow 0$ and $\beta\rightarrow 0$ such that $g_s/\beta$ remains fixed. In this second limit, the string coming from a D3-brane wrapping the blown-down 2-cycle gets tensionless at the same rate than the fundamental string. To elucidate the fate of the bulk theory in this limit, it is useful to study the T-dual of the Taub-Nut geometry associated to this $A_1$ orbifold singularity, following \cite{Gadde:2009dj}. The Type IIA T-dual involves two NS5-branes localized on the dual circle 
of radius $R$ and at a distance $\tau_0=2\pi\beta R$ from each other. T-duality also maps the $N$ D3 branes on the IIB side to two stacks of $N$ D4 branes on the IIA side, each stack ending on
the two NS5 branes and extended along either arc segment of the dual circle.

The dual IIA variables are related to the original CFT gauge couplings by
\beq
\frac1{g_1^2}=\frac{\beta R}{2\pi g_s^A l_s} \, , \quad \frac1{g_2^2}=\frac{(1-\beta) R}{2\pi g_s^A l_s} \, .
\eeq
It is well known that two coincident NS5-branes give rise to a strongly coupled near-horizon background in which the dilaton varies and blows up as the NS5's are approached. Upon decoupling gravity by sending $g_s^A\rightarrow 0$, this yields a strongly coupled theory known as little string theory (see e.g. \cite{Seiberg:1997zk,Ooguri:1995wj,Gregory:1997te,Aharony:1998ub}).  Here, however, we are interested in a double-scaling limit of little string theory (LST) \cite{Giveon:1999px,Giveon:1999tq}, in which one takes simultaneously 
\beq
\label{limitLST}
\tau_0=2\pi \beta R\rightarrow 0 \ , \quad g_s^A\rightarrow 0 \ ,\quad \text{with } \frac{\tau_0}{l_sg_s^A} \text{ fixed}
\eeq
and then send $l_s\rightarrow 0$. The limit \eqref{limitLST} precisely corresponds from the CFT side to taking $g_2\rightarrow 0$ with $g_1$ fixed (i.e. the second limit described below \eqref{eq:quiver-alpha}). 

Let us focus for a moment only on the closed string sector. In this limit, the near-horizon region decouples and is described by the non-critical string background of \cite{Giveon:1999px}. The CFT worldsheet is better described upon a further T-duality to the exact IIB background $\mathbb{R}^{5,1} \times SL(2)_2/U(1)/\mathbb{Z}_2$. The effective string coupling of the resulting theory is given by $g_s^{A}/\tau_0\sim g_1^2$, so there is still one parameter to play with, which corresponds to the leftover gauge coupling on the CFT side. Hence, the theory gets weakly coupled for large $\tau_0$, while it recovers the strongly coupled LST if $\tau_0$ becomes small at a faster rate. 

One can also follow the fate of the D4-branes through the T-duality and double-scaling limit; the branes suspended between the two NS5s become D3 branes localized at the
tip of the cigar, while the semi-infinite D4's become D5 branes extended on the cigar \cite{Gadde:2009dj}. The theory on the worldvolume of the $N$ D3 branes (the “color” branes) gets mapped holographically to $\mathcal N = 2$ $SU(N)$ SYM, coupled to $N_f = 2N$ hypermultiplets arising from the open strings stretched between the D3s and the “flavor” D5's. It was also argued in \cite{Gadde:2009dj} that the dual geometry is an eight dimensional background with only seven geometric dimensions (since one circle is of string size) containing an $AdS_5$ factor and a circle $S^1$ whose size is of order the AdS scale.\footnote{The bulk dual does not admit a low energy Einstein gravity description, since a fraction of the closed string sector of the interpolating theory becomes light in the double-scaling limit, so that the bulk dual of $\mathcal N = 2$ SCQCD contains light (closed string) higher-spin fields with mass of order AdS. In addition, there is a tower of heavy string states that only become light in the weak-coupling limit of $\mathcal N = 2$ SCQCD. The latter are precisely the tower of higher-spin fields that we have studied in this paper.} This sub-critical string background was then conjectured in \cite{Gadde:2009dj} to be dual to the flavor singlet sector of $\mathcal N = 2$ SCQCD. This has been studied in more detail in the very recent work of \cite{Dei:2024frl}.

Following this proposal, we can now see that the weak-coupling limit of $\mathcal N = 2$ SCQCD (which corresponds to a Type 3 limit according to our classification in Table \ref{table:summary}), corresponds to the weak-coupling / tensionless limit of this subcritical dual string. The results of our paper then suggest that the weak-coupling limit of all the other SCFTs which are in the same group than $\mathcal{N}=2$ SCQCD (i.e. that have a Type 3 limit with $\alpha=\sqrt{2/3}$ and, thus,  the same Hagedorn temperature) are also dual to the the tensionless limit of this same non-critical string.

\medskip

Finally, one could envision to learn about the non-critical string associated to a Type 2 limit in a similar way, i.e., by looking for an interpolating model that connects a holographic theory with one of the SCFTs in the second group of Table \ref{table:SCFTs} (i.e., with $\alpha=\sqrt{7/12}$). A promising candidate could be a $\mathcal N=1$ $SU(N)\times SU(N)$ gauge theory with two chirals in the adjoint of each $SU(N)$, one chiral in $(N,\bar N)$, and another one in $(\bar N, N)$. One can check that this theory indeed has $a=c$ at large $N$, so it could have a holographic Einstein gravity description in the bulk. By decoupling one of the gauge factors, the spectrum reduces to $\mathcal N=1$ $SU(N)$ with two chiral multiplets in the adjoint, $N$ in the fundamental, and $N$ in the antifundamental (third row of the second group in Table \ref{table:SCFTs}). A possible obstruction to using this as interpolating model is that $\mathcal N=1$ supersymmetry is not enough to guarantee that both gauge couplings are exactly marginal. Thus, perhaps it is not possible to decouple one of the gauge factors without moving away from the conformal manifold. In this sense, another interesting candidate is a $\mathcal N=2$ $SU(N-1) \times SU(N+1)$ gauge theory with hypermultiplets in the antisymmetric of $SU(N-1)$, the symmetric of $SU(N+1)$, and the bifundamental. This theory, which also satisfies $a=c$ at large $N$, recovers the first and second rows of the second group in Table \ref{table:SCFTs} when decoupling each of the two gauge factors. It would be interesting to check further whether any of these two theories can indeed be used as a valid interpolating model to learn about the bulk description of the non-critical string background arising after taking the free limit. We leave this for future work.

\section{Conclusions} \label{sec:conclusions}

In this work, we have initiated a classification of infinite distance limits in the conformal manifold of CFTs. We have focused on 4d Lagrangian\footnote{The only known non-compact conformal manifolds for $d\geq 2$ occur in 4d SCFTs, where the infinite distance limits correspond to weak-coupling limits in which some vector fields become free. Therefore, the theory has a Lagrangian description in the limit.} SCFTs with a non-compact complex one-dimensional conformal manifold. This requirement narrows down the setup to supersymmetric gauge theories with simple gauge group, which were classified in \cite{Bhardwaj:2013qia,Razamat:2020pra}. Consequently, there is an infinite distance limit in the conformal manifold in which the entire theory becomes free.  Since we are interested in the bulk dual description of these weak-coupling limits, we additionally require the theories to admit a large $N$ limit. We end up with the mini-landscape summarized in Table \ref{table:SCFTs}.

As shown in \cite{Baume:2020dqd,Perlmutter:2020buo} and encoded in the CFT Distance Conjecture \cite{Perlmutter:2020buo}, there is always an infinite tower of HS operators that saturate the unitarity bound at the free point and whose anomalous dimension behaves exponentially on the Zamolodchikov distance. Our classification is based on the parameter $\alpha$, which measures the exponential decay rate of the anomalous dimensions of the higher-spin conserved currents. In the bulk, these conserved currents are dual to the tower of states becoming light predicted by the Distance Conjecture \cite{Ooguri:2006in}. Using $\alpha$ to classify different infinite distance limits is motivated by recent results in the Swampland literature (see e.g. \cite{Etheredge:2022opl,Etheredge:2023odp,Calderon-Infante:2023ler,Etheredge:2023usk,Castellano:2023stg,Castellano:2023jjt,Etheredge:2024tok}) which suggest a correspondence between the value of $\alpha$ and the microscopic nature of the tower through its density of states.

Out of the 21 theories in our mini-landscape, only the 3 values values for $\alpha$ in \eqref{eq:three-alphas} arise at large $N$, as already found in \cite{Perlmutter:2020buo}. We denote them as Type 1, 2 and 3 infinite distance limits, and propose that they correspond to the tensionless limit of three different strings in AdS. Type 1 limits correspond to the tensionless limit of the ten-dimensional Type IIB critical string, while Type 2 and 3 limits seem to be associated to non-critical strings. We provide different pieces of evidence for this result. 

First, we showed that there is a one-to-one correspondence between the value of $\alpha$ and the ratio of the central charges for these theories. This automatically reveals that theories with Type 1 limit have $a=c$ at large $N$ and are therefore holographic. This is consistent with the fact that most of the theories in our mini-landscape with this type of limit are known to admit a Type IIB description in the bulk (see e.g. \cite{Ennes:2000fu}). On the other hand, Type 2 or Type 3 limits cannot be holographic, as they have $a\neq c$ at large $N$.  As a consequence, these putative two new string backgrounds in AdS do not admit a low energy description in terms of weakly-coupled Einstein gravity, which we expect to be due to the existence of additional low-lying higher-spin excitations. 
One concrete example with a Type 3 limit is $\mathcal N=2$ SCQCD, whose dual string background was studied in \cite{Gadde:2009dj}. It was proposed there that the dual description is a non-critical string background yielding a (non-Einstein) eight-dimensional gravitational theory containing an AdS$_5$ factor. Our results imply that all the other SCFTs in the same universality class than $\mathcal N=2$ SCQCD (i.e. exhibiting a Type 3 limit)  also contain the same dual non-critical closed string. 

To further support our proposal,  we studied the large $N$ partition function of these theories at the free point to test the link between the value of $\alpha$ (characterizing the type of limit) and the behaviour of the density of states. We extend previous results in the literature (see e.g. \cite{Sundborg:1999ue,Aharony:2003sx}) by considering any simple gauge group and including matter in all possible representations up to two indices. We showed that UV complete gauge theories with simple gauge group exhibit a Hagedorn-like density of states at large $N$, as befit the spectrum of excitations of a string. The exponential growth of the density of states with the energy is controlled by the so-called Hagedorn temperature. Quite remarkably, we find a universal result for the Hagedorn temperature which only depends on the value of $\alpha$ for the theories in our mini-landscape (and consequently, on the ratio of the central charges). This suggests that the three types of limits characterized by the value of $\alpha$ indeed correspond to the tensionless limit of three different types of bulk strings, providing support to our proposal. Intriguingly, this one-to-one correspondence between $T_H$ and $\alpha$ extends to any supersymmetric gauge theory with simple gauge group, large $N$, and vanishing one-loop beta function; a total of 70 theories according to the classifications in \cite{Bhardwaj:2013qia,Razamat:2020pra}. For those in our mini-landscape (i.e. having a conformal manifold), we also check that this one-to-one correspondence holds even after restricting to the flavor singlet sector of the theory to attain a sparse spectrum at large $N$; both if we  consider the flavor symmetry preserved at the free point or beyond the free point for $\mathcal{N}=2$ theories. 

\medskip

For those theories that are holographic, we also performed a more global analysis of the asymptotic field space of the bulk dual. We studied the behaviour of the towers of states in the $N$-direction (which is not part of the conformal manifold)  and compared the results with the gravity expectations in Section \ref{sec:convex-hulls}. From the bulk side, the exponential mass decay rate of the tower of oscillator modes of the critical string is $1/\sqrt{3}$ in a weakly curved background. We showed how to connect this value with the $1/\sqrt{2}$ found along the conformal manifold.  For this, it is important to notice two things. First, the usual emergent string limit studied in flat space compactifications --for which KK and string excitation modes become light at the same rate-- does not belong to the conformal manifold. In fact, we see that it corresponds to the 't Hooft limit, i.e., $N\to\infty$ with $\lambda=g_{\text{YM}}^2N$ fixed. Secondly, the supergravity approximation breaks down when the tension of the string becomes of order the AdS scale. In this regime, the string propagates in a highly-curved AdS, which changes the exponential decay rate of its excitation modes as predicted by the CFT. The exponential rate along its gradient direction becomes $\sqrt{8/15}$ and reproduces the value of $1/\sqrt{2}$ when projected on the conformal manifold.

Using these results, we built the first convex hull for the Distance Conjecture in AdS/CFT in Section \ref{sec:convex-hull-N=4}, taking into account the HS and the KK towers. Remarkably, the sharpened bound for $\alpha$ put forward in \cite{Etheredge:2022opl} (namely, $\alpha\geq 1/\sqrt{3}$ in five dimensions) holds rather non-trivially even in the strongly curved AdS regime. This result highly depends on the precise way in which the conformal dimension of the higher-spin fields changes from the weakly-curved supergravity regime (large 't Hooft coupling) to the field theory perturbative regime (small 't Hooft coupling). For $\mathcal N=4$ SYM, we used integrability results to follow this variation and show that the sharpened lower bound for the exponential rate is always satisfied. Intriguingly, the main features of this transition seem to be guaranteed by having a bona fide 't Hooft limit, which guarantees that the convex hulls of the towers in these two regimes glue nicely together.

As an interesting spinoff, in Section \ref{sec:scale-separation} we used our results to reveal a tantalizing connection between the absence of scale separation between the AdS and the internal dimensions in the bulk and the sharpened bound for the Distance Conjecture. We illustrated this for holographic 4d CFTs with a conformal manifold and a Type 1 infinite distance limit, showing the the AdS$_5$ bulk dual cannot be scale separated unless the sharpened lower bound for the exponential rate of the tower gets violated. Furthermore, we argued that, if these AdS/CFT pairs were to allow for scale separation, the CFT could not have a bona fide 't Hooft limit either.

\medskip

Our work leaves several interesting questions for future research:

\begin{itemize}
	\item It would be interesting to extend the classification of infinite distance limits beyond our mini-landscape of theories. The most natural next step would be to consider gauge theories with multiple gauge factors. We hope that our results for one dimensional conformal manifolds can be used as building blocks for the higher dimensional case. The other possible generalization would be to consider theories without a large $N$ limit, but then we would lose the gravity interpretation.
	
	\item The one-to-one correspondence between the ratio of central charges and $\alpha$ holds along the overall free limit of any 4d SCFT, even if it has multiple gauge factors. For such limit, we expect that our classification still holds, so it would be interesting to check this expectation by extending the computation of the large $N$ Hagedorn temperature to those cases. Despite the anticipated technical difficulties, a good starting point for this seems to be the classification of $\mathcal N=2$ theories in \cite{Bhardwaj:2013qia}.
	
	\item As remarked in Section \ref{sec:partition-function-intro}, our results for the large $N$ partition function can be used to easily determine the supersymmetric index of these theories and study their BPS spectrum. An interesting question in this context is whether non-holographic theories with $a \neq c$ at large $N$ always feature low-lying BPS higher-spin operators, as found for $\mathcal N=2$ SCQCD in \cite{Gadde:2009dj}.
	
	\item When studying the behavior of the towers of operators with $N$, we introduced the parameter $\alpha_N$ to define some flat coordinates and fixed it by matching with a canonically normalized volume field in the bulk. It would be interesting to be able to fix the value of this parameter purely from the CFT perspective. This would allow to extend our results to other theories, including those with Type 2 and 3 infinite distance limits, which are not holographic. It could also allow us to generalize and strengthen the connection between no separation of scales and the Sharpened Distance Conjecture in Section \ref{sec:scale-separation}.

    \item To learn more about the non-critical string associated to Type 2 limits, it would be useful to find a holographic interpolating model for one of the SCFTs with $\alpha=\sqrt{7/12}$ in Table \ref{table:SCFTs}. We provided a couple of promising candidates with $a=c$ at large $N$ in Section \ref{sec:beyond}, which would be interesting to explore further. The goal would be to find the Type IIB holographic dual of these interpolating models and follow the decoupling limit in the bulk associated to sending one of the boundary gauge couplings to zero. This way, one could try to learn about the resulting non-critical string background, as was done in \cite{Gadde:2009dj} for $\mathcal N=2$ SCQCD (which features instead a Type 3 limit).
	
	\item Perhaps the most fascinating proposal in this paper is the existence of two types of non-critical closed strings in AdS, which characterize the bulk dual of all large $N$ 4d SCFTs with $a\neq c$ and an overall free limit. It would be very exciting to understand them better from top-down and build their worldsheet description.
    This is a difficult task due to the well-known complications related to quantizing a string in AdS background. We hope that recent developments in the understanding of tensionless strings in AdS (see e.g. \cite{Gaberdiel:2018rqv,Eberhardt:2018ouy,Eberhardt:2019ywk,Gaberdiel:2021qbb,Gaberdiel:2021jrv,Gaberdiel:2022iot}) will help in this endeavor.
\end{itemize}

On more general grounds, this work is yet another instance of the fruitful intersection between the Swampland and the bootstrap programs. Studying properties of conformal manifolds has already been quite successful for learning about the Distance Conjecture in AdS/CFT \cite{Baume:2020dqd,Perlmutter:2020buo,Baume:2023msm,Ooguri:2024ofs}. This work shows that the opposite is also possible: Swampland ideas can be used to motivate and uncover new CFT results! It was our experience with the Distance Conjecture in string compactifications what motivated us to look for common properties of the CFTs with the same $\alpha$ at large $N$ in our mini-landscape. Consequently, we uncovered that these theories also share the same ratio between central charges and Hagedorn temperature at large $N$. This suggests that the bulk dual of these three families of SCFTs are described by three different strings in AdS!

All in all, there is much to learn from and about the Distance Conjecture in AdS/CFT. We hope that this work makes a good case for this and encourages new exciting results.

\section*{Acknowledgments}

We would like to thank Jorge Luis Dasilva Gol\'an, Hee-Cheol Kim, Shota Komatsu, Miguel Montero, Kyriakos Papadodimas, Elli Pomoni, Cumrun Vafa, Max Wiesner and Timo Weigand for useful discussions. We also thank the Erwin Schr\"odinger International Institute
for Mathematics and Physics for their hospitality during the programme ``The Landscape vs. the Swampland''. The work of I.V. is supported by the Spanish Agencia Estatal de Investigaci\'on through the grant ``IFT Centro de Excelencia Severo Ochoa'' CEX2020-001007-S, the grant PID2021-123017NB-I00, funded by MCIN/AEI/10.13039/ 501100011033 and by ERDF A way of making Europe, the grant RYC2019-028512-I from the MCI (Spain) and the ERC Starting Grant QGuide101042568 - StG 2021.

\appendix

\section{Single-trace vs Multi-trace Partition Function} \label{app:single-vs-multi}

In Section \ref{sec:Hagedorn}, we used the large $N$ partition function in the free limit to detect the exponential density of states that we associate with the string becoming tensionless. Nevertheless, to compare with the spectrum of excitations of a string, one would like to only consider the minimum energy state of each single particle state. Via AdS/CFT, these are dual to single-trace primary operators at large $N$. For this reason, one might worry that the partition function in \eqref{partion-function} also counts descendants and multi-trace operators. In this appendix we show that restricting to single-trace primary operators yields the same Hagedorn temperature $T_H$, thus leaving our results unchanged. In other words, we argue that only single-trace primary operators can be responsible for the exponential degeneracy at high energies.

As explained in \cite{Aharony:2003sx}, given the single-trace partition function, the full multi-trace partition function is obtained by using the plethystic exponential
\begin{equation}
  Z(x) = \exp \left( \sum_{n=1}^{\infty} \frac{1}{n} Z_{s.t.}(x^n) \right) \, .
\end{equation}
This uses that the multi-trace states are the Fock space of single-trace states, which behave as identical bosonic particles. This equation can be inverted by using the plethystic logarithm (see e.g. \cite{Feng:2007ur}):
\begin{equation} \label{plethystic-logarithm}
  Z_{s.t}(x) = \sum_{n=1}^{\infty} \frac{\mu(n)}{n} \log \left( Z(x^n) \right) \, .
\end{equation}
The single-trace partition function clearly diverges whenever $Z(x^n)$ does. Since $Z(x)$ is monotonically increasing with $x$, which is upper-bounded by one, the first of these divergences as we increase $x$ happens when the $n=1$ term diverges, thus giving the same Hagedorn temperature as for the full partition function. One could entertain the possibility of having an earlier divergence due the sum in \eqref{plethystic-logarithm} not converging. However, this would imply that the density of single-trace states is larger than the total density of states, which is not possible. This shows that the exponential growth of states cannot be due to multi-traces.

Similarly, excluding the descendants cannot change the exponential behavior of the density of states. The reason is that, given a primary, the tower of descendants does not lead to such exponential density of states, but to a polynomial one. Thus, the exponential degeneracy of states in the full spectrum must be due to an exponential number of primaries.

Let us remark that excluding both multi-traces and/or descendants from the density of states will indeed change the exponential law in \eqref{Hagedorn-density} by a polynomial prefactor. As pointed out in Footnote \ref{footnote1}, this type of prefactor are irrelevant for our considerations.

\section{Large $N$ Partition Function with Orthogonality Relations} \label{app:orthogonality-relations}

In this appendix, we re-derive the large $N$ partition function \eqref{SU-partition-function} in the particular case $\tilde z_{F} = \tilde z_{\bar F}=0$ with the method of the orthogonal polynomials. The main observation is that the symmetric polynomials $p_N(u)$ obey the orthogonality relations \cite{Dolan:2007rq,Dolan:2008qi}
\begin{equation}
  \int_{SU(N)} d\mu(u) \, \prod_{n\geq 1} p_{N}(u^{n})^{a_n} p_{N}(u^{-n})^{b_n} = \prod_{n\geq 1} n^{a_n} a_n! \ \delta_{a_n b_n} \, , \quad \ \sum_{n\geq 1} n\, a_n , \sum_{n\geq 1} n\, b_n < N \, .
\end{equation}
In the large $N$ limit, we can ignore the condition on the right and use these relations for any powers $a_n$ and $b_n$.

We now use this to solve the matrix integral for $SU(N)$ when $\tilde z_{F} = \tilde z_{\bar F}=0$. In this case, \eqref{eq:integral-simplified} simplifies to
\begin{equation}
\begin{split}
	  \sum_{n=1}^{\infty} \frac{1}{n} \left( \sum_{R} z_R(x,n) \, \chi_R(u^n) \right) = \sum_{n=1}^{\infty} \frac{1}{n} &\bigg( z_{Ad}(x,n) \Big[ p_N(u^{n}) p_N(u^{-n}) - 1 \Big] \\
	  & + z_{AS}(x,n) \Big[ p_N(u^{n})^2 + p_N(u^{-n})^2 \Big] \bigg) \, .
\end{split}
\end{equation}
Thus, the integrand of the matrix model reads
\begin{equation}
\begin{split}
	  \prod_{n\geq 1} \exp \left\{ \frac{1}{n} \left( \sum_{R} z_R \, \chi_R(u^n) \right) \right\} = \prod_{n\geq 1} &\exp \left\{ - \frac{1}{n} z_{Ad} \right\} \exp \left\{ \frac{1}{n} z_{Ad} \, p_N(u^{n}) p_N(u^{-n}) \right\} \\ &\exp \left\{ \frac{1}{n} z_{AS} \, p_N(u^{n})^2 \right\} \exp \left\{ \frac{1}{n} z_{AS} \, p_N(u^{-n})^2 \right\} 
\end{split}
\end{equation}
Expanding the three last exponentials, this yields
\begin{equation}
  \prod_{n\geq 1} \exp \left\{ -\frac{1}{n} z_{Ad} \right\} \sum_{a_n,b_n,\bar b_n \geq 0} \frac{z_{Ad}^{a_n} \, z_{AS}^{b_n + \bar b_n}}{n^{a_n+b_n+\bar b_n} \, a_n! \, b_n! \, \bar b_n !} \, p_{N}(u^{n})^{a_n+2b_n} p_{N}(u^{-n})^{a_n+2\bar b_n} \, .
\end{equation}

In the large $N$ limit, we use the orthogonality relations above to integrate this expression. This automatically allows us to solve the sum over $\bar b_n$ by simply imposing $\bar b_n = b_n$. After doing so, we end up with
\begin{equation}
  Z(x) \simeq \prod_{n\geq 1} \exp \left\{- \frac{1}{n} z_{Ad} \right\} \sum_{a_n,b_n \geq 0} \frac{(a_n + 2b_n)!}{a_n! \, (b_n!)^2 } \, z_{Ad}^{a_n} \, z_{AS}^{2 b_n} \, .
\end{equation}

For the rest of the derivation, the strategy is to recombine the two left-over sums into known functions. For the sum over $a_n$ we use
\begin{equation}
  (a_n+2b_n)!={a_n+2b_n \choose a_n} a_n! \, (2b_n)! \ , \qquad \sum_{r\geq 0} {r+s \choose s } \, x^r = \frac{1}{(1-x)^{s+1}} \, ,
\end{equation}
yielding
\begin{equation}
  Z(x) \simeq \prod_{n\geq 1} \exp \left\{ -\frac{1}{n} z_{Ad} \right\} \sum_{b_n\geq 0} \frac{(2b_n)!}{(b_n!)^2} \frac{z_{AS}^{2b_n}}{(1-z_{Ad})^{2b_n+1}} \, .
\end{equation}
Similarly, for the last sum we use
\begin{equation}
	{2b_n \choose b_n} = \frac{(2b_n)!}{(b_n!)^2} \, , \qquad \sum_{r\geq 0} {2r \choose r} x^{2r} = \frac{1}{\sqrt{1-4x^2}} \, ,
\end{equation}
to finally arrive to
\begin{equation}
  Z(x) \simeq \prod_{n\geq 1} \exp \left\{- \frac{1}{n} z_{Ad} \right\} \frac{1}{\sqrt{(1-z_{Ad})^2 - 4 z_{AS}^{2}}} \, .
\end{equation}
As advertised, this recovers the result in \eqref{SU-partition-function} for $\tilde z_{F} = \tilde z_{\bar F}=0$.

\section{Flavor Singlet Partition Functions} \label{app:flavor-singlet-computations}

In this appendix, we compute the various large $N$ partition functions with the restriction to flavor singlets used in the main text.

\subsection{Free flavor singlet restriction for $SU(N)$ theories} \label{app:SU-singlets}

To restrict a free $SU(N)$ gauge theory to its flavor singlet sector, we introduce chemical potentials for the flavor charges and integrate over the flavor symmetry group. We can take the large $N$ partition function in \eqref{SU-partition-function} --for which the integral over the gauge group has been already performed-- as starting point. 

As explained in Section \ref{sec:flavor-singlets}, we first plug \eqref{SU-flavor-replacement} into \eqref{SU-partition-function}. Taking into account the definitions in \eqref{z-effectives}, this yields the flavored partition function
\begin{equation} \label{SU-Z-flavored}
\begin{split}
  Z(x,u,w)\simeq \prod_{n=1}^{\infty} \exp& \bigg\{ \frac{1}{n} \left( y_1 - z_{Ad} \right) \bigg\} \ \frac{1}{\sqrt{(1-z_{Ad})^2-4z_{AS}^{2}}} \\
  \exp \bigg\{ \frac{1}{n} &\bigg( y_{AS} \big( (\rho_n + \sigma_n)^2 + (\bar \rho_n + \bar \sigma_n)^2 \big) \\ 
  & + y_{Ad} (\rho_n + \sigma_n ) ( \bar\rho_n + \bar\sigma_n ) + y_F ( \rho_n + \sigma_n + \bar\rho_n + \bar\sigma_n ) \bigg) \bigg\} \, ,
\end{split}
\end{equation}
where we have defined
\begin{equation} \label{y-defs}
\begin{array} {llr}
	y_{AS} = \dfrac{1}{4} \dfrac{z_{AS}\,z_c^{2}}{(1-z_{Ad})^2-4z_{AS}^{2}} \, , & \quad \quad &
	y_{Ad} = \dfrac{1}{4} \dfrac{(1-z_{Ad})\,z_c^{2}}{(1-z_{Ad})^2-4z_{AS}^{2}} \, , \\ \\
	y_{F} = \dfrac{1}{2} \dfrac{(1-z_{Ad}+2z_{AS})\, z_{\delta F} \,z_c}{(1-z_{Ad})^2-4z_{AS}^{2}} \, , & \quad \quad &
	y_1 = \dfrac{(1-z_{Ad}+2z_{AS})\, z_{\delta F}^2 }{(1-z_{Ad})^2-4z_{AS}^{2}} \, .
\end{array}
\end{equation}
To ease notation, and anticipating the large $n_{F}$ and $n_{\bar F}$ limit, we have already replaced $p_N(v^n) \to \rho_n$ and $p_N(w^n)\to \sigma_n$ with $\rho$ and $\sigma$ being the density of eigenvalues for $v$ and $w$, respectively (c.f. equation \eqref{angles}-\eqref{largeN-replacements}).

To get the partition function restricted  to flavor singlets, we now have to integrate over the flavor group $SU(n_F)\times SU(n_{\bar{F}})$ flavor group. In the large $n_F$ and $n_{\bar F}$ limit, we do this using the density of eigenvalues already introduced above. As we learned in Section \ref{sec:SU-case}, the integration measures introduce an extra factor
\begin{equation}
   \exp \left\{ \sum_{n=1}^{\infty} \frac{1}{n} \bigg( \rho_n \bar\rho_n + \sigma_n \bar\sigma_n \bigg) \right\} 
\end{equation}
in the integrand. Similarly, the integral over $v$ and $w$ becomes (c.f. equation \eqref{replacement-integral-SU})
\begin{equation}
  \int d[\rho] d[\sigma]  = \prod_{n=1}^{\infty} \frac{1}{n^2 \, \pi^2} \int d^2\rho_n d^2\sigma_n \, .
\end{equation}
All this leads to the following expression for the flavor singlets partition function:
\begin{equation} \label{s.t.-Z}
\begin{gathered}
	Z_{f.s.}(x) \simeq \prod_{n=1}^{\infty} \exp \bigg\{ \frac{1}{n} \left( y_1 - z_{Ad} \right) \bigg\} \, \frac{1}{\sqrt{(1-z_{Ad})^2-4z_{AS}^{2}}} \\ 
	\times \frac{1}{n^2 \, \pi^2} \int d^2\rho_n d^2\sigma_n \, e^{-S_{\text{eff}}} \, ,
\end{gathered}
\end{equation}
where the effective action reads
\begin{equation}
\begin{split}
	S_{\text{eff}} = \frac{1}{n} &\bigg( \rho_n \bar\rho_n + \sigma_n \bar\sigma_n - y_{AS} \big( (\rho_n + \sigma_n)^2 + (\bar \rho_n + \bar \sigma_n)^2 \big) \\
	&- y_{Ad} (\rho_n + \sigma_n ) ( \bar\rho_n + \bar\sigma_n ) - y_F ( \rho_n + \sigma_n + \bar\rho_n + \bar\sigma_n ) \bigg) \, .
\end{split}
\end{equation}
The integral can be shown to yield
\begin{equation}
  \frac{1}{n^2 \, \pi^2} \int d^2\rho_n d^2\sigma_n \, e^{-S_{\text{eff}}} = \exp \left\{ \frac{1}{m} \frac{2 y_{F}^{2}}{1-2y_{Ad}-4y_{AS}} \right\} \, \frac{1}{\sqrt{(1-y_{Ad})^2-16y_{AS}^{2}}} \, .
\end{equation}
Plugging this back into \eqref{s.t.-Z}, unwrapping the definitions in \eqref{y-defs}, and simplifying, we finally get
\begin{equation}
\begin{split}
  Z_{f.s.}(x) \simeq & \exp \bigg\{ \sum_{n=1}^{\infty} \frac{1}{n} \bigg( \frac{z_{\delta F}^{2}}{1-z_{Ad}-2z_{AS}-\frac{1}{2} z_{c}^{2}} -z_{Ad} \bigg) \bigg\} \\
  & \prod_{n=1}^{\infty} \frac{1}{\sqrt{\left( 1-z_{Ad}-\frac{1}{2} z_c^{2}\right)^2 - 4 z_{AS}^{2}}}
\end{split}
\end{equation}
Notice that this partition function does no longer blow up in the $N\to\infty$ limit, as long as $n_{Ad}$, $n_{S}$ and $n_{A}$ stay order one. This shows that restricting to the flavor singlet sector indeed leads to a sparse spectrum in the large $N$ limit.

As used in the main text, we see that this partition function blows up for $x_H$ satisfying
\begin{equation}
  z_{Ad}(x_H,n) + z_{S}(x_H,n) + z_{A}(x_H,n) + \frac{1}{2} z_c(x_H,n)^{2} = 1 \, ,
\end{equation}
where we have already plugged the definition of $z_{AS}$ in \eqref{z-effectives}.

\subsection{Free flavor singlet restriction for $USp(2N)$ and $SO(2N)$ theories} \label{app:USp-SO-singlets}

In this section, we deal with the restriction to the flavor singlet sector for theories with $USp(2N)$ and $SO(2N)$ gauge groups. As we did for $SU(N)$ and described in \ref{sec:flavor-singlets} we introduce chemical potentials for the flavor charges and integrate over the flavor group. The starting point are the large $N$ partition functions in \eqref{USp-partition-function} and \eqref{SO-partition-function}. Notice that both of them take the form
\begin{equation} \label{USp/SO-partition-function}
  Z(x) \simeq \exp \left\{ \sum_{n=1}^{\infty} \frac{1}{n} \left( \frac{ \tilde z_F^2 }{ 2\left( 1-z_S-z_A \right) } -z_{1} \right) \right\} \ \prod_{n=1}^{\infty} \frac{1}{\sqrt{1-z_{S}-z_{A}}} \, ,
\end{equation}
with $z_1 = z_A$ for $USp(2N)$ and $z_1 = z_S$ for $SO(2N)$.

As described in the main text, we introduce the chemical potentials by doing the replacement in \eqref{z-flavor-USp-SO}. Upon also using the definitions in \eqref{eff-zF-USp} and \eqref{eff-zF-SO} --that only differ in the form of $z_{\delta F}$-- we get
\begin{equation} \label{USp/SO-Z-flavored}
\begin{split}
  Z(x,u,w)\simeq \prod_{n=1}^{\infty} \exp& \bigg\{ \frac{1}{n} \left( y_1 - z_{1} \right) \bigg\} \ \prod_{n=1}^{\infty} \frac{1}{\sqrt{1-z_{S}-z_{A}}} \\
  \exp \bigg\{ \frac{1}{n} &\bigg( y_{Ad} \, \rho_n \bar\rho_n + y_{AS} ( \rho_n^2 + \bar\rho_n^2 ) + y_F ( \rho_n + \bar\rho_n ) \bigg) \bigg\} \, ,
\end{split}
\end{equation}
where we have defined
\begin{equation} \label{y-defs-2}
\begin{array} {lcr}
	y_{Ad} = \dfrac{z_c^{2}}{4(1-z_{S}-z_{A})} \, , & \quad &
	y_{AS} = \dfrac{z_c^{2}}{8(1-z_{S}-z_{A})} \, , \\ \\
	y_{F} = \dfrac{z_c \, z_{\delta F}}{2(1-z_{S}-z_{A})} \, , & \quad &
	y_1 = \dfrac{z_{\delta F}^{2}}{2(1-z_{S}-z_{A})} \, .
\end{array}
\end{equation}
To ease notation, and anticipating the large $n_{F}$ limit, we have already replaced $p_N(v^n) \to \rho_n$ with $\rho$ the density of eigenvalues for $v$ as in equations \eqref{angles}-\eqref{largeN-replacements}.

We now integrate over the $SU(n_F)$ flavor group. In the large $n_F$ limit, we do this again following the method in Section \ref{sec:SU-case}. The integration measure introduces the extra factor
\begin{equation}
   \exp \left\{ \sum_{n=1}^{\infty} \frac{1}{n} \, \rho_n \bar\rho_n  \right\} 
\end{equation}
in the integrand and the integral over $v$ becomes (c.f. equation \eqref{replacement-integral-SU})
\begin{equation}
  \int d[\rho]  = \prod_{n=1}^{\infty} \frac{1}{n \, \pi} \int d^2\rho_n \, .
\end{equation}
All this leads to the following expression for the flavor singlets partition function:
\begin{equation} \label{s.t.-Z-2}
\begin{gathered}
	Z_{f.s.}(x) \simeq \prod_{n=1}^{\infty} \exp \bigg\{ \frac{1}{n} \left( y_1 - z_{1} \right) \bigg\} \ \prod_{n=1}^{\infty} \frac{1}{\sqrt{1-z_{S}-z_{A}}} \\ 
	\times \frac{1}{n \, \pi} \int d^2\rho_n \, e^{-S_{\text{eff}}} \, ,
\end{gathered}
\end{equation}
where the effective action reads
\begin{equation}
	S_{\text{eff}} = \frac{1}{n} \bigg( (1-y_{Ad}) \, \rho_n \bar\rho_n - y_{AS} ( \rho_n^2 + \bar\rho_n^2 ) - y_F ( \rho_n + \bar\rho_n ) \bigg) \, .
\end{equation}
The integral can be solved explicitly, yielding
\begin{equation}
  \frac{1}{n \, \pi} \int d^2\rho_n \, e^{-S_{\text{eff}}} = \exp \left\{ \frac{1}{n} \frac{y_{F}^{2}}{1-y_{Ad}-2y_{AS}} \right\} \, \frac{1}{\sqrt{(1-y_{Ad})^2-4 y_{AS}^{2}}} \, .
\end{equation}
Notice that the integral is completely analogous to that in \eqref{SU-Z-integral}. The only modifications are setting replacing $z_{Ad}\to y_{Ad}$, $z_{AS}\to y_{AS}$, $\tilde z_{F},\tilde z_{\bar F} \to y_{F}$ and ignoring the $\rho_n$-independent term that we have already brought outside of the integral. Taking this into account, one can check that the result obtained here then agrees with that in \eqref{SU-partition-function}.

Plugging this back into \eqref{s.t.-Z-2}, unwrapping the definitions in \eqref{y-defs-2}, and simplifying, we finally get
\begin{equation}
\begin{split}
  Z_{f.s.}(x) \simeq & \exp \bigg\{ \sum_{n=1}^{\infty} \frac{1}{n} \bigg( \frac{1}{2} \frac{z_{\delta F}^{2}}{1-z_{S}-z_{A}-\frac{1}{2} z_{c}^{2}} - z_{1} \bigg) \bigg\} \\
  & \prod_{n=1}^{\infty} \frac{1}{ \sqrt{ 1-z_{S}-z_{A}-\frac{1}{2} z_c^{2} }}
\end{split}
\end{equation}
Notice that this partition function does no longer blow up in the $N\to\infty$ limit, as long as $n_{S}$ and $n_{A}$ stay order one. This shows that restricting to the flavor singlet sector indeed leads to a sparse spectrum in the large $N$ limit.

Finally, we see that this partition function blows up for $x_H$ satisfying
\begin{equation}
  z_S(x_H,n) + z_A(x_H,n) + \frac{1}{2} z_c(x_H,n)^{2} = 1 \, .
\end{equation}
as used in the main text.

\subsection{$\mathcal N =2$ flavor singlet restriction for $SU(N)$ theories} \label{app:N=2-singlets}

In this section, we restrict the $\mathcal N=2$ $SU(N)$ theories to the singlet sector of the flavor symmetry that is preserved at any point in the conformal manifold. The computations are totally analogous to those in the two previous sections, but using \eqref{N=2-flavor-replacement} to introduce the chemical potentials into the partition function in \eqref{SU-partition-function}.

Also taking into account the definitions in \eqref{z-effectives}, this yields the flavored partition function
\begin{equation} \label{SU-Z-flavored-2}
\begin{split}
	  Z(x,u,w)\simeq \prod_{n=1}^{\infty} \exp& \bigg\{ \frac{1}{n} \left( y_1 - z_{Ad} \right) \bigg\} \ \prod_{n=1}^{\infty} \frac{1}{\sqrt{(1-z_{Ad})^{2}-4z_{AS}^{2}}} \\
  \exp \bigg\{ \frac{1}{n} &\bigg( y_{Ad} \, \rho_n \bar\rho_n + y_{AS} ( \rho_n^2 + \bar\rho_n^2 ) + y_F ( \rho_n + \bar\rho_n ) \bigg) \bigg\} \, ,
\end{split}
\end{equation}
where we have defined
\begin{equation} \label{y-defs-3}
\begin{array} {lcr}
	y_{Ad} = \dfrac{(1-z_{Ad})\, z_c^{2}}{(1-z_{Ad})^{2}-4z_{AS}^{2}} \, , & \quad &
	y_{AS} = \dfrac{z_{AS}\, z_c^{2}}{(1-z_{Ad})^{2}-4z_{AS}^{2}} \, , \\ \\
	y_{F} = \dfrac{(1-z_{Ad}+2z_{AS}) \, z_{\delta F} \, z_c}{(1-z_{Ad})^{2}-4z_{AS}^{2}} \, , & \quad &
	y_1 = \dfrac{(1-z_{Ad}+2z_{AS}) \,z_{\delta F}^{2}}{(1-z_{Ad})^{2}-4z_{AS}^{2}} \, .
\end{array}
\end{equation}
To ease notation, and anticipating the large $n_{F}$ limit, we have already replaced $p_N(v^n) \to \rho_n$ with $\rho$ the density of eigenvalues for $v$ as in equations \eqref{angles}-\eqref{largeN-replacements}.

In the large $n_F$ limit, we integrate over the $SU(n_F)$ flavor group using the method in Section \ref{sec:SU-case}. The integration measure introduces the extra factor (c.f. equation \eqref{measure-SU})
\begin{equation}
   \exp \left\{ -\sum_{n=1}^{\infty} \frac{1}{n} \, \rho_n \bar\rho_n  \right\} 
\end{equation}
in the integrand and the integral over $v$ becomes (c.f. equation \eqref{replacement-integral-SU})
\begin{equation}
  \int d[\rho]  = \prod_{n=1}^{\infty} \frac{1}{n \, \pi} \int d^2\rho_n \, .
\end{equation}
All this leads to the following expression for the flavor singlets partition function:
\begin{equation} \label{s.t.-Z-3}
\begin{gathered}
	Z_{f.s.}(x) \simeq \prod_{n=1}^{\infty} \exp \bigg\{ \frac{1}{n} \left( y_1 - z_{Ad} \right) \bigg\} \ \prod_{n=1}^{\infty} \frac{1}{\sqrt{(1-z_{Ad})^{2}-4z_{AS}^{2}}} \\ 
	\times \frac{1}{n \, \pi} \int d^2\rho_n \, e^{-S_{\text{eff}}} \, ,
\end{gathered}
\end{equation}
where the effective action reads
\begin{equation}
	S_{\text{eff}} = \frac{1}{n} \bigg( (1-y_{Ad}) \, \rho_n \bar\rho_n - y_{AS} ( \rho_n^2 + \bar\rho_n^2 ) - y_F ( \rho_n + \bar\rho_n ) \bigg) \, .
\end{equation}
This is precisely the same integral as in the previous section. We recall the result here:
\begin{equation}
  \frac{1}{n \, \pi} \int d^2\rho_n \, e^{-S_{\text{eff}}} = \exp \left\{ \frac{1}{n} \frac{y_{F}^{2}}{1-y_{Ad}-2y_{AS}} \right\} \, \frac{1}{\sqrt{(1-y_{Ad})^2-4 y_{AS}^{2}}} \, .
\end{equation}
Plugging this back into \eqref{s.t.-Z-3} and unwrapping the definitions in \eqref{y-defs-3} we finally get
\begin{equation}
\begin{split}
  Z_{f.s.}(x) \simeq & \exp \bigg\{ - \sum_{n=1}^{\infty} \frac{1}{n} \bigg( \frac{(1-z_{Ad}-2z_{AS}-2z_{c}^{2}) \, z_{\delta F}^{2}}{ (1-z_{Ad}-2 z_{AS}) (1-z_{Ad}-2z_{AS}-z_{c}^{2}) } + z_{Ad} \bigg) \bigg\} \\
  & \prod_{n=1}^{\infty} \frac{1}{ \sqrt{ (1-z_{Ad}- z_c^{2})^{2}-4z_{AS}^{2} }}
\end{split}
\end{equation}
Notice that this partition function does no longer blow up in the $N\to\infty$ limit, as long as $n_{Ad}$, $n_{S}$ and $n_{A}$ stay order one. Thus, this flavor singlet sector is sparse spectrum in the large $N$ limit.

As used in the main text, we see that this partition function blows up for $x_H$ satisfying
\begin{equation}
  z_{Ad}(x_H,n) + z_{S}(x_H,n) + z_{A}(x_H,n) + z_c(x_H,n)^{2} = 1 \, ,
\end{equation}
where we have already plugged the definition of $z_{AS}$ in \eqref{z-effectives}. Indeed, imposing the vanishing of the beta function and rewriting everything in terms of $\alpha$ as in Section \ref{sec:SU-case}, this reduces to \eqref{SU-N=2-prescription}.

Finally, if we restrict to a theory with $n_S=n_A=0$, and thus with $z_{AS}=z_{\delta F}=0$, the partition function simplifies to
\begin{equation}
  Z_{f.s.}(x) \simeq  \exp \bigg\{ - \sum_{n=1}^{\infty} \frac{z_{Ad}}{n}  \bigg\} \, \prod_{n=1}^{\infty} \frac{1}{1-z_{Ad}- z_c^{2}} \, .
\end{equation}
As also mentioned in Section \ref{sec:flavor-singlets}, this partition function has exactly the same functional form as the superconformal index of the flavor singlet sector of $\mathcal N=2$ SCQCD computed in \cite{Gadde:2009dj}.

\subsection{$\mathcal N =2$ flavor singlet restriction for $USp(N)$ and $SO(2N)$ theories} \label{app:N=2-singlets-USp-SO}

In this section, we restrict the $\mathcal N=2$ $USp(2N)$ and $SO(2N)$ theories to the singlet sector of the flavor symmetry that is preserved at any point in the conformal manifold. The computations are totally analogous to those in sections \ref{sec:USp-case} and \ref{sec:SO-case}.

The large $N$ partition function for $USp(2N)$ and $SO(2N)$ theories can be written as
\begin{equation}
  Z(x) \simeq \exp \left\{ \sum_{n=1}^{\infty} \frac{1}{n} \left( \frac{ z_F^2 + 2 z_{\delta F} z_F + z_{\delta F}^2 }{ 2\left( 1-z_S-z_A \right) } -z_{1} \right) \right\} \ \prod_{n=1}^{\infty} \frac{1}{\sqrt{1-z_{S}-z_{A}}} \, ,
\end{equation}
where we have replaced the definition in \eqref{eff-zF-USp} (or \eqref{eff-zF-SO}) and defined
\begin{equation}
    z_{1}(x,n) = 
    \begin{cases} 
    z_{A}(x,n) & \text{for } USp(2N) \, , \\
    z_{S}(x,n) & \text{for } SO(2N) \, .
    \end{cases}
\end{equation}
Further replacing \eqref{SO-USp-N=2-prescription}, the flavored partition function reads
\begin{equation}
\begin{split}
	  Z(x,v)\simeq \prod_{n=1}^{\infty} \exp& \bigg\{ \frac{1}{n} \left( y_1 - z_{1} \right) \bigg\} \ \prod_{n=1}^{\infty} \frac{1}{\sqrt{1-z_{S}-z_{A}}} \\
  \exp \bigg\{ \frac{1}{n} &\bigg( 2 \, y_{AS}\, \rho_n^2 + 2\, y_F\, \rho_n \bigg) \bigg\} \, ,
\end{split}
\end{equation}
where we have defined
\begin{equation} \label{y-defs-4}
	y_{AS} =  \frac{z_c^2}{1-z_{S}-z_{A}}  \, , \quad
	y_{F} =  \frac{z_c \, z_{\delta F}}{1-z_{S}-z_{A}}  \, , \quad
	y_1 =  \frac{z_{\delta F}^2}{2(1-z_{S}-z_{A}))}  \, .
\end{equation}
Anticipating the large $n_F$ limit, we have replaced $p_{n_F/2}(v^n) \to \rho_n$ following \eqref{angles}-\eqref{largeN-replacements}, already taking into account that $\rho_n = \bar\rho_n$.

To integrate over the flavor groups, we introduce the measures in \eqref{measure-USp} and \eqref{measure-SO}), which we rewrite as
\begin{equation}
    \exp \left\{ -\sum_{n=1}^{\infty} \frac{1}{n} \left( 2 \, \rho_n^2 + 2\,  y_{\delta F}\, \rho_n \right)  \right\} \, .
\end{equation}
For this, we have defined
\begin{equation}
    y_{\delta F}(x,n) = 
    \begin{cases} 
    0 & \text{if } n \text{ odd} \, , \\
    \mp 1 & \text{if } n \text{ even} \, ,
    \end{cases}
\end{equation}
with the minus sign for the $SO(n_F)$ flavor group of the $USp(2N)$ theory and the plus sign for the $USp(n_F)$ flavor group of the $SO(2N)$ one. In both cases the integral over $v$ becomes (c.f. equation \eqref{replacement-integral-USp} or \eqref{replacement-integral-SO}) 
\begin{equation}
    \int d[\rho] = \prod_{n=1}^{\infty} \sqrt{\frac{2}{n \, \pi}} \int_{-\infty}^{\infty} d\rho_n \, .
\end{equation}
Introducing all this, in the $n_F\to \infty$ limit the flavor singlets partition function takes the form
\begin{equation} \label{s.t.-Z-4}
\begin{gathered}
	Z_{f.s.}(x) \simeq \prod_{n=1}^{\infty} \exp \bigg\{ \frac{1}{n} \left( y_1 - z_{1} \right) \bigg\} \ \prod_{n=1}^{\infty} \frac{1}{\sqrt{1-z_{S}-z_{A}}} \\ 
	\times \sqrt{\frac{2}{n \, \pi}} \int_{-\infty}^{\infty} d\rho_n \, e^{-S_{\text{eff}}} \, ,
\end{gathered}
\end{equation}
where the effective action reads
\begin{equation}
	S_{\text{eff}} = \frac{1}{n} \bigg( 2(1-y_{AS}) \,\rho_n^2 + 2\, \tilde y_{F}\, \rho_n \bigg) \, ,
\end{equation}
and, to shorten notation, we have defined $\tilde y_{F} = y_F + y_{\delta F}$. The integral can be solved directly, yielding
\begin{equation}
    \sqrt{\frac{2}{n \, \pi}} \int_{-\infty}^{\infty} d\rho_n \, e^{-S_{\text{eff}}} = \exp \left\{ \frac{1}{n} \frac{\tilde y_{F}^{2}}{2(1-y_{AS})} \right\} \, \frac{1}{\sqrt{1- y_{AS}}}
\end{equation}
Plugging this result back into \eqref{s.t.-Z-4} and unwrapping the definitions in \eqref{y-defs-4} we finally get
\begin{equation}
\begin{split}
  Z_{f.s.}(x) \simeq & \exp \bigg\{ \sum_{n=1}^{\infty} \frac{1}{n} \bigg( \frac{ z_{\delta F}^2 + 2 \, z_c \, z_{\delta F} \, y_{\delta F} + (1-z_S-z_A) y_{\delta F}^2 }{ 2 (1-z_S-z_A-z_c^2) } - z_{1} \bigg) \bigg\} \\
  & \prod_{n=1}^{\infty} \frac{1}{ \sqrt{ 1-z_S-z_A-z_c^2 }}
\end{split}
\end{equation}
As expected, this flavor singlet partition function does no longer blow up as $N\to\infty$, as long as $n_{S}$ and $n_{A}$ stay order one, thus achieving a sparse spectrum in this limit.

This flavor singlet partition function blows up for $x_H$ such that
\begin{equation}
    z_S(x_H,n)+z_A(x_H,n)+z_c(x_H,n)^2 = 1\, .
\end{equation}
Imposing the vanishing of the beta function and writing everything in terms of $\alpha$ as in sections \ref{sec:USp-case} and \ref{sec:SO-case}, this reduces to the result used in the main text, i.e., equation \eqref{USp-SO-N=2-prescription}.

\bibliographystyle{JHEP}
\bibliography{references}
\end{document}